\documentclass[11pt,a4paper]{article}

\usepackage[utf8]{inputenc}
\usepackage[margin=1in]{geometry}
\usepackage{authblk}
\usepackage[style=apa,uniquename=false,backend=biber]{biblatex}
\usepackage[american]{babel}
\usepackage{csquotes}
\DeclareLanguageMapping{american}{american-apa}
\bibliography{bibliography.bib,Alexander.bib}

\usepackage{amsmath,amsfonts,amssymb,bm,bbm,amsthm}
\usepackage[colorinlistoftodos,textsize=small]{todonotes}
\usepackage{graphicx}
\usepackage{todonotes}
\usepackage{xcolor} % color math symbols, delete later
\usepackage{verbatim}

\usepackage{nicefrac}

\theoremstyle{definition} % no italiced theorem style
\newtheorem{theorem}{Theorem}[section]
\newtheorem{prop}{Proposition}

\newtheorem{lemma}[prop]{Lemma}

\newtheoremstyle{case}{}{}{}{}{}{:}{ }{}
\theoremstyle{case}

\usepackage[colorinlistoftodos,textsize=small]{todonotes}
\usepackage[colorlinks=true, allcolors=blue, bookmarks=false]{hyperref}

\usepackage{listings}
\lstdefinestyle{mystyle}{
  language=R,                     % the language of the code
  basicstyle=\ttfamily,       % the size of the fonts that are used for the code
  numbers=none,                   % where to put the line-numbers
  backgroundcolor=\color{lightgray},  % choose the background color. You must
  columns=fullflexible,
  showspaces=false,               % show spaces adding particular underscores
  showstringspaces=false,         % underline spaces within strings
  showtabs=false,                 % show tabs within strings adding particular underscores
  tabsize=2,                      % sets default tabsize to 2 spaces
  breaklines=true,                % sets automatic line breaking
  keywordstyle=\color{blue},      % keyword style
  commentstyle=\color{darkgray},   % comment style
  alsoletter={._},
  otherkeywords={!,!=,~,$,*,\&,\%/\%,\%*\%,\%\%,<-,<<-,/}
}

\newcommand{\dx}[1]{\enspace \mathrm{d}{#1}}
\newcommand{\prior}[1]{\pi\left({#1}\right)}

\newcommand{\FBeta}[2]{\Bc\left({#1},\ {#2}\right)}

\newcommand{\FHypGeo}[4]{\,_2F_1\left({#1},{#2};{#3};{#4}\right)}
\newcommand{\vecB}[1]{\boldsymbol{#1}}

\newcommand{\simplex}[1]{\mathbb{S}^{#1}}

\newcommand*\samethanks[1][\value{footnote}]{\footnotemark[#1]}

\newcommand{\numberthis}{\addtocounter{equation}{1}\tag{\theequation}}
\newcommand{\FD}[1]{\todo[inline, color=pink]{ \textbf{FD}: #1 }}
\newcommand{\DB}[1]{\todo[inline, color=orange]{ \textbf{DB}: #1 }}

\newcommand{\revise}{\textcolor{black}}
\newcommand{\refEq}[1]{Eq.~{\eqref{#1}}}

\newcommand{\refSec}[1]{Section~{\ref{#1}}}
\newcommand{\refLem}[1]{Lemma~{\ref{#1}}}

\newcommand{\refThm}[1]{Theorem~{\ref{#1}}}

\newcommand{\refApp}[1]{Appendix~{\ref{#1}}}

\newcommand{\bibOrder}[1]{}

\newcommand{\field}[1]{\mathbb{#1}} % the font for a mathematical field is blackboard
\newcommand{\R}{\field{R}} % the field of the reals
\newcommand{\N}{\field{N}} % the field of the natural numbers
\newcommand{\Bc}{\mathcal{B}}

\newcommand{\Hc}{\mathcal{H}}

\newcommand{\Mc}{\mathcal{M}}

\newcommand{\Nc}{\mathcal{N}}

\newcommand{\Oc}{\mathcal{O}}

\newcommand{\Pf}{\mathbb{P}}

\newcommand{\Uc}{\mathcal{U}}

\newcommand{\der}{\mathrm{d}}

\newcommand{\iid}{\textnormal{iid}}

\newcommand{\Beta}{\textnormal{Beta}}

\newcommand{\Dir}{\textnormal{Dir}}

\newcommand{\inDist}{\overset{\textnormal{d}}{\rightarrow}}

\newcommand{\iidSim}{\overset{\iid}{\sim}}

\newcommand{\la}{\langle} % Inner product: left
\newcommand{\ra}{\rangle} % Inner product: right

\newcommand{\var}{\textnormal{Var}}

\newcommand{\BF}{\textnormal{BF}}

\newcommand{\iiFi}[4]{ \ensuremath{ \, {_{2} F}_{1} \left ( #1, #2 \, ; \, #3 \, ; \,  #4 \right )}}

\definecolor{nRed}{RGB}{220,0,12}
\definecolor{nOrange}{RGB}{254,166,0}
\definecolor{nBlue}{RGB}{40,96,180}

\date{}
\title{Default Bayes Factors for Testing the (In)equality of Several Population Variances}
\author[1]{Fabian Dablander\thanks{These authors share first authorship.}}
\author[1]{Don van den Bergh\samethanks[1]}
\author[1]{Eric-Jan Wagenmakers}
\author[1,2]{Alexander Ly}
\affil[1]{Department of Psychological Methods, University of Amsterdam}
\affil[2]{Centrum Wiskunde \& Informatica}

\begin{document}
\maketitle

\begin{abstract}
\noindent Testing the (in)equality of variances is an important problem in many statistical applications. We develop default Bayes factor tests to assess the (in)equality of two or more population variances, as well as a test for whether the population variances equal a specific value. The resulting test can be used to check assumptions for commonly used procedures such as the \( t \)-test or ANOVA, or test substantive hypotheses concerning variances directly. We show that our Bayes factor fulfills a number of desiderata. Researchers may have directed hypotheses such as \( \sigma_{1}^{2} > \sigma_{2}^{2} \), \revise{they may} want to extend \( \mathcal{H}_{0} \) to have a null-region, or \revise{wish} to combine hypotheses about equality with hypotheses about inequality, for example \( \sigma_{1}^{2} = \sigma_{2}^{2} > (\sigma_{3}^{2}, \sigma_{4}^{2}) \). We extend our Bayes factor test to allow for these deviations from our proposed default and illustrate it on a number of practical examples. Our procedure is implemented in the R package \textit{bfvartest}.
\end{abstract}

%\tableofcontents

\section{Introduction}
Testing the (in)equality of variances is important in many sciences and applied contexts. In engineering, for example, researchers may want to assess whether a new, cheaper measurement instrument achieves the same precision as the gold standard \parencite{sholts2011comparison}. In genetics and medicine, scientists are not only interested in studying the genetic effect on the mean of a quantitative trait, but also on its variance \parencite{pare2010use}. In economics and archeology, ideas such as that increased economic production should reduce variability in products directly lead to statistical hypotheses on variances \parencite{kvamme1996alternative}. In a court of law, one may be interested in reducing unwanted variability in civil damage awards and may want to compare how different interventions reduce this variability \parencite{saks1997reducing}. In psychology, educational researchers may be interested in studying how the variance in pupil's mathematical ability changes across school grades \parencite{aunola2004developmental}.

While there exist several classical \( p \)-value tests for assessing the (in)equality of population variances \parencite[e.g.,][]{levene1961robust, brown1974robust, gastwirth2009impact}, testing such hypotheses has received little attention from a Bayesian perspective. Such a perspective, however, would offer practitioners the possibility (a) to quantify evidence in favor of the null hypothesis \parencite[e.g.,][]{morey2016philosophy}, (b) allow one to incorporate prior knowledge \parencite[e.g.,][]{o2006uncertain}, (c) to use sequential sampling designs which in many cases is more cost-effective \parencite[e.g., than a fixed-\( N \) design, see][]{stefan2019tutorial}, and (d) to translate \revise{substantive} predictions more easily into statistical hypotheses by specifying equality and inequality constraints \parencite[e.g.,][]{boing2018automatic,hoijtink2008bayesian}.

In light of these benefits and recent recommendations to go beyond \( p \)-value testing \parencite{wasserstein2016asa}, we develop default Bayes factor tests \parencite[e.g.,][]{consonni2018prior, jeffreys1939theory, ly2016harold, ly2016evaluation} for the (in)equality of several population variances. Our work is inspired by \textcite[][pp. 222-224]{jeffreys1939theory}, who developed a test for the ``agreement of two standard errors''. Equipped with our procedure, researchers are able to state graded evidence both for the case of testing assumptions of other tests (e.g., the equality of variances assumption in the Student's \( t \)-test), as well as testing order-constrained hypotheses on variances directly.

%This paper is structured as follows. In the first part, we derive a default prior for the \( K = 2 \) group case and discuss sensible choices for the scale of the prior. \revise{We describe a one-sample test that follows directly from our two-sample test, and we illustrate our procedure on three real-world examples, extending it to allow order-constrained (directed) and interval null hypotheses. In the second part, we generalize the Bayes factor to \( K > 2 \) groups, propose an efficient procedure to evaluate (in)equality constraints based on bridge sampling \parencite[e.g.,][]{meng1996simulating, gronau2017tutorial}, and compare our method to a fractional Bayes factor procedure proposed by \textcite{boing2018automatic}.} We apply the \( K > 2 \) method to two data sets from archeology and educational psychology. All derivations and proofs are given in the appendix.
%
%

This paper is structured as follows. In \refSec{secNotation}, we introduce the problem setup and propose the default Bayes factor. In \refSec{secPropertiesBf}, we elaborate on the desiderata that the proposed Bayes factor adheres to. In \refSec{secSpecialCases}, we discuss the special case with \( K=2 \) groups, including directed and interval Bayes factors, and compare our method to a fractional Bayes factor procedure proposed by \textcite{boing2018automatic}. We illustrate our default Bayes factor test and deviations from it on a number of practical examples in \refSec{secExamplesKIsTwo}. We conclude in \refSec{secConclusion}. All derivations and proofs can be found in the appendix.

\section{Default Bayes Factor for $K$ Groups}
\label{secNotation}
\subsection{Notation and Problem Setup}
The problem of testing the (in)equality of variances can be equivalently expressed in terms of variances \( \sigma_{j}^{2} \) or precisions \( \tau_{j} = \sigma_{j}^{-2} \). For the data we assume that \( Y_{ji} \overset{\textnormal{iid}}{\sim} \mathcal{N} ( \mu_{j} , \tau_{j}^{-1}) \), where \( i \in [n_{j}] \) and \( j \in [K] \) with the rectangular brackets embracing an integer denoting the set of positive integers up to and including that integer, e.g., \( [K] := \{ 1, 2, \ldots, K-1, K \} \subset \N \).

As the \( K \) groups are assumed to be independent of each other, the data \( y^{[K]} \) can be sufficiently summarized by the sample means \( \vecB{\bar{y}} = ( \bar{y}_{1}, \ldots, \bar{y}_{K}) \), where \( \bar{y}_{j} = \frac{1}{n_{j}} \sum_{i=1}^{n_{j}} y_{ji} \) and the (unbiased) sample variances \( \vecB{s^{2}} = (s_{1}^{2}, \ldots, s_{K}^{2}) \), where \( s_{j}^{2} = \frac{1}{\nu_{j}} \sum_{i=1}^{n} (y_{ji} - \bar{y}_{j})^{2} \) and where \( \nu_{j} = n_{j} - 1 \) is the degree of freedom of group \( j \). As a convention, we denote \( K \)-dimensional vectors in bold, whereas an arrow is used to denote a \( K-1 \) dimensional vector, e.g., \( \vecB{s^{2}} = ( \vec{s^{2}}, s_{K}^{2}) \). A subscript \( + \) is used to denote summation over the vector's elements, e.g., \( \vecB{\tau}_{+} = \sum_{j=1}^{K} \tau_{j} \), whereas \( \vec{\vartheta}_{+} = \sum_{j=1}^{K-1} \vartheta_{j} \), since \( \vec{\vartheta} \in \R^{K-1} \).

The null hypothesis $\mathcal{H}_0$ states that all precisions are the same, while the alternative hypothesis $\mathcal{H}_1$ includes at least one inequality. Formally, we compare
\begin{align}
\label{eqProblem0}
\Hc_{0}: \tau_{j}  &= \tau_{k} \text{ for all } j, k \in [K], \\
\Hc_{1}: \tau_{j}  &\neq \tau_{k} \text{ for some } j \neq k \in [K] ,
\end{align}
regardless of the nuisance parameters \( \vecB{\mu} = (\mu_{1}, \mu_{2}, \ldots, \mu_{K}) \in \R^{K} \). The null hypothesis restricts the \( K \) precisions to a single but unknown precision, whereas the alternative allows all precisions to vary freely. Including the means, the null model has \( K+1 \) free parameters, whereas the alternative model has \( 2 K \) free parameters.

We rephrase the model comparison by generalizing the reparametrization proposed by \textcite[][pp. 222-224]{jeffreys1939theory}; see also \refApp{sec:jeffreys-parameterization}. More specifically, in the alternative model we reparametrize the \( K \) precisions \( \vecB{\tau} \) in terms of an average precision \( \bar{\vecB{\tau}} = \tfrac{1}{K} \vecB{\tau}_{+} \) and \( K-1 \) proportions \( \vec{\vartheta}  \) with \( \vartheta_{j} = \tfrac{\tau_{j}}{ \vecB{\tau}_{+}} \). Note that this reparametrization is invertible as it should be. In this parametrization the hypotheses translate into %
%
%the model comparison of interest translates into comparing %null and alternative hypotheses imply that
%
\begin{align}
\label{eqProblem}
\Hc_{0}: \vartheta_{j}  &= \tfrac{1}{K} \text{ for all } j \in [K-1], \\
\Hc_{1}: \vartheta_{j}  &\neq \tfrac{1}{K} \text{ for some } j  \in [K-1] ,
\end{align}
regardless of the values of the nuisance parameter \( \vecB{\mu} \in \R^{K} \) and the average precision \( \bar{\vecB{\tau}} > 0 \), which are common to both models.

From a Bayesian perspective, we assess the relative merits of $\mathcal{H}_0$ and $\mathcal{H}_1$ by virtue of how well they predict the data, that is, by their respective marginal likelihoods. The ratio of marginal likelihoods is known as the Bayes factor \parencite{kass1995bayes}, and \revise{its specification} requires assigning priors to \revise{both the free parameters of the null and the alternative model}. For the models being compared this implies one prior on the \( 2 K \) free parameters of the alternative model, and another prior on the \( K+1 \) free parameters of the null model. To simplify matters, we mimic the nesting of the null model into the alternative model and choose \( \pi_{1}(\vecB{\mu}, \bar{\vecB{\tau}}, \vec{\vartheta}) = \pi_{0}(\vecB{\mu}, \bar{\vecB{\tau}}) \pi_{1}(\vec{\vartheta}) \). The Bayes factor we propose is constructed from a right Haar prior \( \pi_{0}(\vecB{\mu}, \bar{\vecB{\tau}}) \propto \bar{\vecB{\tau}}^{-1} \) on the common parameters and from a (proper) Dirichlet prior \( \pi_{1}(\vec{\vartheta}) \) on the test-relevant parameters \( \vec{\vartheta} \) with hyperparameters \( \vecB{u} \), where \( u_{j} > 0 \) for all \( j \in [K] \). %

In the remainder of this section we show that this choice of priors results in a Bayes factor that is analytic. In \refSec{secPropertiesBf} we show that the proposed Bayes factor fulfills certain Bayesian model comparison desiderata.

\subsection{The Proposed Bayes Factor}
The choice for \( \pi_{0}(\vecB{\mu}, \bar{\vecB{\tau}}) \propto \bar{\vecB{\tau}}^{-1} \) is based on the observation that the hypotheses to be tested are invariant under (1) scalar multiplications of all the data points, and (2) location shifts of the data points of each sample/group.%
\footnote{The nesting \( \pi_{1}(\vecB{\mu}, \bar{\vecB{\tau}}, \vec{\vartheta}) = \pi_{0}(\vecB{\mu}, \bar{\vecB{\tau}}) \pi_{1}(\vec{\vartheta}) \) makes the use of the improper priors \( \pi_{0}(\vecB{\mu}, \bar{\vecB{\tau}}) \propto \bar{\vecB{\tau}}^{-1} \) permissible as a limit of proper priors with normalization constants cancelling due to their appearances in both the numerator and denominator of the Bayes factor \parencite[see also][]{hendriksen2021optional, ly2016evaluation, robert2016expected}.} %
The derivations in \refApp{appDerivationBf} show that with \( \pi_{0}(\vecB{\mu}, \bar{\vecB{\tau}}) \propto \bar{\vecB{\tau}}^{-1} \) on the nuisance parameters, the Bayes factor simplifies to
\begin{align}
\label{eqBfGeneral}
\BF_{10}(y^{[K]}) & =\frac{\int \limits_{\Theta} \left ( \int \limits_{\R_{>0}} \int \limits_{ \R^{K} } f( y^{[K]} \, | \, \vecB{\mu}, \bar{\vecB{\tau}}, \vec{\vartheta}) \pi_{0}(\vecB{\mu}, \bar{\vecB{\tau}}) \der \vecB{\mu} \der \bar{\vecB{\tau}} \right )  \pi_{1}(\vec{\vartheta}) \der \vec{\vartheta}}{ \int \limits_{\R_{>0}} \int \limits_{ \R^{K} } f( y^{[K]} \, | \, \vecB{\mu}, \bar{\vecB{\tau}}, \vec{\vartheta} = \tfrac{1}{K}) \pi_{0}(\vecB{\mu}, \bar{\vecB{\tau}}) \der \vecB{\mu} \der \bar{\vecB{\tau}} } = \int_{\Theta} h( \vecB{s^{2}} \, | \, \vec{\vartheta}) \pi_{1}(\vec{\vartheta}) \der \vec{\vartheta} ,
\end{align}
where \( \R_{>0} \) denotes the positive reals, \( \Theta := \{ \vec{\theta} \in \R^{K-1} \, | \, \vec{\theta}_{+} < 1 \} \subset \R_{>0}^{K-1} \), and where we refer to \( h( \vecB{s^{2}} \, | \, \vec{\vartheta}) \) as the reduced likelihood, which is given by
\begin{align}
\label{eqReducedLikelihood}
h( \vecB{s^{2}} \, | \, \vec{\vartheta}) := \Big ( 1 + \sum_{j=1}^{K-1}  \tfrac{\nu_{j} s_{j}^{2}}{\nu_{K} s_{K}^{2}} \Big )^{\tfrac{\vecB{\nu}_{+}}{2}}  \Big [ \prod_{j=1}^{K-1} \vartheta_{j}^{\tfrac{\nu_{j}}{2}} \Big ] (1 - \vec{\vartheta}_{+} )^{\tfrac{\nu_{K}}{2}} \Big ( 1 - \sum_{j=1}^{K-1} [ 1 - \tfrac{\nu_{j} s_{j}^{2}}{\nu_{K} s_{K}^{2}}] \vartheta_{j} \Big )^{- \tfrac{\vecB{\nu}_{+}}{2}},
\end{align}
where \( \vecB{\nu}_{+} = \sum_{j=1}^{K} \nu_{j} \), and \( \vec{\vartheta}_{+} := \sum_{j=1}^{K-1} \vartheta_{j} \). Note that, for any proper prior \( \pi_{1}(\vec{\vartheta}) \), the nesting and the choice \( \pi_{0}(\vecB{\mu}, \bar{\vecB{\tau}}) \propto \bar{\vecB{\tau}}^{-1} \) leads to a measurement invariant Bayes factor, as desired. This is because \( h( \vecB{s^{2}} \, | \, \vec{\vartheta}) \) and therefore \( \BF_{10}(y^{[K]}) = \BF_{10}( \vecB{s^{2}}) \) only depend on the data via the ratios of sums of squares \( \tfrac{ \nu_{j} s_{j}^{2}}{ \nu_{K} s_{K}^{2}} \), and because each \( s_{k}^{2} \) is invariant under location shifts within sample/group \( k \).

The Dirichlet prior \( \pi_{1}(\vec{\vartheta}) \) on the test-relevant parameters is inspired by the form of \( h( \vecB{s^{2}} \, | \, \vec{\vartheta}) \) and makes the proposed Bayes factor analytic. By definition of the integral form of the type D Lauricella function, the proposed Bayes factor is %
\begin{align}
\label{eqBfMulti}
\BF_{10}( \vecB{s^{2}} ) & =  \frac{\Bc ( \tfrac{\vecB{\nu}}{2}  + \vecB{u}) }{\Bc ( \vecB{u})} \Big ( 1 + \sum_{j=1}^{K-1}  \tfrac{\nu_{j} s_{j}^{2}}{\nu_{K} s_{K}^{2}} \Big )^{\tfrac{\vecB{\nu}_{+}}{2}}  F_{D} \Big ( \tfrac{\vecB{\nu}_{+}}{2} \, ; \, \tfrac{ \vec{\nu}}{2} + \vec{u} \, ; \, \tfrac{\vecB{\nu}_{+}}{2} + \vecB{u}_{+}  \, ; \, \vec{1} - \tfrac{ \overrightarrow{\nu s^{2}} }{ \nu_{K} s_{K}^{2}} \Big ) ,
\end{align}
where \( \Bc ( \vecB{u}) = \frac{ \Gamma( u_{1}) \cdots \Gamma(u_{K}) }{ \Gamma ( u_{+} ) }  \) is the multivariate beta function, \( \vec{1}= ( 1, \ldots, 1) \in \R^{K-1} \), \( \overrightarrow{\nu s^{2}} = (\nu_{1} s_{1}^{2}, \ldots, \nu_{K-1} s_{K-1}^{2}) \) is the \( K-1 \) vector of sums of squares, and where \( F_{D} \) is a type D Lauricella function which has the integral representation $F_{D}(a \, ; \, \vec{b} \, ; \, d \, ; \, \vec{x} ) = \frac{\Gamma\left(d\right)}{\Gamma\left(a\right)\Gamma\left(d - a\right)} \int_{0}^{1} t^{a-1} (1 - t)^{d-a-1} (1 - x_{1}t)^{-b_{1}} \cdots (1- x_{K-1} t)^{-b_{K-1}} \der t$ whenever $d > a$, which holds trivially since $u > 0$ always.
Observe that, with \refEq{eqBfMulti} at hand, we also have an analytic marginal posterior for \( \vec{\vartheta} \), namely,
\begin{align}
\label{eqPosteriorTheta}
\pi_{1} ( \vec{\vartheta} \, | \, y^{[K]}) = \frac{ \Big [ \prod_{j=1}^{K-1} \vartheta_{j}^{\tfrac{\nu_{j}}{2}} \Big ] (1 - \vec{\vartheta}_{+} )^{\tfrac{\nu_{K}}{2}} \Big ( 1 - \sum_{j=1}^{K-1} [ 1 - \tfrac{\nu_{j} s_{j}^{2}}{\nu_{K} s_{K}^{2}}] \vartheta_{j} \Big )^{- \tfrac{\vecB{\nu}_{+}}{2}}}{ \Bc ( \tfrac{\vecB{\nu}}{2}  + \vecB{u})  F_{D} \Big ( \tfrac{\vecB{\nu}_{+}}{2} \, ; \, \tfrac{ \vec{\nu}}{2} + \vec{u} \, ; \, \tfrac{\vecB{\nu}_{+}}{2} + \vecB{u}_{+}  \, ; \, \vec{1} - \tfrac{ \overrightarrow{\nu s^{2}} }{ \nu_{K} s_{K}^{2}} \Big )} .
\end{align}
The proposed Bayes factor can be computed from the sample variances and sample sizes directly. This makes it possible to re-evaluate the published literature without the need to have access to the raw data, as shown in \refSec{secExamplesKIsTwo}. In the next section, we show that the proposed Bayes factor fulfills a number of desiderata.

\section{Properties of the Proposed Bayes Factor}
\label{secPropertiesBf}
An important result of this paper is that our proposed Bayes factor fulfills a number of desiderata \parencite{bayarri2012criteria, consonni2018prior,jeffreys1939theory, ly2016harold,ly2016evaluation}. More specifically, we show that the proposed Bayes factor has the finite-sample properties of being (i) labelling invariant, (ii) (exactly) predictively matched, and (iii) information consistent. It also has the asymptotic properties of being (iv) model selection consistent and (v) limit and across-sample consistent. Information consistency requires \( u_{j} \leq 1/2 \) for \( j \in [K] \) while labelling invariance requires \( u_{i} = u_{j} \) for all \( i, j \in [K] \), suggesting the default choice of \( u_{j}=1/2 \) for all \( j \in [K] \).

%\refThm{thmLabellingInvariance} states that choosing
%\refThm{thmPredictiveMatching} below states that for (i) the right Haar priors on \( \mu_{j} \propto 1 \) for \( j \in [K] \), \( \bar{\vecB{\tau}} \propto \bar{\vecB{\tau}}^{-1} \) and a proper prior on \( \vec{\vartheta} \) suffices. \refThm{thmInformationConsistent} states that for (ii) a Dirichlet prior with parameters \( \vecB{u} = (u_{1}, \ldots, u_{K}) \) with \( u_{j} \leq 1/2 \) for \( j \in [K] \) suffices. \refThm{thmModelSelectionConsistency} and \refThm{thmAcrossSampleConsistency} requires the sample variance \( S_{k}^{2} \) to be \( \sqrt{n}_{k} \)-consistent, and we show the rate of convergence with the dependence on \( K \) made explicit. Labelling invariance combined with these results suggest the default choice \( u_{j}=1/2 \) for all \( j \in [K] \).

\subsection{Labelling Invariance}
A Bayes factor is labelling invariant if it is independent of the arbitrary choice of which group is labelled \( K \).

\begin{theorem}[Labelling invariance] \label{thmLabellingInvariance}
The proposed Bayes factor with \( u_{i} = u_{j} \) for all \( i, j \in [K] \) is labelling invariant. \( \hfill \diamond \)
\end{theorem}

\begin{proof}
See \refApp{appLabellingInvariant}.
\end{proof}

\subsection{Predictive Matching}
A Bayes factor is (exactly) predictively matched if it equals 1 for all data sets of insufficient size, that is, \( \BF_{10}(y^{[K]}) = 1 \) for all \( y^{[K]} \) with \( \vecB{n} = (n_{1}, \ldots, n_{K}) \) smaller than the minimal sample sizes \parencite{bayarri2012criteria}.
% For these uninformative data sets nothing is learned, because the posterior model probabilities then equal the prior model probabilities.
The insufficient sizes are: (a) \( n_{1} = \ldots = n_{K}= 1 \) as then \( \nu_{j} s_{j}^{2} = 0 \) for all \( j \in [K] \) regardless of the observations, and (b) \( n_{k} = 2 \) for some \( k \in [K] \) and \( n_{j} = 1 \) for all \( j \in [K] \setminus \{ k \} \), in which case there is no other sample variance to compare \( s_{k}^{2} \) to.

\begin{theorem}[Predictive matching]
\label{thmPredictiveMatching}
A Bayes factor constructed from the pair of priors \( \pi_{1}(\vecB{\mu}, \bar{\vecB{\tau}}, \vec{\vartheta}) = \pi_{0}(\vecB{\mu}, \bar{\vecB{\tau}}) \pi_{1}(\vec{\vartheta}) \) and \( \pi_{0}(\vecB{\mu}, \bar{\vecB{\tau}}) \propto \bar{\vecB{\tau}}^{-1} \) with \( \pi_{1}(\vec{\vartheta}) \) proper is predictively matched. This holds for our proposed Bayes factor. \( \hfill \diamond \)
\end{theorem}

\begin{proof}
See \refApp{appPredictiveMatching}.
\end{proof}

\subsection{Information Consistency}
\begin{sloppypar}
Information consistency implies that for all data sets of sufficient size, that is, fixed \( \vecB{n}=(n_{1}, \ldots, n_{K}) \) with at least two indexes \( j \neq k \in [K] \) such that \( n_{j}, n_{k} \geq 2 \), the Bayes factor in favor of the alternative over the null should tend to infinity whenever it becomes abundantly clear that the null cannot hold true. This occurs in the limit \( s_{j}^{2} / s_{K}^{2} \rightarrow 0 \), that is, when the observed variance \( s_{K}^{2} \) is of a much higher order than another sample variance \( s_{j}^{2} \).
\end{sloppypar}

\begin{theorem}[Information consistency]
\label{thmInformationConsistent}
The proposed Bayes factor is information consistent if \( u_{j} \leq 1/2 \) for \( j \in [K] \). \( \hfill \diamond \)
\end{theorem}

\begin{proof}
See \refApp{appInformationConsistency}.
\end{proof}

\subsection{Model Selection Consistency}
A Bayes factor is model selection consistent if it selects the correct model as \( \vecB{n} \rightarrow \infty \), that is, if
\begin{align}
 \BF_{10} (Y^{[K]}, \vecB{n}) \overset{\Pf}{\rightarrow} 0 \textnormal{ if } \Pf \in \Mc_{0} , \textnormal{ and }  \BF_{01} (Y^{[K]}, \vecB{n}) \overset{\Pf}{\rightarrow} 0 \textnormal{ if } \Pf \in \Mc_{1} ,
\end{align}
where \( \Pf \) refers to the data generating distribution, and where  \( X_{n} \overset{\Pf}{\rightarrow} X \) denotes convergence in probability, that is, \( \lim_{n \rightarrow \infty} \Pf ( | X_{n} - X | > \epsilon ) = 0  \) for all \( \epsilon > 0 \).

To state the theorem and to allow the \( K \) sample sizes go to infinity independently of each other, we let \( n_{K}  := n \) and \( n_{j} := c_{j} n \) for \( c_{j} > 0 \), \( j \in [K] \), thus, \( c_{K} = 1 \) by definition. To also allow the (data-governing) variances to differ arbitrarily as well, we let \( \gamma_{j} \) be the relative size of the variance \( \sigma_{j}^{2} \) with respect to \( \sigma_{K}^{2} \), that is, \( \sigma_{j}^{2} := \gamma_{j} \sigma_{K}^{2} \) where \( \gamma_{j} > 0 \) for \( j \in [K] \), thus, \( \gamma_{K}=1 \) by definition. Note that the null hypothesis is equivalent to \( \vecB{\gamma} = \vecB{1} \in \R^{K} \), whereas under the alternative there exists at least one \( j \in [K] \) such that \( \gamma_{j} \neq 1 \).

\begin{theorem}[Model selection consistency]
\label{thmModelSelectionConsistency} The proposed Bayes factor is model selection consistent. Furthermore, let \( Y_{ji} \iidSim \Nc(\mu_{j}, \sigma_{j}^{2}) \) where \( \sigma_{j}^{2} = \gamma_{j} \sigma_{K}^{2} \) for \( i \in [n_{j}]  \), \( n_{j} = c_{j} n \), and \( n_{K}=n \) for \( j \in [K] \), then as all the sample sizes tend to infinity, the Bayes factor behaves as
\begin{align}
\label{eqBfModelSelectionConsistency1}
\BF_{10}(\vecB{s}^{2}, n) & = C_{0}(K, \vecB{c}, \vecB{u} \, | \, \vecB{\gamma}) n^{\tfrac{1-K}{2}} \big ( \tfrac{ \la \vecB{c}, \vecB{\gamma} \ra}{ \vecB{c}_{+} } \big ) ^{ \tfrac{ \vecB{c}_{+}}{2} n } \Big ( \prod_{j=1}^{K-1} \gamma_{j}^{- \tfrac{c_{j}}{2} n} \Big ) %\\
\exp ( V(n)),
\end{align}
where \( \la \vecB{c} , \vecB{\gamma} \ra := \sum_{j=1}^{K} c_{j} \gamma_{j} \), \( V(n)= \Oc_{P}(n^{-1/2}) \) under the null and \( V(n) = \Oc_{P}(n^{1/2}) \) under the alternative, and where
\begin{align}
\label{eqC0Constant}
C_{0}(K, \vecB{c}, \vecB{u} \, | \, \vecB{\gamma}) & = \frac{(4 \pi)^{\tfrac{K-1}{2}}  \vecB{c}_{+}^{\tfrac{1}{2}}  \Big ( \prod_{j=1}^{K-1} \gamma_{j}^{- u_{j}} \Big ) }{\Bc ( \vecB{u})  \Big ( \prod_{j=1}^{K-1} c_{j}^{\tfrac{1}{2}} \Big ) (\vecB{c}_{+} - \sum_{j=1}^{K-1} \tfrac{ c_{j} \gamma_{j} -1 }{ \gamma_{j}})^{\vecB{u}_{+}}} . %\\
\end{align}
This means that under the alternative, \( \Hc_{1} : \gamma_{j} \neq 1 \) for some \( j \in [K-1] \), we have that
\begin{align*} \label{eq:convergence-I}
\log ( \BF_{10} ( \vecB{s}^{2}, n) ) & = \log \big ( C_{0}(K, \vecB{c}, \vecB{u} \, | \, \vecB{\gamma})  \big ) + \tfrac{1-K}{2} \log ( n ) \\
+ & \Big ( \vecB{c}_{+} \log \big ( \tfrac{ \la \vecB{c}, \vecB{\gamma} \ra}{ \vecB{c}_{+} } \big ) - \sum_{j=1}^{K-1} c_{j} \log (\gamma_{j}) \Big ) \frac{n}{2} + \Oc_{P}(n^{1/2}). \numberthis
\end{align*}
Under the null, \( \Hc_{0}: \vec{\gamma} = \vec{1} \), this simplifies drastically, and the logarithm of the Bayes factor then behaves as
\begin{align*} \label{eq:convergence-II}
\log ( \BF_{10}(\vecB{s}^{2}, n) ) & = \tfrac{1-K}{2} \Big ( \log( n) - \log ( 4 \pi) \Big ) + \tfrac{1}{2} \Big ( \log ( \vecB{c}_{+}) - \sum_{j=1}^{K-1} \log(c_{j}) \Big )  \\
& - \vecB{u}_{+} \log(K) - \log \Bc( \vecB{u}) + \Oc_{P}( n^{-1/2}) . \numberthis
\end{align*}
Hence, \( \BF_{10}(\vecB{s}^{2}, n) \) converges relatively slowly to zero under the null compared to the exponential decay of \( \BF_{01}(\vecB{s}^{2}, n) \) under the alternative. \( \hfill \diamond \)
\end{theorem}

\begin{proof}
See \refApp{appModelSelectionConsistency}.
\end{proof}

\subsubsection{Illustrating the Rate of Convergence}
\label{secSimulation}
\begin{figure}[!h]
   \centering
   \includegraphics[width=\textwidth]{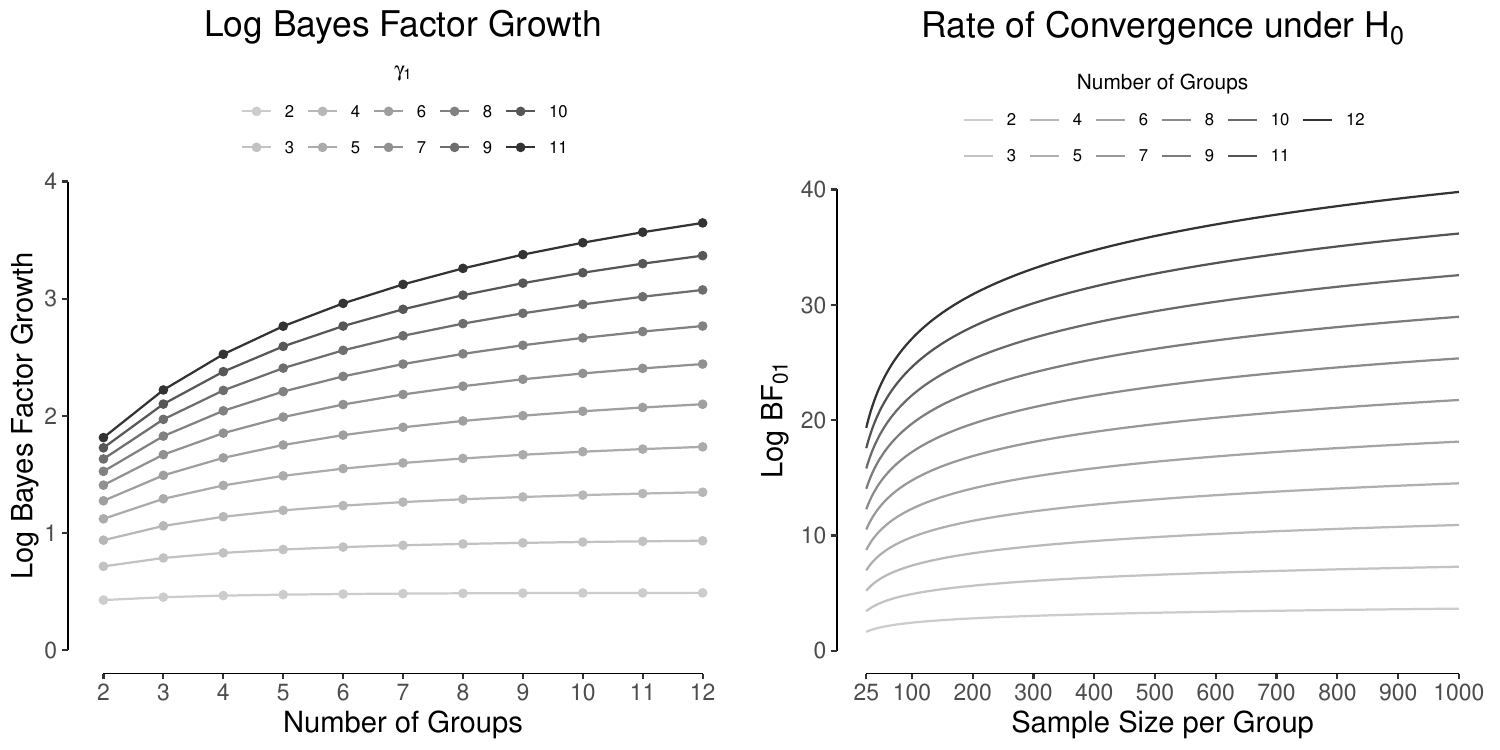}
   \caption{\textcolor{black}{Left: Shows the rate of the linear growth of the log Bayes factor under \( \mathcal{H}_{1} \) for increasing $\gamma_1$ and number of groups. Right: Shows how $\log ( \BF_{01}  )$ grows as a function of $n$ when \( \mathcal{H}_{0} \) is true for different number of groups $K$.}}
   \label{fig:assessing-convergence}
\end{figure}

We illustrate the rate of convergence of our default Bayes factor by visualizing Equations \eqref{eq:convergence-I} and \eqref{eq:convergence-II} as a function of $K \in [2, 12]$ and $\gamma_1 \in [2, \ldots, 11]$ with $\gamma_2 = \ldots = \gamma_K = 1$ and $\sigma_K^2 = 1$. Equation \eqref{eq:convergence-I} shows that under the alternative the asymptotic behavior of $\log (\BF_{10})$ is mostly linear in \( n \). The left panel in Figure \ref{fig:assessing-convergence} shows the slope of this linear increase — termed the log Bayes factor growth — as a function of $K$ and $\gamma_1$. We arrive at this slope by computing Equation \eqref{eq:convergence-I} for a large number of $n$ and regressing the result on $n$. When \( \mathcal{H}_{1} \) is true, the rate of convergence of the Bayes factor is exponential, and so the log Bayes factor grows linearly. We visualize the slope of how the log Bayes factor grows across the number of groups, with larger values indicating more rapid exponential growth. We find that, as the number of groups increases, the log Bayes factor grows more quickly. This increase is also dependent on $\gamma_1$; for larger values, the Bayes factor grows more quickly with increasing number of groups.

The right panel in Figure \ref{fig:assessing-convergence} illustrates $\log ( \BF_{01}  )$ as a function of the sample size per group for different number of groups $K$ under the null hypothesis, using Equation \eqref{eq:convergence-II}. In contrast to the scenario when \( \mathcal{H}_{1} \) is true, the rate of convergence when \( \mathcal{H}_{0} \) is true is no longer exponential \parencite[see also][]{johnson2010use, jeffreys1961theory, bahadur2009optimality}. %This rate of convergence can be improved by using non-overlapping interval null hypotheses, as we have outlined in Section \ref{secInterval} for the \( K = 2 \) group case. Encouragingly, we see that the rate of convergence increases as the number of groups increases. For practical purposes in settings where $K > 2$, if \( \mathcal{H}_{0} \) is indeed true, then one would find substantial evidence in favor of it already for small sample sizes; for example, $\text{BF}_{01} = 74.67$ and $\text{BF}_{01} = 608.81$ for $n = 50$ with $K = 3$ and $K = 4$ number of groups, respectively. The comparatively slower rate of convergence under \( \mathcal{H}_{0} \) therefore poses no problems in practice.

%
%The nice thing about this result is that it also makes the dependence on \( K \) explicit, see the illustrations in \refSec{secSimulation}.

%In this subsection we study the large-sample behavior of the Bayes factor with only one of the sample sizes going to infinity. Under labelling invariance we can consider without loss of generality the situation where the first \( K-1 \) samples are fixed as \( n_{K} \rightarrow \infty \). We assume that \( S_{K}^{2} \) is a \( \sqrt{n_{K}} \)-consistent estimator for the data-governing variance \( \sigma_{0}^{2} \) of the \( K \)th group, which by Chebyshev's inequality is certainly the case when \( Y_{Ki} \sim \Nc( \mu_{K}, \sigma_{K}^{2}) \).

\subsection{Limit and Across-Sample Consistency}
A Bayes factor is limit consistent if it remains bounded as long as not all $n_j \rightarrow \infty$ for $j \in [K]$ \parencite[][Ch. 6]{ly2018phd}. A Bayes factor is across-sample consistent if the limit of the \( K \)-sample Bayes factor as a function of the fixed observations of the groups \( i \in [K-1] \) results in a \( K-1 \) sample Bayes factor \parencite[][Ch. 4]{pena2018phd}. Note that we can consider without loss of generality the situation where the first \( K-1 \) samples are fixed as \( n_{K} \rightarrow \infty \) because of labelling invariance. For the following, we assume that \( S_{K}^{2} \) is a \( \sqrt{n_{K}} \)-consistent estimator for the data-governing variance \( \sigma_{0}^{2} \) of the \( K \)th group, which by Chebyshev's inequality is certainly the case when \( Y_{Ki} \sim \Nc( \mu_{K}, \sigma_{0}^{2}) \).

We call the \( K \)-sample Bayes factor \( \BF_{10}^{[K]}( \vec{s^{2}}, S_{K}^{2}) \) \emph{across-sample consistent} if, as \( n_{K} \rightarrow \infty \), it converges in probability under \( \sigma_{0}^{-2} \) to a \( K-1 \) Bayes factor \( \BF_{10 \, ; \, \sigma_{0}^{2}}^{[K-1]}( y^{[K-1]}) \), comparing the hypotheses
\begin{align}
\label{eqProblemKMinusOne}
\Hc_{0 \, ; \, \sigma_{0}^{2}}^{[K-1]}: \tau_{j}  &= \sigma_{0}^{-2} \text{ for all } j \in [K-1] \\
\Hc_{1 \, ; \, \sigma_{0}^{2}}^{[K-1]}: \tau_{j}  &\neq \sigma_{0}^{-2} \text{ for some } j \in [K-1] .
\end{align}
Here the null hypothesis states that the \( K-1 \) precisions are all equal to the known constant \( \sigma_{0}^{-2} \), whereas the alternative states that at least one precision is unequal to \( \sigma_{0}^{-2} \).

The theorem below implies that the proposed Bayes factor converges in probability to a lower dimensional Bayes factor \( \BF_{10 \, ; \, \sigma_{0}^{2}}^{[K-1]}(\vec{s^{2}}) \) that is based on uniform priors on the nuisance parameters \( \vec{\mu} \in \R^{K-1} \), and an inverse Dirichlet distribution on the precisions \( \vec{\tau} = (\tau_{1}, \ldots, \tau_{K-1}) \in \R^{K-1} \) scaled by \( 1/\sigma_{0}^{-2} \), that is,
\begin{align}
\label{eqScaledInverseDirichletPrior}
\pi_{\sigma_{0}^{2}}(\vec{\tau} \, | \, \Mc_{1}^{[K-1]}) = \frac{(\sigma_{0}^{2} )^{K-1} \prod_{j=1}^{K-1} (\sigma_{0}^{2} \tau_{j})^{u_{j} - 1}}{ \Bc ( \vec{u}, w)  (1+ \sigma_{0}^{2} \vec{\tau}_{+})^{\vec{u}_{+} + w}} ,
\end{align}
where we wrote \( w = u_{K} \) so the statement only involves vectors of length \( K-1 \). The integral representation of the multivariable generalisation of Tricomi's confluent hypergeometric function of the second kind \( \Uc \), see for instance \parencite{ng2011dirichlet,phillips1988characteristic}, shows that the resulting \( K-1 \) sample Bayes factor is given by %has the representation %
\begin{align}
\nonumber
\BF_{10 \, ; \, \sigma_{0}^{2}}^{[K-1]}(\vec{s^{2}}) & = \frac{ \int \Big ( \prod_{j=1}^{K-1} \tau_{j}^{\tfrac{ \nu_{j}}{2}} \Big ) \exp ( - \tfrac{1}{2} \sum_{j=1}^{K-1} \nu_{j} s^{2}_{j} \tau_{j} ) \pi_{\sigma_{0}^{2}}(\vec{\tau} \, | \, \Mc_{1}^{[K-1]}) \der \vec{\tau}   }{ (\sigma_{0}^{2})^{ - \tfrac{ \vec{\nu}_{+}}{2}} \exp ( - \tfrac{ (\overrightarrow{\nu s^{2}})_{+} }{2 \sigma_{0}^{2} }  ) } , \\
\label{eqKMinusOneTricomi}
& = \frac{ \Big ( \prod_{j=1}^{K-1} \Gamma( \tfrac{\nu_{j}}{2} + u_{j}) \Big ) \Uc \Big ( \tfrac{ \vec{\nu}}{2} + \vec{u} \, ; \, \tfrac{ \vec{\nu}_{+}}{2} - u_{K} + 1 \, ; \, \tfrac{ \overrightarrow{\nu s^{2}} }{2 \sigma_{0}^{2}} \Big )}{ \Bc ( \vec{u}, w) \exp ( - \tfrac{ (\overrightarrow{\nu s^{2}})_{+} }{2 \sigma_{0}^{2} }  ) }  ,
\end{align}
where \( \overrightarrow{\nu s^{2}} = ( \nu_{1} s_{1}^{2}, \ldots, \nu_{K-1} s_{K-1}^{2}) \) denotes the vector of sums of squares, \( (\overrightarrow{\nu s^{2}})_{+} = \sum_{j=1}^{K-1} \nu_{j} s_{j}^{2} \), and \( \vec{\nu}_{+} := \sum_{j=1}^{K-1} \nu_{j} \), as before.

\begin{theorem}[Limit and Across-Sample \( \sqrt{n}_{K} \)-consistency]
\label{thmAcrossSampleConsistency}
\begin{sloppypar}
If \( S_{K}^{2} \) is an \( \sqrt{n_{K}} \)-consistent estimator for \( \sigma_{0}^{2} \), then the Bayes factor \( \BF_{10}^{[K]}(\vec{s^{2}}, S_{K}^{2} ) \) is a \( \sqrt{n}_{K} \)-consistent estimator of the \( K-1 \)-sample Bayes factor \( \BF_{10 \, ; \, \sigma_{0}^{2}}^{[K-1]}(\vec{s^{2}}) \) given in \refEq{eqKMinusOneTricomi}. Furthermore, if \( Y_{Ki} \sim \Nc ( \mu_{K}, \sigma_{0}^{2}) \), then \( \sqrt{n}_{K} ( S_{K}^{2} - \sigma_{0}^{2}) \) is asymptotically normal, and consequently so is the \( K \)-sample Bayes factor, that is, %as \( n_{K} \rightarrow \infty \) %
\begin{align}
\sqrt{n_{K}} \Big ( \BF_{10}^{[K]}(\vec{s^{2}}, S_{K}^{2}) - \BF_{10 \, ; \, \sigma_{0}^{2}}^{[K-1]}(\vec{s^{2}}) \Big ) \inDist \Nc \Big (0, 2 \sigma_{0}^{4} \breve{T}_{1}^{2}  \Big )
\end{align}
where \( \breve{T}_{1} \) is given by \refEq{eqT1} in the appendix. \( \hfill \diamond \)
\end{sloppypar}
\end{theorem}

\begin{proof}
See \refApp{appAcrossSampleConsistency}.
\end{proof}

\section{Special Cases and Deviations from the Default} \label{secSpecialCases}
The comparison of $K = 2$ groups occurs frequently in practice and we discuss the Bayes factor for this special case in the following section. We also consider three modifications of the default choice in order to incorporate a subject assessment of the test-relevant parameter, and to accommodate directed tests and interval Bayes factors.

\subsection{The Bayes Factor for \( K=2 \) Groups}
\label{secBfKIsTwo}
For the \( K = 2 \) group case, the null model of equal precisions has three parameters \( (\mu_{1}, \mu_{2}, \bar{\tau}) \) whereas the alternative has four \( (\mu_{1}, \mu_{2}, \bar{\tau}, \vartheta) \). The comparison of interest is then between \( \Hc_{0}: \vartheta  = \tfrac{1}{2} \) and \( \Hc_{1}: \vartheta \neq \tfrac{1}{2} \). In this case, the proposed Bayes factor simplifies to
\begin{align}
\label{eqBfKTwo}
\BF_{10}(\vecB{s^{2}}) & = \tfrac{\Bc (\tfrac{\nu_{1}}{2} + u_{1}, \tfrac{\nu_{2}}{2} + u_{2}) }{\Bc ( u_{1}, u_{2})} \big ( 1 + \tfrac{\nu_{1} s_{1}^{2}}{\nu_{2} s_{2}^{2}} \big )^{\tfrac{\nu_{1} + \nu_{2}}{2}} {_{2} F}_{1} \big ( \tfrac{\nu_{1} + \nu_{2}}{2}, \tfrac{ \nu_{1} + 2 u_{1}}{2}  \, ; \, \tfrac{\nu_{1} + \nu_{2} + 2 (u_{1} + u_2)}{2}  \, ; \, \tfrac{ \nu_{2} s_{2}^{2} - \nu_{1} s_{1}^{2}}{\nu_{2} s_{2}^{2}} \big ) ,
\end{align}
where \( {_{2} F}_{1} \) refers to the Gaussian or ordinary hypergeometric function, \revise{which has the integral representation ${_{2} F}_{1} ( a, b \, ; \, c \, ; \, z) = \frac{\Gamma\left(c\right)}{\Gamma\left(b\right)\Gamma\left(c - b\right)} \int_0^1 t^{b - 1} (1 - t)^{c - b - 1} (1 - tz)^{-a} \mathrm{d}t$, with $\text{Re}(c) > \text{Re}(b) > 0$ \parencite[][eq. 15.3.1]{abramowitz1972handbook}.} Observe that across-sample consistency implies that for \( Y_{2i} \iidSim \Nc(\mu_{2}, \sigma_{0}^{2}) \) and \( n_{2} \rightarrow \infty \), the two-sample Bayes factor is a \( \sqrt{n}_{2} \)-consistent estimator of the one-sample Bayes factor
\begin{align}
\label{eqBfOneSample}
\BF^{[1]}_{10 \, ; \, \sigma_{0}^{2}}(s_{1}^{2}) = \frac{ \Gamma ( \tfrac{ \nu_{1} }{2} + u_{1}) \Uc \left ( \tfrac{\nu_{1}}{2} + u_{1} \, ; \, \tfrac{\nu_{1}}{2} - u_{2} + 1 \, ; \, \tfrac{ \nu_{1} s_{1}^{2}}{2 \sigma_{0}^{2}} \right )}{ \Bc ( u_{1}, u_{2}) \exp ( - \tfrac{ \nu_{1} s_{1}^{2} }{2 \sigma_{0}^{2}})} .
\end{align}
This Bayes factor compares the alternative hypothesis \( \Hc_{1 \, ; \, \sigma_{0}^{2}}^{[1]}: \tau_{1}  \neq \sigma_{0}^{-2}  \) to the null hypothesis \( \Hc_{0 \, ; \, \sigma_{0}^{2}}^{[1])}: \tau_{1} = \sigma_{0}^{-2}  \) with \( \sigma_{0}^{2} \) known. Here \( \Uc(a \, ; \, b \, ; \, z) = \frac{1}{\Gamma\left(a\right)} \int_0^{\infty} e^{-zt} t^{a - 1}(1 + t)^{b - a - 1} \der t \) is the (one-dimensional) Tricomi's confluent hypergeometric function of the second kind \parencite[][Eq. 13.2.5]{abramowitz1972handbook}.

%\FD{Think about changing $\delta$ to $\phi$, would also require changing the R package.}

\subsection{Prior elicitation}
\label{secPriorElicitation}
\begin{figure}
   \centering
   \includegraphics[width = 1\textwidth]{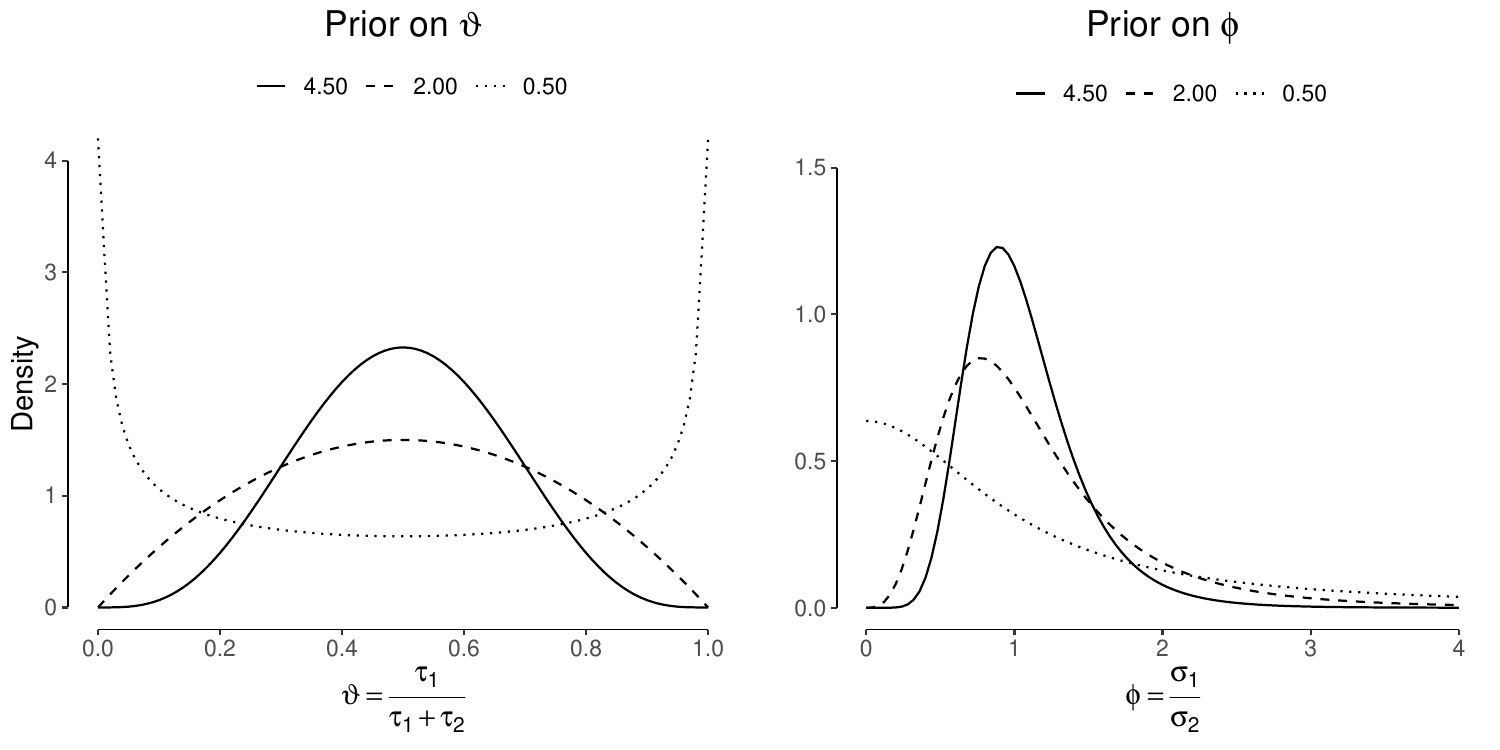}
   \caption{Prior on \( \vartheta \) (left) and induced prior on \( \phi \) (right) for \( u := u_1 = u_2 \in \{4.50, 2.00, 0.50\} \); see \refSec{secPriorElicitation} for the rationale behind these values.}
   \label{figPriorThetaAndPhi}
\end{figure}

For prior elicitation, it is arguably more intuitive to express the prior on the test-relevant parameter in terms of the ratio of the standard deviations, \( \phi = \frac{\sigma_2}{\sigma_1} = \sqrt{\frac{\vartheta}{1 - \vartheta}} \), thus, \( \int_{0}^{1} \der \vartheta = \int_{0}^{\infty} 2 \phi (1+\phi^{2})^{-2} \der \phi \). The prior \( \vartheta \sim \Beta(u_{1}, u_{2}) \) underlying \refEq{eqBfKTwo} induces a generalized beta prime distribution on \( \phi \) with density
\begin{align}
\pi( \phi \, ; \, u_{1}, u_{2}) = \frac{ 2 \phi^{2 u_{1} - 1} (1 + \phi^{2})^{-(u_{1} + u_{2})}}{\Bc ( u_{1}, u_{2}) }.
\end{align}
Figure \ref{figPriorThetaAndPhi} visualizes the prior \revise{assigned to} \( \vartheta \) and \( \phi \) for various values of \( u: = u_{1} = u_{2} \). %
A statistician may now elicit a researcher's prior beliefs in terms of (a ratio of) standard deviations conditional on the alternative holding true. For example, if the researcher believes that the probability of one standard deviation being twice as large or twice as small as the other does not exceed 95\%, then she should choose \( u = 4.50 \)\revise{. Note that the resulting Bayes factor is not information consistent anymore.} It is also interesting to note that on this scale \( \phi \) the \( m \)th raw moment is given by \( \frac{ \Gamma ( \tfrac{m}{2} + u_1) \Gamma ( u_{2} - \tfrac{m}{2})}{\Gamma ( u_{1}) \Gamma(u_{2})} \). Hence, it has no finite mean whenever \( u_{2} \leq 1/2 \). A change of variables shows that the posterior distribution in terms of \( \phi \) is given by:
\begin{align}
\pi(\phi \, | \, \vecB{y}^{(2)}) = \frac{ 2 \phi^{\nu_{1} + 2 u_{1} -1} (1 + \phi^{2})^{- (u_{1} + u_{2}) } ( 1 + \tfrac{ \nu_{1} s_{1}^{2} }{ \nu_{2} s_{2}^{2}} \phi^{2})^{- \tfrac{ \nu_{1} + \nu_{2} }{2}}}{ \Bc( \tfrac{ \nu_{1}}{2} + u_{1}, \tfrac{ \nu_{2}}{2} + u_{2}) \iiFi{ \tfrac{ \nu_{1}+ \nu_{2}}{2} }{\tfrac{ \nu_{1}}{2} + u_{1}}{ \tfrac{ \nu_{1} + \nu_{2}}{2} + u_{1} + u_{2}}{1 - \tfrac{\nu_{1} s_{1}^{2}}{\nu_{2} s_{2}^{2}}}} .
\end{align}

\subsection{Interval Bayes Factors}
\label{secInterval}
Researchers may wish to extend the sharp null hypothesis $\vartheta = \nicefrac{1}{2}$ to include a null-region around the point null value. If the null-region overlaps with the prior under the alternative, this leads to an (inconsistent) peri-null Bayes factor \parencite[e.g.,][]{ly2021bayes, morey2011bayes}. If the null-region does not overlap with the prior under the alternative, that is, if we compare the hypotheses:
\begin{align}
{\Hc}_{0} &: \phi \in [a, b] \\
{\Hc}_{1} &: \phi \not \in [a, b],
\end{align}
then this yields a non-overlapping interval-null Bayes factor \parencite[e.g.,][]{berger1987testing, rousseau2007approximating}. The null-region is usually informed by the problem at hand, as we will see later on an example. For a potential default approach to specify the non-overlapping interval bounds, see Appendix \ref{app:bayes-factor-interval}.

% \DB{So the beginning of the section starts with ``... with $a$, $b$ perhaps suggested by the problem at hand.''. However, the remainder of the section first introduces a reparametrization of $\vartheta$ called $\delta\in\R$. Sure, apparently $\delta$ has a known distribution given the prior on $\vartheta$, but what is the point of that? That the region $\epsilon$ must be smaller than some value otherwise the behavior is different from \refEq{eqBfKTwo}?}
%
%
%For the normal likelihood, say, \( \bar{X} \sim \Nc( \delta, \sigma^{2}/n) \) and a symmetric unimodal prior \( \pi(\delta) \) around the null value, \textcite{berger1987testing} showed that the Bayes factor comparing \( \breve{\Hc}_{1} \) to \( \breve{\Hc}_{0} \) with the priors truncated accordingly approximates the standard (point-null) Bayes factor well if \( \epsilon \) is proportional to the standard error. To derive the standard error we
%
%For the standard error, we mimic \text{berger1987testing} by

\subsection{Directed Bayes Factors}
Researchers sometimes desire to quantify evidence in favor of hypotheses such as \( \Hc_{-}: \sigma_{1}^{2} > \sigma_{2}^{2} \), or \( \Hc_{+} : \sigma_{1}^{2} < \sigma_{2}^{2} \). More generally, let \( \mathcal{H}_{r} \) denote such an order-constrained \revise{or directed} hypothesis. \revise{Since \( \sigma_{1}^{2} = (2 \vartheta \bar{\vecB{\tau}})^{-1} \) and \( \sigma_{2}^{2} = (2 (1 - \vartheta) \bar{\vecB{\tau}})^{-1} \), we have that \( \sigma_{1}^{2} > \sigma_{2}^{2} \) implies \( \vartheta < \nicefrac{1}{2} \). We therefore restrict the beta prior on \( \vartheta \) accordingly in the calculation of the the marginal likelihood for \( \mathcal{H}_{r} \) \parencite[see also][]{ly2016harold}, which can then be used to calculate directed Bayes factors.}

%Note that \( \Hc_{1} : \Hc_{-} \cup \Hc_{+} \), and that a symmetry of the prior implies \( 2 \BF_{10}(\vecB{s^{2}}) = \BF_{+0}(\vecB{s^{2}}) + \BF_{-0}(\vecB{s^{2}}) \), thus, \( \BF_{+0}(\vecB{s^{2}}) \leq 2 \BF_{10}(\vecB{s^{2}}) \). Hence, the (asymptotic) behavior under \( \Hc_{+} \) of \( \BF_{+0}(\vecB{s^{2}}) \) can be derived from that of the standard Bayes factor \( \BF_{10}(\vecB{s^{2}}) \).

%\AL{Perhaps add the argument that \( 2 \BF_{10}(\vecB{s^{2}}) = \BF_{+0}(\vecB{s^{2}}) + \BF_{-0}(\vecB{s^{2}}) \) implies that \( \BF_{+0}(y) = 2 \BF_{10}(y) \times \text{posterior mass at the restriction} \), i.e., Fubini. Then mention bridge sampling for the larger dimensional case.}

In the more general $K > 2$ group case, we can similarly specify equality or inequality constraints by encoding them in the prior distribution on $\vec{\vartheta}$. An example of such a constrained hypotheses is given by:
\begin{equation*}
    \mathcal{H}_r: \vartheta_1 = \vartheta_2 > (\vartheta_3, \vartheta_4, \vartheta_5 = \vartheta_6) > \vartheta_7 \enspace,
\end{equation*}
which incorporates two equality constraints ($\vartheta_1 = \vartheta_2$ and $\vartheta_5 = \vartheta_6$), several order constraints (e.g., $\vartheta_1 > \vartheta_3$, $\vartheta_1 > \vartheta_4$, $\vartheta_3 > \vartheta_7$, $\vartheta_4 > \vartheta_7$), and no constraints between the $\vartheta_3$, $\vartheta_4$, $\vartheta_5 = \vartheta_6$ (and therefore also the standard deviations and variances). Note that while this hypothesis is formulated in terms of the parameter $\mathbf{\vartheta}$, it has immediate implications for the precisions and thus for the standard deviations and variances. \revise{We could also directly formulate the hypotheses on the variances or standard deviations, for example, with $(\sigma_1 = \sigma_2) > \sigma_3$ implying that $(\vartheta_1 = \vartheta_2) < \vartheta_3$. This flexibility allows researchers to translate substantive predictions directly into statistical hypotheses.}

We compute \revise{Bayes factors including} mixed hypotheses \revise{such as $\mathcal{H}_r$} as follows. First, we introduce a new \revise{auxiliary} hypothesis \revise{$\mathcal{H}_a$} which does not include order-constraints. In our example, this yields:
\begin{equation*}
\mathcal{H}_a: \vartheta_1 = \vartheta_2, \vartheta_3, \vartheta_4, \vartheta_5 = \vartheta_6, \vartheta_7 \enspace .
\end{equation*}
We estimate the \revise{(auxiliary)} Bayes factor \revise{$\text{BF}_{ra}$} by dividing the proportion of samples $\mathbf{\vartheta}$ that respect the order-constraints in $\mathcal{H}_r$ in the posterior by the proportion of samples that respect it in the prior \parencite{klugkist2005bayesian}. \revise{Separately, we then estimate the Bayes factor in favor of $\mathcal{H}_a$ over $\mathcal{H}_1$ (or $\mathcal{H}_0$) using bridge sampling \parencite{meng1996simulating, gronau2017tutorial}. Combining these two Bayes factors yields the desired Bayes factor in favor of $\mathcal{H}_r$ over $\mathcal{H}_1$ (or $\mathcal{H}_0$), that is, $\text{BF}_{r1} = \text{BF}_{ra} \times \text{BF}_{a1}$.} The R package \textit{bfvartest}, which is available from \url{https://github.com/fdabl/bfvartest}, implements this and all other procedures described above; see  Appendix \ref{sec:analysis-code} for how to use the package.

\subsection{Comparison to a Fractional Bayes Factor}
\color{black}
%The search for automatic and objective Bayesian model selection has a long history \parencite{berger2006case}. It is well known that Bayesian testing requires careful construction of the prior since testing --- in contrast to estimation --- is greatly influenced by the prior \parencite{degroot1982lindley}. Using uninformative priors for test-relevant parameters is therefore ill-advised \parencite{lindley1997some, jeffreys1939theory}
One alternative to choosing the prior based on desiderata, as done in this paper, is to use the data to inform the prior. \textcite{o1995fractional} proposed the \textit{fractional} Bayes factor, which uses a fraction $b = \nicefrac{m_0}{n}$ of the entire likelihood to construct a prior, where $m_0$ is the size of the minimal training sample and $n$ is the sample size. \textcite{boing2018automatic} developed a fractional Bayes factor for testing the (in)equality of several population variances. Here, we compare our proposed default Bayes factor to their fractional Bayes factor.

\begin{figure}[!h]
   \centering
   \includegraphics[width = 0.85\textwidth]{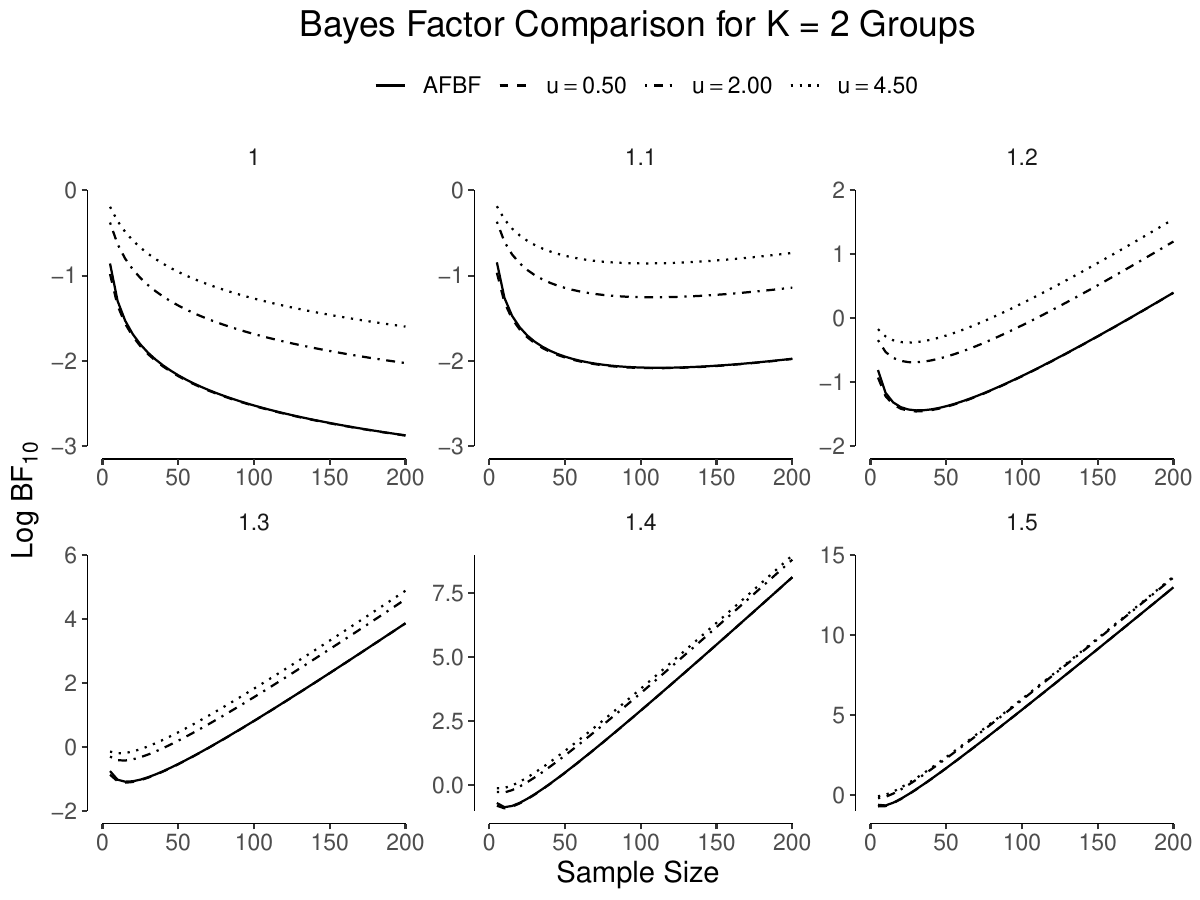}
   \caption{Comparison of the Bayes factor proposed by \textcite{boing2018automatic} and our Bayes factor for $K = 2$ groups as a function of $n := n_1 = n_2$, prior specification $u := u_1 = u_2$, and effect size $\phi = \{1, 1.1, 1.2, 1.3, 1.4, 1.5\}$.}
   \label{fig:compare-bfsk2}
\end{figure}

Since the likelihood is the same, the key difference between the two Bayes factors is in their respective prior specification. As we are concerned with hypotheses that can feature both inequality and equality constrains, we need to introduce additional notation. Let $\mathcal{H}_r$ denote a hypothesis with $q^E_r$ equality and $q^I_r$ inequality constraints on $K$ population variances, such that there are $J_r = K - q^E_r$ unique variances $\vec{\sigma}^2_r = (\sigma_1^2, \ldots, \sigma^2_{J_r})$. Further, let $K_j$ be the number of populations sharing the unique variance $\sigma^2_j$, and $n_{j_k}$ be the sample size of the $k^{\text{th}}$ population sharing the unique variance $\sigma^2_j$. \textcite{boing2018automatic} use population-specific fractions given by $b_{j_k} = \nicefrac{2}{n_{j_k}}$, where $m_0 = 2$ is the minimal training sample size for the automatic prior to be proper; it is in this sense that their Bayes factor relies on minimal prior information. They calculate the marginal likelihood for hypothesis $\mathcal{H}_r$ as:
\begin{equation}
    p(y^{[K]} \mid \mathcal{H}_r) = \frac{\int_{\Omega_t}\int_{\mathbb{R}^K}f(y^{[K]}; \vecB{\mu}, \vec{\sigma_r}^2)\pi(\vecB{\mu}, \vec{\sigma_r}^2)\mathrm{d}\vecB{\mu}\mathrm{d}\vec{\sigma_r}^2}{\int_{\Omega_t^a}\int_{\mathbb{R}^K}f(y^{[K]}; \vecB{\mu}, \vec{\sigma_r}^2)^{\vecB{b}}\pi(\vecB{\mu}, \vec{\sigma_r}^2) \mathrm{d}\vecB{\mu}\mathrm{d}\vec{\sigma_r}^2} \enspace,
\end{equation}
%\DB{So $\vec{y}$ means in Lexi's notation: $y_1, \ldots, y_{K-1}$, so this should probably be $y^{[K]}$. The same goes for stuff like $\vec{b}$ below.}
where $\vecB{b}$ is the vector of population-specific fractions, $\pi(\vecB{\mu}, \vec{\sigma}^2_r) \propto \prod_{i=1}^{J_r} \sigma^{-2}_i$ is the Jeffreys prior, $\Omega_t$ specifies the region of integration depending on the inequality constraints in $\mathcal{H}_t$, and $\Omega_t^a$ is the adjusted integration region given by:
\begin{equation}
    \Omega_t^a = \left\{\vec{\sigma_r}^2: \bm{R}^I[a_1 \sigma_1^2 \hdots a_{J_r}\sigma_{J_r}^2] > \vec{0}\right\} \enspace ,
\end{equation}
where $\bm{R}^I$ encodes the inequality constraints among the $J_r$ unique variances, and where $a_j = \nicefrac{K_j}{2\sum_{k = 1}^{K_j}\left(1 - \frac{s_{j_k}^2}{n_{j_k}}\right)}$. \textcite{boing2018automatic} show that this setup leads to the following expression for the marginal likelihood of $\mathcal{H}_r$:
\begin{align*} \label{eq:AFBF}
    p(y^{[K]} \mid \mathcal{H}_r) &= \frac{\int_{\Omega_r} \prod_{j = 1}^{J_r} \text{Inv-Gamma}\left(\sigma^2_j ; \frac{\sum_{k = 1}^{K_j} n_{j_k} - K_j}{2}, \frac{\sum_{k = 1}^{K_j} \left(n_{j_k} - 1\right)s_{j_k}^2}{2} \right)\mathrm{d}\sigma^2_j}{\int_{\Omega_r} \prod_{j = 1}^{J_r} \text{Inv-Gamma}\left(\frac{K_j}{\sum_{k = 1}^{K_j} \left(2 - \frac{1}{n_{j_k}}\right)s_{j_k}^2} \sigma^2_j ; \frac{K_j}{2}, \frac{K_j}{2} \right)\mathrm{d}\sigma^2_j} \pi^{\frac{-\sum_{j = 1}^{J_r} \sum_{k = 1}^{K_j} (n_{j_k} - 2)}{2}} \\
    & \left(\prod_{j = 1}^{J_r}\prod_{k = 1}^{K_j} \left(\frac{n_{j_k}}{2}\right)^{\frac{1}{2}}\right) \prod_{j = 1}^{J_r} \frac{\Gamma\left(\frac{\sum_{k = 1}^{K_j} n_{j_k} - K_j}{2}\right)\left(\sum_{k = 1}^{K_j} \left(2 - \frac{1}{n_{j_k}}\right)s_{j_k}^2\right)^{\frac{K_j}{2}}}{\Gamma\left(\frac{K_j}{2}\right)\left(\sum_{k = 1}^{K_j} (n_{j_k} - 1)s_{j_k}^2\right)^{\frac{\sum_{k = 1}^{K_j} n_{j_k} - K_j}{2}}}\enspace , \numberthis
\end{align*}
where $\text{Inv-Gamma}(x; \alpha, \beta)$ is the density of the inverse Gamma distribution, and the ratio of the two integrals gives the probability that the constraints hold in the posterior divided by the probability that they hold in the prior. This ratio equals 1 when testing hypotheses without order-constraints, i.e., $\Omega_t^{\alpha} = \Omega_t$. From Equation (\ref{eq:AFBF}) it follows that the prior distribution assigned to $\sigma_j^2$ under hypothesis $\mathcal{H}_r$ is given by:
\begin{align*}
    \sigma^2_j &\sim \text{Inv-Gamma}\left(\frac{K_j}{2}, \frac{\sum_{k = 1}^{K_j}\left(2 - \frac{1}{n_{j_k}}\right) s_{j_k}^2}{2}\right) \enspace ,
\end{align*}
where $n_{j_k}$ and $s_{j_k}^2$ are the sample size and the sum of squares of the $k^{\text{th}}$ group sharing population variance $\sigma_j^2$. Note that, in contrast to our proposed default prior, the prior for the fractional Bayes factor proposed by \textcite{boing2018automatic} depends on the data. Similarly, our prior specification results in a joint distribution on $\vecB{\sigma^2}$ that cannot be factorized, that is, it results in a dependent prior, where the dependency is created through the weights $\vec{\vartheta}$. The prior specification by \textcite{boing2018automatic} induces a Dirichlet prior on $\vec{\vartheta}$ with $u = \nicefrac{K_j}{2}$ and a non-standard prior on $\bar{\tau}$ (it follows a Gamma distribution if and only if all sample sizes and sum of squares are equal). Figure \ref{fig:compare-bfsk2} shows our default Bayes factor and the fractional Bayes factor for $K = 2$, sample sizes $n := n_1 = n_2 \in [5, \ldots, 200]$, and different values of $\phi = \{1, 1.2, 1.3, 1.4, 1.5\}$. While our proposed default Bayes factor and the fractional Bayes factor differ, they show very similar results for $u = \nicefrac{1}{2}$. %the prior on the test-relevant parameter $\vec{\vartheta}$ is the same in case one compares the hypothesis with all equality constraints ($\mathcal{H}_0$) to the hypothesis with all inequality constraints ($\mathcal{H}_1$) because then $u = \nicefrac{1}{2}$. Further assuming equal sample sizes, the two Bayes factors yield virtually identical results, as can be seen in Figure \ref{fig:compare-bfsk2}, which show this for $K = 2$ of sample sizes $n := n_1 = n_2 \in [5, \ldots, 200]$ and for different values of $\phi = \{1, 1.2, 1.3, 1.4, 1.5\}$.

There an interesting discrepancy between the two Bayes factors when testing directed hypotheses. In case there is overwhelming evidence for the hypothesis that $\mathcal{H}_r: \sigma_1^2 > \hdots > \sigma_K^2$, the Bayes factor in favor of it over $\mathcal{H}_1: \sigma_1^2 \neq \hdots \neq \sigma_K^2$ reaches the bound $K!$. However, in case there are the same $J$ equalities in both hypotheses, the fractional Bayes factor does not reach the bound of $(K - J)!$, while our proposed default Bayes factor does. This is because \textcite{boing2018automatic} set $b_{j_k} = \nicefrac{2}{n_{j_k}}$ for all groups. While this is desirable in the sense that one thus uses the same `minimal' amount of information under each hypothesis, this results in a different shape parameter of the inverse gamma prior distribution, and the bound is therefore not reached, which can be considered a shortcoming of the fractional Bayes factor.% (B{\"o}ing-Messing \& Mulder, personal communication) \EJ{So we confirmed they are OK with us stating it like this?}.

\section{Practical Examples}
\label{secExamplesKIsTwo}
In the following sections we apply our proposed Bayes factor test on a number of examples.

\subsection{Sex Differences in Personality}
\begin{figure}
   \centering
   \includegraphics[width = 1\textwidth]{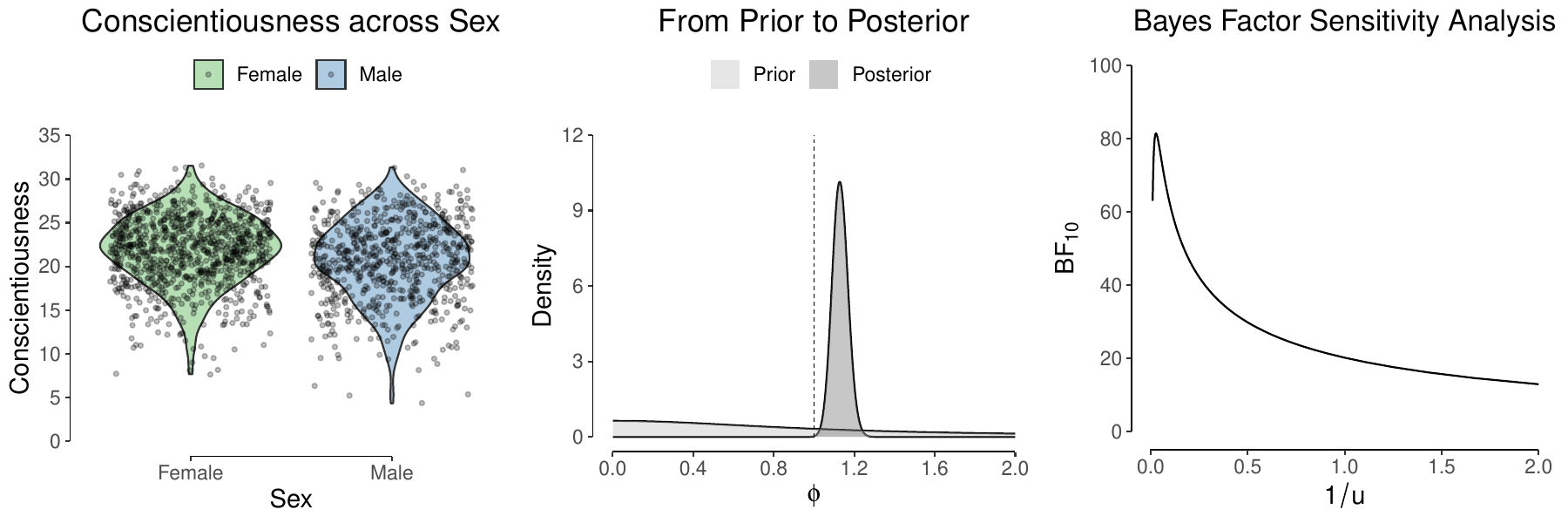}
   \caption{Left: Peer-rated conscientiousness of Estonian men and women. Middle: Prior and posterior of $\phi$ (with $u = \nicefrac{1}{2}$). Right: Bayes factor sensitivity analysis for $u \in [\nicefrac{1}{2}, 100]$.}
   \label{fig:sex-differences}
\end{figure}

%Our second example is taken from the literature on sex differences in personality. \textcite{borkenau2013sex} find that men exhibit 1.21 times as much variance as women on informant reported Openness to Experience in a sample of $N = 3046$. Figure \ref{fig:empirical-example} (a) illustrates the learning process via sequential analysis of the data, and Figure \ref{fig:empirical-example} (b) shows the prior and posterior distribution over $\vartheta$.
%\AL{I don't see the Bayes factor being reported. With \( u_{1} = u_{2}=1/2 \) I get \( \BF_{10}(\vecB{s^{2}}) = 12.9822 \). This is the most analysis, so it'll be good to report this one first.}
There is a rich history of research and theory about differences in variability between men and women, going back at least to Charles Darwin \parencite{darwin1871descent}. \textcite{borkenau2013sex} studied whether men and women differ in the variability of personality traits. Here, we focus on peer-rated conscientiousness in Estonian women and men ($s_f^2 = 15.6$, $s_m^2 = 19.9$, $n_f = 969$, $n_m = 716$). The left panel in Figure \ref{fig:sex-differences} visualizes the raw data, and the middle panel shows the prior (using $u = \nicefrac{1}{2}$) and the posterior distribution for the effect size $\phi$. The default Bayes factor yields $\BF_{10} = 12.98$ in favor of a difference in variance, and the right panel shows a \revise{sensitivity analysis to the specification of $u$ in the default Bayes factor (note that the $x$-axis scale is $1/u$)}; as expected, a smaller value of $u$ corresponds to a wider prior of $\phi$ under \( \mathcal{H}_{1} \)  and decreases the predictive performance of \( \mathcal{H}_{1} \) compared to \( \mathcal{H}_{0} \). Nevertheless, across the range of $u$ visualized in Figure \ref{fig:sex-differences}, there is strong evidence that Estonian men show larger variability in conscientiousness than Estonian women. %Here, we focus on the data on people's Conscientiousness rated by their peers in Estonia ($s_f^2 = 15.6$, $s_m^2 = 20$, $n_f = 969$, $n_m = 716$), Belgium ($s_f^2 = 13.3$, $s_m^2 = 17.3$, $n_f = 232$, $n_m = 59$), Germany ($s_f^2 = 15.4$, $s_m^2 = 19.7$, $n_f = 169$, $n_m = 134$), and Czech Republic ($s_f^2 = 18.2$, $s_m^2 = 21.8$, $n_f = 416$, $n_m = 416$). In particular, we analyze the data from the Estonian sample first, and use the posterior on $\vartheta$ as our prior when testing hypotheses in the other samples. Figure \ref{fig:empirical-example} (a) illustrates the learning process via sequential analysis of the data from Estonia, and Figure \ref{fig:empirical-example} (b) shows the prior and posterior distribution over $\vartheta$. We find strong evidence that the variability in male Conscientiousness is higher than the variability in female Conscientiousness ($\text{BF}_{10} = 96$). We use this posterior distribution as our prior distribution for the Flemish sample. Specifically, we used maximum likelihood estimation to fit a Beta distribution to the posterior samples of $\vartheta$, which resulted in parameters $u_1 = 95$ and $u_2 = 50$. Using this informed prior lead to overwhelming evidence for \( \mathcal{H}_{1} \) in the Flemish sample ($\text{BF}_{10} = 6868$), strong evidence in favor of \( \mathcal{H}_{0} \) in the Czech sample ($\text{BF}_{01} = 135.3$), and equivocal evidence in the German sample ($\text{BF}_{01} = 1.4$). In the next section, we show how our Bayes factor can be generalized to \( K > 2 \) groups.

%\AL{Prior and posterior plot: Perhaps it's better to show things in terms of \( \delta \in \R \) or simply in \( \vartheta \). The fact that \( \phi \in (0, \infty) \) skews the picture due to thresholding at 0.}

\subsection{Testing Against a Single Value}
\label{secExSingleValue}
%\AL{The paper is titled default Bayes factors, but that's not being illustrated here. Furthermore, I'm not sure if the one-sample \refEq{eqBfOneSample} is actually exemplified. I suggest to (1) show the default two-sided results, (2) the approximation of \( \BF_{10} \) to \( \BF_{10}^{(1)} \) for \( \nu_{2} \) ``large'' and (3) show the one-sided result. The one-sided result \( \leq \) two times the two-sided result. Worst case scenario, (4) show a robustness analysis.}

%\AL{The standard two-sided one-sample Bayes factor yields \( \BF^{(1)}_{10}(s^{2}) = 0.410497 \) and with ``\( \nu_{2} \)''\( = 100 \), I get \( \BF_{10}(s^{2}, \sigma^{2}) = 0.41178 \), which is quite accurate. With \( u_{1} = u_{2} = 2.16 \), I get \( \BF^{(1)}_{10}(s^{2}) = 0.810625 \) and \( \BF_{10}(s^{2}, \sigma^{2}) = 0.808186 \). The story resulting in \( u_{1} = u_{2} = 2.16 \) is completely arbitrary. Any other numbers lead to a different result. This whole cutting up the prior undermines the title of the paper, and it is better just to report the default things.}

%\FD{Updated this section, but did not incorporate the approximation (since we relegated it into an appendix).}

%\FD{I get the same result as Lexi with our R package.}

Polychlorinated biphenyls (PCB), which are used in the manufacture of large electrical transformers and capacitors, are hazardous contaminants when released into the environment. Suppose that the Environmental Protection Agency is testing a new device for measuring PCB concentration (in parts per million) in fish, requiring that the instrument yields a variance of less than \( 0.10 \) (a standard deviation \( \sigma_0 \leq 0.32 \))\revise{, thus \( \phi > 1 \). This suggests the use of a directed Bayes factor.} Seven PCB readings on the same sample of fish are subsequently performed, yielding a sample standard deviation of \( s = 0.22 \) and a sample effect size of \(\hat{\phi} = \frac{\sigma_0}{s} = 1.42 \) \parencite[see][p. 420]{mendenhall2016statistics}. We compare the following hypotheses
\begin{align*}
\mathcal{H}_{0}&: \phi = 1 \\
\mathcal{H}_{+}&: \phi > 1 ,
\end{align*}
which yields \( \text{BF}_{+0} = 0.51 \) for the default value $u = \nicefrac{1}{2}$, a value slightly higher than for an undirected test, \( \text{BF}_{10} = 0.41 \). To illustrate prior elicitation, assume that the makers of the new device are highly confident, assigning \( 50\% \) probability to the outcome that the new device reduces the required standard deviation at least by half. Defining \( \phi = \frac{\sigma_0}{\sigma_{\text{device}}} \), this formally translates into \( \pi(\phi \in [2, \infty]) = \nicefrac{1}{2} \), which is fulfilled by a (truncated) prior with \( u = 2.16 \). Using this prior specification results in \( \text{BF}_{+0} = 0.83 \).

\subsection{Comparing Measurement Precision}
\label{secInterval-example}
%\FD{Alex Kuiper from IBIS at UvA sent me a data set but it requires more complicated modeling than simply comparing two variances. We need a real-world example here instead of this made-up one.}
%\AL{Same as before. Instead of (3) above you can show the non-overlapping interval Bayes factor with \( \epsilon = (1+c)/(\sqrt{n} c) \), which will be quite similar to the default Bayes factor. It will illustrate whether the approximations used in \refSec{secInterval} yield relatively good results. This whole cutting up the prior undermines the title of the paper, and it is better just to report the default things.}
%\FD{I suggest to not do what Lexi suggests; still unclear to me what his non-overlapping Bayes factor is good for.}

In paleoanthropology, researchers study the anatomical development of modern humans. An important problem in this area is to adequately reconstruct excavated skulls. \textcite{sholts2011comparison} compared the precision of coordinate measurements of different landmark types on human crania using a 3D laser scanner and a 3D digitizer. They reconstructed five excavated skulls and found --- for landmarks of Type III, that is, the smooth part of the forehead above and between the eyebrows --- an average \revise{(across skulls)} standard deviation of $0.98$ for the Digitizer ($n_1 = 990$) and an average standard deviation of $0.89$ for the Laser ($n_2 = 990$). We define $\phi = \frac{\sigma_{\text{Digitizer}}}{\sigma_{\text{Laser}}}$ and observe that the sample effect size is $1.10$. We demonstrate two tests. First, we test whether the Laser has a lower standard deviation than the Digitizer, writing:
\begin{align*}
  \mathcal{H}_{0}&: \phi = 1 \\
  \mathcal{H}_+&: \phi > 1 \enspace .
\end{align*}
The default Bayes factor in favor of \( \mathcal{H}_{1} \) is $\text{BF}_{+0} = 4.93$ --- about double the undirected Bayes factor $\text{BF}_{+0} = 2.47$ ---  indicating moderate evidence for the hypothesis that a 3D Laser is a more precise tool for measuring Type III landmarks on the excavated human scull compared to a 3D Digitizer. Second, in this specific scenario, \revise{a researcher might treat the Digitizer as being equally as precise as the Laser when its standard deviation differs by a maximum of 10\%. She might then choose to} compare the following non-overlapping hypotheses:
\begin{align*}
  \mathcal{H}_{0}^{'} &: \phi \in [0.90, 1.10] \\
  \mathcal{H}_{+}^{'} &: \phi > 1.10 \enspace .
\end{align*}
The Bayes factor with $u = \nicefrac{1}{2}$ in favor of $\mathcal{H}'_0$ is $\text{BF}_{0+}^{'} = 7.03$, indicating moderate support for the hypothesis that the Laser and the Digitizer have about equal performance. \revise{In general, we recommend researchers use the default Bayes factor unless substantive prior knowledge or particular circumstances justify a different test.}

\subsection{The ``Standardization'' Hypothesis in Archeology}
%\AL{Again I'd recommend to report the default analysis first. By symmetry you'll get \( \BF_{r0} \leq 8 \BF_{10} \).}
%\FD{I have adjusted this section. I think Lexi means $K! = 3! = 6$ as theoretical maximum, not $8$.}

Economic growth encourages increased specialization in the production of goods, which leads to the ``standardization'' hypothesis: increased production of an item would lead to it becoming more uniform. \textcite{kvamme1996alternative} sought to test this hypothesis by studying chupa-pots, a type of earthenware produced by three different Philippine communities: the \textit{Dangtalan}, where ceramics are primarily made for household use; the \textit{Dalupa}, where ceramics are traded in a non-market based barter economy; and the \textit{Paradijon}, which houses full-time pottery specialists that sell their ceramics to shopkeepers for sale to the general public. Thus, there is an increased specialization across these three communities. \textcite{kvamme1996alternative} use circumference, height, and aperture as measures for the chupa-pots; here, we focus on the latter two. The authors test whether the standard deviations across these three groups are different, comparing:
\begin{align*}
    \mathcal{H}_{0} &: \sigma_1 = \sigma_2 = \sigma_3 \\
    \mathcal{H}_1   &: \sigma_1 \neq \sigma_2 \neq \sigma_3 \enspace ,
\end{align*}
where $\sigma_1$, $\sigma_2$, and $\sigma_3$ correspond to the standard deviations of chupa-pots in the Dangtalan, Dalupa, and Paradijon communities, respectively. Since our Bayes factor test only requires summary statistics, we can test these hypotheses using the data from Table 4 in \textcite{kvamme1996alternative}. The authors observed $n = 55$ pots from the Dangtalan community with a standard deviation in aperture of $12.74$; $n = 171$ pots from the Dalupa community with a standard deviation of $8.13$; and $n = 117$ pots from the Paradijon community with a standard deviation of $5.83$. Using our default prior choice of $u = \nicefrac{1}{2}$, we find overwhelming evidence for a difference in the standard deviations of the aperture measurements, $\log(\text{BF}_{10}) = 20$. Note that we can formulate a stronger statistical hypothesis based on the substantive ``standardization'' hypothesis, namely that the standard deviations in aperture \textit{increase} from the Paradijon to the Dangtalan community, $\mathcal{H}_r: \sigma_1 > \sigma_2 > \sigma_3$. This yields even stronger evidence, $\log(\text{BF}_{r0}) = 21.80$, such that the Bayes factor in favor of $\mathcal{H}_r$ compared to $\mathcal{H}_1$ is very close to its theoretical maximum, $\text{BF}_{r1} = 5.98 \approx 3!$. If we were to use height instead of aperture measurements of the pots, which yield standard deviations of $9.60$, $7.23$, and $7.81$, respectively, the evidence in favor of $\mathcal{H}_1$ and $\mathcal{H}_r$ compared to $\mathcal{H}_0$ would be much weaker, $\text{BF}_{10} = 2.27$ and $\text{BF}_{r0} = 2.87$, respectively.

\begin{figure}[!h]
   \centering
   \includegraphics[width = \textwidth]{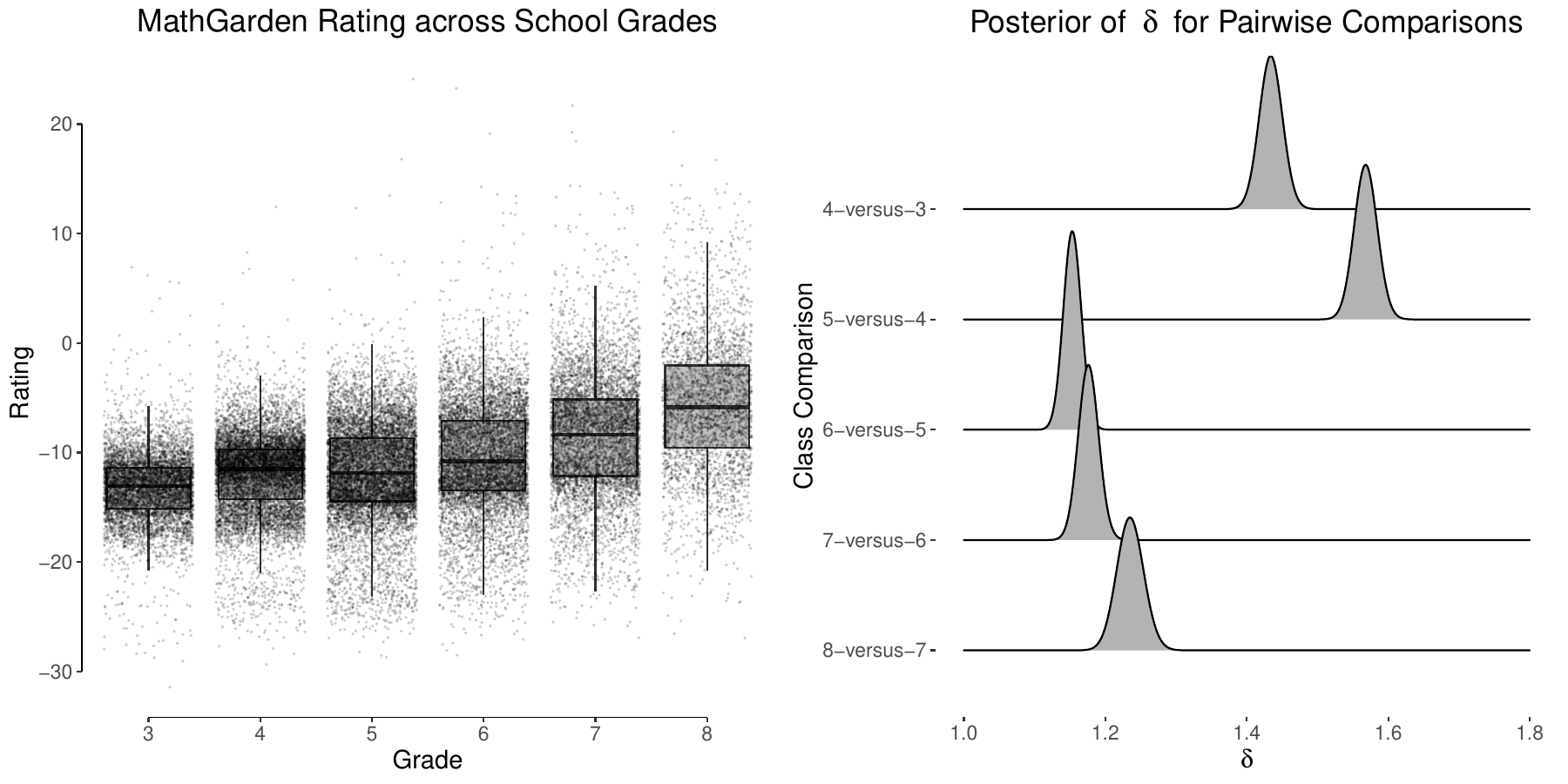}
   \caption{Left: Shows MathGarden rating scores across school grades. Right: Shows posterior of $\phi$ for pairwise \revise{consecutive} class comparisons. Virtually all probability mass is assigned to $\phi > 1$, implying that, indeed, the variance increases with every school grades.}
   \label{fig:math-garden-example}
\end{figure}

\subsection{Increased Variability in Mathematical Ability}
%\AL{Again default first. By symmetry \( \BF_{r0} \leq 64 \BF_{10} \). Perhaps mention bridge sampling here to estimate the posterior mass of the restriction.}
%\FD{Adjusted this slightly. I think he again means $6! = 720$ rather than $2^6 = 64$.}

\textcite{aunola2004developmental} find that the variance in mathematical ability increases across school grades. Using large-scale data from Math Garden, an online learning platform in the Netherlands \parencite{brinkhuis2018learning}, we assess the evidence for this hypothesis using our Bayes factor test. Math Garden assigns each pupil a rating, similar to an ELO score used in chess, and which increases if the pupil solves problems correctly. We have data from $n = 41,801$ different pupils across school grades 3 -- 8, which is visualized in the left panel of Figure \ref{fig:math-garden-example}. From grade 3 upwards, the standard deviations of the Math Garden ratings are $3.08, 3.69, 4.62, 4.97, 5.39$, and $5.99$, for respective sample sizes of $6,410$, $9,395$, $9,160$, $7,549$, $6,007$, and $3,280$. Following \textcite{aunola2004developmental}, we wish to compare the following three hypotheses:
\begin{align*}
    \mathcal{H}_{0}&: \sigma_i = \sigma_j \hspace{1em} \forall (i, j) \\
    \mathcal{H}_{1}&: \sigma_i \neq \sigma_j \hspace{1em} \forall (i, j) \\
    \mathcal{H}_r&: \sigma_i > \sigma_j \hspace{1em} \forall (i > j) \enspace .
\end{align*}
Using the default choice $u = \nicefrac{1}{2}$, we find overwhelming support in favor of a difference in the standard deviations, $\log(\text{BF}_{10}) = 1660.53$. As is suggested by the raw data visualized in the left panel of Figure \ref{fig:math-garden-example}, we also find overwhelming support for an increase in variability with increased school grade, $\log(\text{BF}_{r0}) = 1667.11$. The order-constrained hypothesis again strongly outperforms the unrestricted hypothesis, yielding evidence close to its theoretical maximum, $\text{BF}_{r1} = 719.69 \approx 6!$. The right panel in Figure \ref{fig:math-garden-example} shows the posterior distribution of $\phi$ for pairwise comparisons across school grades.

\section{Conclusion} \label{secConclusion}
In this paper, we \revise{proposed} a default Bayes factor test for assessing the (in)equality of several population variances and showed that it fulfills a number of desiderata for Bayesian model comparison \parencite[e.g.,][]{bayarri2012criteria,consonni2018prior, jeffreys1939theory,ly2016harold,ly2018phd,pena2018phd}. In addition, we extended the Bayes factor test to cover the $K-1$-sample case, non-overlapping interval nulls, and mixed restrictions for the \( K > 2 \) case. The proposed procedure allows researchers to inform their statistical tests with prior knowledge. It also generalizes Jeffreys's test for the agreement of two standard errors \parencite[][pp. 222-224]{jeffreys1939theory}; see \refApp{sec:jeffreys-parameterization}.

A limitation of the proposed methodology is that it assumes that the data follow a Gaussian distribution, which might not always be adequate in practical applications. A potential extension would be to use a $t$-distributions with a small number of degrees of freedom \revise{$\nu \geq 3$}, so as to better accommodate outliers, and then test whether the scales of these \( t \)-distributions differ. Another future avenue is to allow for data from the same unit, that is, allow for correlated observations or dependent groups. Similarly, researchers may wish to not only compare specific (substantive) hypotheses, but instead test all possible equalities. This is an important yet difficult challenge since the number of equalities grows extremely quickly with the number of groups. \textcite{gopalan1998bayesian} use a Dirichlet process prior on this large model space and use a stochastic search algorithm to estimate posterior probabilities for all possible equalities. We leave combining this approach with our default Bayes factor test for future work. For the present, we believe that our work provides an elegant Bayesian complement to popular classical tests for assessing the (in)equality of several independent population variances, ready for routine applications.

\paragraph{Author Contributions.} FD and DvB proposed the study. They both worked out the initial derivations and proofs for the deterministic \( K = 2 \) case with the help of AL. FD wrote the first draft of the manuscript and analyzed the data. FD developed the software package with the help of DvB. AL extended the results to the \( K \geq 2 \) case and provided the proofs shown in Appendices B and C. FD, DvB, and AL wrote the manuscript. EJW provided detailed feedback on the manuscript and guidance throughout. All authors read and approved the submitted version of the paper. They also declare that there were no conflicts of interest.

\paragraph{Acknowledgements.} The authors would like to thank Victor Pe\~{n}a for inspiring discussions on across-sample consistency and the editor Michele Guindani and two anonymous reviewers for their remarks on a previous version of the manuscript.

\paragraph{Funding.} FD, DvB, EJW, and AL were supported by a Vici grant no. C.2523.0278.01.

%for the \( K = 2 \) group case, which also led us to derive a one-sample Bayes factor. We have extended our procedure to allow non-overlapping hypotheses to more efficiently gather evidence in favor of the null, as well as allow researchers to translate their theoretical predictions into order-constrained or mixed hypotheses on variances for \( K > 2 \) groups. We have illustrated our method on five data sets, and we have made it available in an easy to use R package called \textit{BayesLevene}.

%Our procedure generalizes the approach of \textcite{boing2018automatic} by allowing researchers to inform their statistical tests with prior knowledge. In conjunction with testing order-constrained or mixed hypotheses, this leads to more powerful testing. When choosing a minimally informative prior --- and thus actually expressing strong beliefs about the size of the effect --- we have shown that the two testing procedures converge. Using our Bayes factor test, researchers can avoid making such implicit assumptions by explicitly specifying a more informative prior.

\printbibliography
\newpage
\appendix
\section{Jeffreys's Bayes Factor for the Agreement of Two Standard Errors} \label{sec:jeffreys-parameterization}
Our work was inspired by \textcite[][pp. 222-224]{jeffreys1939theory}, who developed a test for the ``agreement of two standard errors''. Specifically, let $\sigma_1$ and $\sigma_2$ be the standard errors for the two groups, respectively. Jeffreys estimates the standard errors by the expectation of the respective sum of squares, $(n_1 - 1)\sigma^{2}_{1}$ and $(n_2 - 1)\sigma^{2}_{2}$, where $n_1$ and $n_2$ are the respective sample sizes. Under the null hypothesis, the expectations are pooled such that $\lambda = (n_1 + n_2 - 2)\sigma_1^2$, where $\sigma_1^2 = \sigma_2^2$. Under the alternative hypothesis, we have $\lambda = (n_1 - 1)\sigma_1^2 + (n_2 - 1)\sigma_2^2$, which can be written as a mixture such that $(n_1 - 1)\sigma^{2}_{1} = \vartheta \lambda$ and $(n_2 - 1)\sigma^{2}_{2} = (1 - \vartheta) \lambda$. Because $\lambda$ is common to both models, we can assign it an improper prior and integrate it out. The test-relevant parameter is $\vartheta \in [0, 1]$, which Jeffreys assigns a uniform prior. After Laplace-approximating the integral under the alternative, Jeffreys arrives at the (approximate) Bayes factor:
\begin{equation}
\text{BF}^J_{01} = \frac{(N - 2)^{3/2}}{2\sqrt{\pi(n_1 - 1)(n_2 - 1)}} \exp \left (2 \frac{n_2 - n_1}{N - 2} z - \frac{(n_1 - 1)(n_2 - 1)}{N - 2}z^2\right) \enspace ,
\end{equation}
where $N = n_1 + n_2$ and $z = \log \left(\frac{s_1}{s_2}\right)$, and where $s_1$ and $s_2$ are the sample standard deviations.

As a side note, we first attempted a parameterization that, unbeknownst to us, Jeffreys substituted for his 1939 averaging idea in the third edition of the \textit{Theory of Probability} \parencite{jeffreys1961theory}: $\sigma_1^2 = \sigma_2^2 e^{\xi}$. We abandoned this idea because we could not generalize it to \( K > 2 \) groups and instead adopted Jeffreys's original averaging idea.

\begin{figure}
   \centering
   \includegraphics[width = 0.85\textwidth]{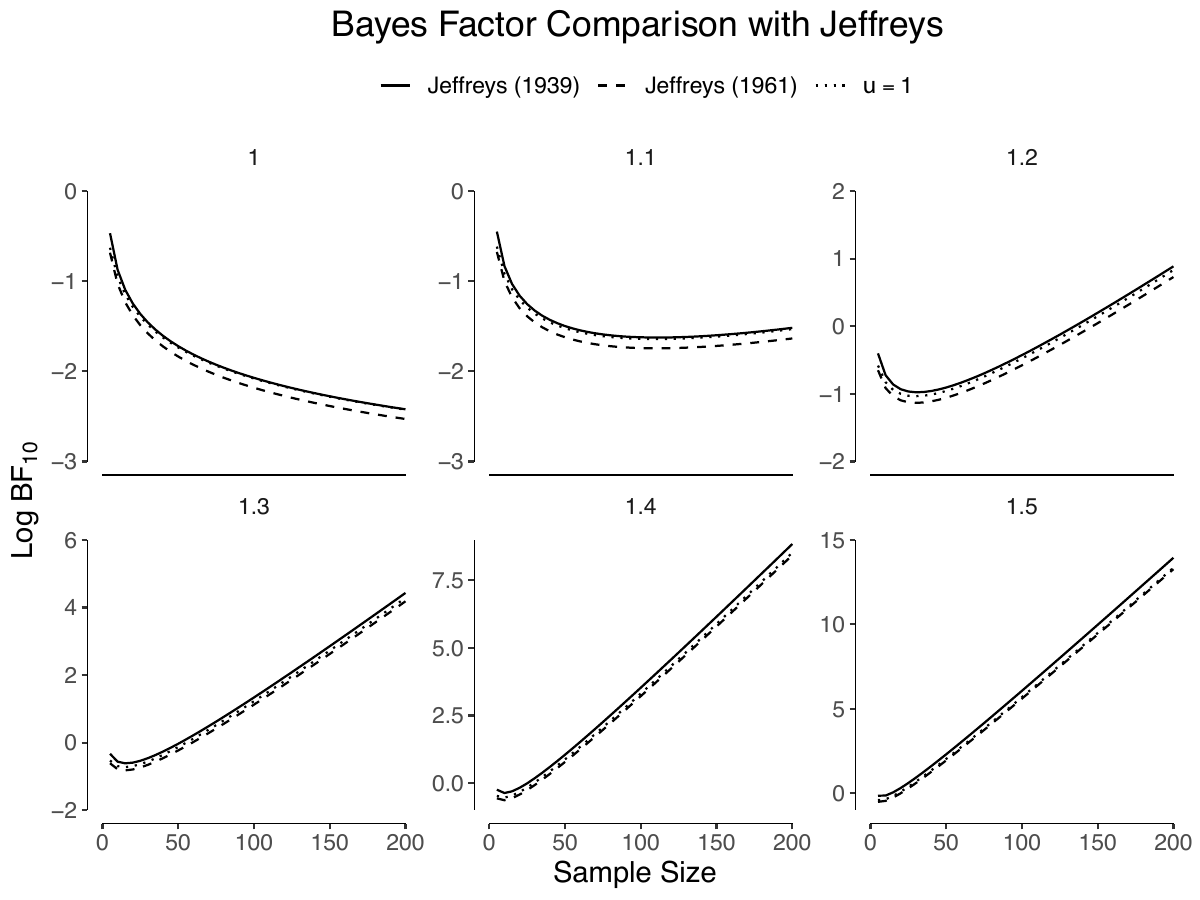}
   \caption{Comparison of the Bayes factor proposed by \textcite{jeffreys1939theory} and our Bayes factor with $u = 1$ for \( K = 2 \) groups as a function of the sample size and the effect size $\phi = \{1, 1.1, 1.2, 1.3, 1.4, 1.5\}$. }
   \label{fig:compare-jeffreys}
\end{figure}

Figure \ref{fig:compare-jeffreys} shows that our Bayes factor with $u = 1$ matches Jeffreys's 1939 Bayes factor very closely, as is expected from the uniform prior on $\vartheta$. The error is due to his approximate solution. For completeness, we also show Jeffreys's 1961 Bayes factor, which is not limit consistent. It strikes us as a curiosity that Jeffreys would develop a test for the standard error instead of the population variance. Since the standard error decreases with the (square root of) the sample size, applying Jeffreys's test to data of unequal group sizes confounds the result (if we were to take his test as a test concerning equality of variances). Formally, both Bayes factors Jeffreys derived are not limit consistent because if we gather infinite data for only one group, \revise{the Bayes factor in favor of \( \mathcal{H}_{1} \)} will go to infinity instead of converging to a bound \parencite[][ch. 6]{ly2018phd}. \revise{For} our Bayes factor, we adopt Jeffreys's averaging idea \revise{to parameterize the problem}, but we focus on the population \revise{precisions} instead of the standard errors.

\color{black}
\section{Derivation of the proposed Bayes factor}
\label{appDerivationBf}

\subsection{Integrating out the nuisance parameters}
Let \( Y_{ji} \overset{\textnormal{iid}}{\sim} \mathcal{N} ( \mu_{j} , \tau_{j}^{-1}) \), where \( i=1, 2, \ldots, n_{j} \) and \( j \in [K] \). For both the null and the alternative models we integrate the nuisance parameters \( \mu_{j} \)s out with respect to the right Haar priors \( \mu_{j} \propto 1 \). This implies that for the observations \( y^{ \{ j \} } \) from the \( j \)th group consisting of \( n_{j} \) observations the likelihood function is %can be summarised by \( \bar{y}_{j}, s_{j}^{2} \) and its likelihood is
\begin{align}
f( y^{ \{ j \} } \, | \, \tau_{j}) & := \int f( y^{ \{ j \} } \, | \, \mu_{j}, \tau_{j}) \pi(\mu_{j}) \der \mu_{j}, \\
& =  (2 \pi)^{-\tfrac{n_{j}}{2}} \tau_{j}^{\tfrac{n_{j}}{2}} \exp(- \tfrac{1}{2} \nu_{j} s_{j}^{2} \tau_{j}) \int \exp ( - \tfrac{n}{2} \tau_{j} ( \bar{y}_{j} - \mu_{j})^{2}) \der \mu_{j} , \\
\label{eqMarginalTau}
& =  (2 \pi)^{-\tfrac{\nu_{j}}{2}} n_{j}^{-\tfrac{1}{2}} \tau_{j}^{\tfrac{\nu_{j}}{2}} \exp(- \tfrac{1}{2} \nu_{j} s_{j}^{2} \tau_{j}) .
\end{align}
%
%The next step requires choosing priors on \( \vec{\tau} = (\tau_{1}, \ldots, \tau_{K}) \) for the null and alternative model.
%
%\section{Reparametrisation and priors on the remaining nuisance parameter}
%For the priors on the precisions, we reparametrise them in terms of the average precision \( \bar{\vecB{\tau}} \) and the proportion \( \vartheta \) of the total precision
%%
%\begin{align}
%\bar{\vecB{\tau}} = \frac{1}{K} \tau_{+}, \quad \vartheta_{j} =  \frac{ \tau_{j}}{ \tau_{+} } \text{ for } k \in [K],
%\end{align}
%%
%where we used the subscript \( + \) to denote summing over all elements, e.g., \( \tau_{+} := \sum_{j=1}^{K} \tau_{j} \). The null hypothesis \( \tau_{i} = \tau_{j} \) for all \( i, j \in [K] \) implies that \( \vartheta_{j} = 1/K \).
%
For data from the \( K \) samples combined, i.e., \( y^{[K]} \), and the parametrisation \( \tau_{j} = \vartheta_{j} \bar{\vecB{\tau}} K \) this yields
\begin{align}
\label{eqLikelihoodFull}
f( y^{[K]} \, | \, \vec{\vartheta}, \bar{\vecB{\tau}}) = (2^{-1} K)^{-\tfrac{\vecB{\nu}_{+}}{2}} C(n) \Big [ \prod_{j=1}^{K} \vartheta_{j}^{\tfrac{\nu_{j}}{2}} \Big ]  \bar{\vecB{\tau}}^{\tfrac{\vecB{\nu}_{+}}{2}} \exp \big ( - 2^{-1} K \bar{\vecB{\tau}} \sum_{j=1}^{K} \vartheta_{j} \nu_{j} s_{j}^{2} \big )  ,
\end{align}
where \( C(n) = (2 \pi)^{- \vecB{\nu}_{+}/2} (n_{1} \ldots n_{K})^{1/2} \) and \( \vecB{\nu}_{+} = \sum_{j=1}^{K} \nu_{j} \). A natural prior on the nuisance parameter \( \bar{\vecB{\tau}} \) is \( \pi(\bar{\vecB{\tau}}) \propto \bar{\vecB{\tau}}^{-1} \) and a standard gamma integral leads to the marginalized likelihood
\begin{align}
\label{eqMarginalisedLikelihood}
\tilde{h}( y^{[K]} \, | \, \vec{\vartheta}) & = \int f( y^{[K]} \, | \, \vec{\vartheta}, \bar{\vecB{\tau}}) \pi(\bar{\vecB{\tau}}) \der \bar{\vecB{\tau}} = C(n) \Gamma \big ( \frac{\vecB{\nu}_{+}}{2} \big ) \Big [ \prod_{j=1}^{K} \vartheta_{j}^{\tfrac{\nu_{j}}{2}} \Big ]  \Big ( \sum_{j=1}^{K} \vartheta_{j} \nu_{j} s_{j}^{2} \Big )^{- \tfrac{\vecB{\nu}_{+}}{2}} .
\end{align}
Since \( \vartheta_{j} > 0 \) and \( \sum_{j=1}^{K} \vartheta_{j} = 1 \) the vector \( \vecB{\vartheta}:=(\vartheta_{1}, \ldots, \vartheta_{K}) \) can be fully described by \( K-1 \) free parameters. Any \( \vartheta_{j} \) can be singled out in the following, but for concreteness, we do so for the \( K \)th one. To rewrite the marginalized likelihood \( \tilde{h}( y^{[K]} \, | \, \vec{\vartheta}) \) in terms of the \( K-1 \) proportions \( \vartheta \), note that
\begin{align}
\sum_{j=1}^{K} \vartheta_{j} \nu_{j} s_{j}^{2} & = \vartheta_{1} \nu_{1} s_{1}^{2} + \vartheta_{2} \nu_{2} s_{2}^{2} + \ldots + \vartheta_{K-1} \nu_{K-1} s_{K-1}^{2} + \big ( 1 - \sum_{j=1}^{K-1} \vartheta_{j} \big )  \nu_{K} s_{K}^{2} \\
& = \nu_{K} s_{K}^{2} - \sum_{j=1}^{K-1} [ \nu_{K} s_{K}^{2} - \nu_{j} s_{j}^{2}] \vartheta_{j} ,
\end{align}
which implies that
\begin{align}
\label{eqMarginalisedLikelihood2}
\Big ( \sum_{j=1}^{K} \vartheta_{j} \nu_{j} s_{j}^{2} \Big )^{- \tfrac{\vecB{\nu}_{+}}{2}}   = (\nu_{K} s_{K}^{2})^{- \tfrac{\vecB{\nu}_{+}}{2}} \Big ( 1 - \sum_{j=1}^{K-1}  [ 1 - \tfrac{\nu_{j} s_{j}^{2}}{\nu_{K} s_{K}^{2}}] \vartheta_{j} \Big )^{- \tfrac{\vecB{\nu}_{+}}{2}} .
\end{align}
This leads to
\begin{align}
\tilde{h}( y^{[K]} \, | \, \vec{\vartheta}) & = C(n) \Gamma \big ( \frac{\vecB{\nu}_{+}}{2} \big ) (\nu_{K} s_{K}^{2})^{- \tfrac{\vecB{\nu}_{+}}{2}}  \Big [ \prod_{j=1}^{K} \vartheta_{j}^{\tfrac{\nu_{j}}{2}} \Big ] \Big ( 1 - \sum_{j=1}^{K-1} [ 1 - \tfrac{\nu_{j} s_{j}^{2}}{\nu_{K} s_{K}^{2}})] \vartheta_{j} \Big )^{- \tfrac{\vecB{\nu}_{+}}{2}} ,
\end{align}
which will be used to derive desiderata on the prior on the test relevant parameters. To highlight the fact that \( \vec{\vartheta} \) is effectively \( K-1 \) dimensional, we can replace \( \Big [ \prod_{j=1}^{K} \vartheta_{j}^{\tfrac{\nu_{j}}{2}} \Big ] = \Big [ \prod_{j=1}^{K-1} \vartheta_{j}^{\tfrac{\nu_{j}}{2}} \Big ] (1- \vec{\vartheta}_{+})^{\tfrac{\nu_{K}}{2}} \), where \( \vec{\vartheta}_{+} := \sum_{j=1}^{K-1} \vartheta_{j} \).

\subsection{Deriving the proposed Bayes factors}
\label{app:bayes-factor}
The marginalized likelihood fully specifies the marginal likelihood of the null, as the plugin \( \vartheta_{j} = 1/K \) yields
\begin{align}
\label{eq:marginal-m0}
p( y^{[K]} \, | \, \Mc_{0}) & = %C(n) \Gamma \big ( \tfrac{\vecB{\nu}_{+}}{2} \big ) \Big ( \sum_{j=1}^{K} \nu_{j} s_{j}^{2} \Big )^{- \tfrac{\vecB{\nu}_{+}}{2}} =
C(n) \Gamma \big ( \frac{\vecB{\nu}_{+}}{2} \big ) (\nu_{K} s_{K}^{2})^{-\tfrac{\vecB{\nu}_{+}}{2}} \Big ( 1 + \sum_{j=1}^{K-1}  \tfrac{\nu_{j} s_{j}^{2}}{\nu_{K} s_{K}^{2}} \Big )^{- \tfrac{\vecB{\nu}_{+}}{2}} .
\end{align}
We let \( h( y^{[K]} \, | \, \vec{\vartheta}) = \frac{\tilde{h}( y^{[K]} \, | \, \vec{\vartheta})}{\tilde{h}( y^{[K]} \, | \, \vec{\vartheta}=\tfrac{1}{K})}  \) be the reduced likelihood, see \refEq{eqReducedLikelihood}, and the Bayes factor is then
\begin{align}
\BF_{10}(y^{[K]}) & =  \Big ( 1 + \sum_{j=1}^{K-1}  \tfrac{\nu_{j} s_{j}^{2}}{\nu_{K} s_{K}^{2}} \Big )^{\tfrac{\vecB{\nu}_{+}}{2}}  \\
\times & \int \Big [ \prod_{j=1}^{K-1} \vartheta_{j}^{\tfrac{\nu_{j}}{2}} \Big ] (1 - \vec{\vartheta}_{+} )^{\tfrac{\nu_{K}}{2}} \Big ( 1 - \sum_{j=1}^{K-1} [ 1 - \tfrac{\nu_{j} s_{j}^{2}}{\nu_{K} s_{K}^{2}}] \vartheta_{j} \Big )^{- \tfrac{\vecB{\nu}_{+}}{2}} \pi_{1}(\vec{\vartheta}) \der \vec{\vartheta} ,
\end{align}
where \( \vec{\vartheta} \in \R^{K-1} \), and the integral is over the \( K-1 \) simplex. A natural prior for \( \vec{\vartheta} \) would be a Dirichlet prior with hyperparameters \( \vecB{u} \), where \( \vecB{u} = ( u_{1}, \ldots, u_{K-1}, u_{K}) \) with non-negative components. For \( \nu_{j} \geq 1  \) for all \( j \in [K] \) and by definition of the multivariate integral representation of the type D Lauricella function of \( K-1 \) variables \parencite{lauricella1893sulle}, this Bayes factor is analytic and given by %
\begin{align}
%\label{eqBfMulti}
\BF_{10}(y^{[K]}) & =  \tfrac{\Bc ( \tfrac{\vec{\nu}}{2}  + \vec{u}) }{\Bc ( \vec{u})} \Big ( 1 + \sum_{j=1}^{K-1}  \tfrac{\nu_{j} s_{j}^{2}}{\nu_{K} s_{K}^{2}} \Big )^{\tfrac{\vecB{\nu}_{+}}{2}}  F_{D} \Big ( \tfrac{\vecB{\nu}_{+}}{2} \, ; \, \tfrac{ \vec{\nu}}{2} + \vec{u} \, ; \, \tfrac{\vecB{\nu}_{+}}{2} + u_{+}  \, ; \, \vec{1} - \tfrac{ \overrightarrow{\nu s^{2}} }{ \nu_{K} s_{K}^{2}} \Big )
%
%\tfrac{\nu_{1}}{2} + u_{1}, \ldots, \tfrac{\nu_{K-1}}{2} + u_{K-1} \, ; \,  \tfrac{\vecB{\nu}_{+}}{2} + u_{+} \, ; \, 1- \tfrac{\nu_{1} s_{1}^{2}}{\nu_{K} s_{K}^{2}}, \ldots, 1- \tfrac{\nu_{K-1} s_{K-1}^{2}}{\nu_{K} s_{K}^{2}} \Big )
\end{align}
%
%\begin{sloppypar}
where \( \Bc ( \vec{u}) = \frac{ \Gamma( u_{1}) \cdots \Gamma(u_{K}) }{ \Gamma ( u_{+} ) }  \) is the multivariate beta function, \( \vec{1}= ( 1, \ldots, 1) \in \R^{K-1} \) and where \( \overrightarrow{\nu s^{2}} = (\nu_{1} s_{1}^{2}, \ldots, \nu_{K-1} s_{K-1}^{2}) \) is the \( K-1 \) vector of sums of squares. %, and \( \vec{1} = (1, \ldots, 1) \in \R^{K-1} \).
%\end{sloppypar}

%The type D Lauricella function in the Bayes factor and its properties are helpful in the analysis, but for computational purposes, we recommend the simpler one-dimensional integral representation
%%
%\begin{align}
%\frac{F_{D}(a \, ; \, \vec{b} \, ; \, d \, ; \, \vec{x} )}{\Bc(a, d-a)}  & = \frac{F_{D} (a \, ; \, b_{1}, \ldots, b_{K-1} \, ; \, d \, ; \, x_{1}, \ldots, x_{K-1})}{\Bc(a, d-a)} \\
%& =  \int_{0}^{1} t^{a-1} (1 - t)^{d-a-1} (1 - x_{1}t)^{-b_{1}} \cdots (1- x_{K-1} t)^{-b_{K-1}} \der t
%\end{align}
%%
%whenever \( d > a \).
%
%To see that the Bayes factor is in fact analytic we note that for \( \vec{x} = (x_{1}, \ldots, x_{K-1}) \) such that \( | x_{k} | < 1 \), the type D Lauricella function %
%%%
%%\begin{align}
%%F_{D}(a \, ; \, \vec{b} \, ; \, d \, ; \, \vec{x}) =F_{D} (a \, ; \, b_{1}, \ldots, b_{K-1} \, ; \, d \, ; \, x_{1}, \ldots, x_{K-1})
%%\end{align}
%%%
%has the series representation
%%
%\begin{align}
%F_{D}(a \, ; \, \vec{b} \, ; \, d \, ; \, \vec{x}) := \sum_{i_{1}, \ldots, i_{K-1}=0}^{\infty} \frac{ (a)_{\vec{i}_{+}} (b_{1})_{i_{1}} \ldots (b_{K-1})_{i_{K-1}}}{(d)_{\vec{i}_{+}} i_{1}! \cdots i_{K-1}!} x_{1}^{i_{1}} \cdots x_{K-1}^{i_{K-1}} .
%\end{align}
%%
%Note that this series representation is permissible for the Bayes factor \refEq{eqBfMulti} by taking the largest sums of squares as the \( K \)th term.
%
%Simpler go for \( F_{D} \) %Butler and Wood
%\( F_{B} \) Lauricella p. 146 (p. 36 and p. 37).

\subsection{Default approach to non-overlapping Bayes factors}
\label{app:bayes-factor-interval}
Note that non-overlapping hypotheses can also be expressed in terms of the location parameter \( \delta = - \log ( \frac{\vartheta}{1 - \vartheta}) \in \R \), which transforms the point null hypothesis \( \Hc_{0} : \vartheta = 1/2 \) to \( \Hc_{0}: \delta = 0 \) yielding a comparison between %and the following two non-overlapping hypotheses can be considered
\begin{align}
\breve{\mathcal{H}}_{0}: |\delta| < \epsilon  \text{ and } \breve{\mathcal{H}}_{1}: |\delta| > \epsilon ,
\end{align}
where \( \epsilon \) defines the half width of the null-region. \textcite{berger1987testing} showed that for the location problem \( \bar{X} \sim \Nc( \delta, \sigma^{2}/n) \) with a unimodal and symmetric prior, and \( \epsilon \leq \tfrac{\sigma}{2 \sqrt{n}} \), the standard (point null) Bayes factor characterizes the behavior of the null-region Bayes factor comparing \( \breve{\Hc}_{1} \) to \( \breve{\Hc}_{0} \) with the priors truncated accordingly.

Note that the prior \( \vartheta \sim \Beta(u, u) \) underlying \refEq{eqBfKTwo} induces a type III generalized logistic distribution on \( \delta \) with density %
\begin{align}
\pi(\delta) = \tfrac{1}{\Bc ( u, u)} e^{-\delta u} ( 1 + e^{-\delta})^{- 2 u }.
\end{align}
This prior is unimodal and symmetric around \( \Hc_{0}: \delta = 0 \). In terms of \( \delta \) the marginalized likelihood \( \tilde{h}(y^{[K]} \, | \, \vartheta) \), see appendix \refEq{eqMarginalisedLikelihood2}, is
\begin{align}
\tilde{h}(y^{[K]} \, | \, \delta) \propto \exp ( - n g(\delta) ), \text{ where } g(\delta) \approx \tfrac{c}{2} \delta + \tfrac{ 1 + c}{2} \log ( 1 + \tfrac{ s_{1}^{2}}{s_{2}^{2}} c e^{-\delta}) ,
\end{align}
whenever \( n_{1} = c n \) and \( n_{2} = n \). Sufficiently large \( n \) combined with a Taylor expansion of \( g(\delta) \) at its maximum point, that is, at \( \hat{\delta} = \log ( \tfrac{s_{1}^{2}}{s_{2}^{2}}) \), yields the approximation
\begin{align}
\tilde{h}(y^{[K]} \, | \, \delta) \propto \exp \Big ( - \tfrac{n c}{4 (1+c)} \big ( \delta - \log ( s_{1}^{2} /s_{2}^{2}) \big )^{2} \Big ) .
\end{align}
Hence, one way to take a null interval is by setting \( \epsilon \leq \tfrac{ (1 + c)}{\sqrt{n} c} \). The resulting null-region Bayes factor will then behave similarly to \refEq{eqBfKTwo}.

\section{Properties of the proposed Bayes factor}
\label{app:desiderata}
%We now show that the proposed Bayes factor \refEq{eqBfMulti} has the finite-sample properties of being (1) predictively matched and (2) information consistent, and that it has the asymptotic properties of being (3) across sample consistent and (4) model selection consistent. For (1) the right Haar priors on \( \mu_{j} \propto 1 \), \( \bar{\vecB{\tau}} \propto \bar{\vecB{\tau}}^{-1} \) and a proper prior on \( \vartheta \) suffices. For (2) a Dirichlet prior with parameters \( \vec{u} = (u_{1}, \ldots, u_{K}) \) with \( u_{j} \leq 1/2 \) for \( j \in [K] \) suffices. The conditions for (1) and (2) suffice for the large sample properties (3) and (4). % do not require TODO

\subsection{Labelling Invariant}
\label{appLabellingInvariant}
\begin{proof}[Proof of labelling invariance, \refThm{thmLabellingInvariance}]
The goal is to show that the integral of the reduced likelihood times prior remains the same after applying the permutation \( \varrho \) that swaps the labels \( K \) for an arbitrary \( i \in [K-1] \). For this integral to remain the same, it suffices to show that the reduced likelihood \( h( \vecB{s^{2}} \, | \, \vec{\vartheta})  \) and its permuted version
\begin{align}
h(\varrho( \vecB{s^{2}}) \, | \, \vec{\vartheta}) & = \Big ( 1 + \tfrac{\nu_{K} s_{K}^{2}}{\nu_{i} s_{i}^{2}} + \sum_{j \in [K-1] \setminus \{ i \}}  \tfrac{\nu_{j} s_{j}^{2}}{\nu_{i} s_{i}^{2}} \Big )^{\tfrac{\vecB{\nu}_{+}}{2}}  \Big [ \prod_{j \in [K-1] \setminus \{ i \} } \vartheta_{j}^{\tfrac{\nu_{j}}{2}} \Big ] \\
\times & \vartheta_{i}^{\tfrac{\nu_{K}}{2}}  (1 - \vec{\vartheta}_{+} )^{\tfrac{\nu_{i}}{2}} \Big ( 1 -  \vec{\theta}_{+} + \tfrac{\nu_{K} s_{K}^{2}}{\nu_{i} s_{i}^{2}} \vartheta_{i}  + \sum_{j \in [K-1] \setminus \{ i \} } \tfrac{\nu_{j} s_{j}^{2}}{\nu_{i} s_{i}^{2}} \vartheta_{j} \Big )^{- \tfrac{\vecB{\nu}_{+}}{2}} ,
\end{align}
are conditionally symmetric. This means that as a function of \( \vartheta_{i} \) with all other coordinates fixed, i.e., \( \vartheta_{j} \) for \( j \in [K-1] \setminus \{ i \} \), the reduced likelihood and its permuted version are symmetric around \( \breve{\vartheta}_{-i} : = \tfrac{1}{2} \big ( 1 - \sum_{j \in [K-1] \setminus \{ i \}} \vartheta_{j} \big ) \).

This can be shown by studying the functions \( g(x) \) and \( g_{\varrho}(-x) \), where \( g(x) \) is the composition of \( x \mapsto \vartheta_{i} = \breve{\vartheta}_{-i}  + x \) and \( \vartheta_{i} \mapsto h( \vecB{s^{2}} \, | \, \vec{\vartheta}) \), whereas \( g_{\varrho}(-x) \) is the composition of \( x \mapsto \vartheta_{i} = \breve{\vartheta}_{-i}  - x \) and \( \vartheta_{i} \mapsto h( \varrho(\vecB{s^{2}}) \, | \, \vec{\vartheta}) \). A straightforward, but tedious computation then shows that \( g( x ) = g_{\varrho}(- x) \) for all \( x \in (0, \breve{\vartheta}_{-i}) \). For the Bayes factor to be labelling invariant, we thus require the prior to be symmetric in the similar fashion. For the Dirichlet prior this implies \( u_{i} = u_{K} \), and for this to hold for all pairs of permutations, we require \( u_{j} = u \) for all \( j \in [K] \).
\end{proof}

\subsection{Predictive Matching}
\label{appPredictiveMatching}
%
%
%\begin{theorem*}
%The proposed Bayes factor \refEq{eqBfMulti} is predictively matched. \( \hfill \diamond \)
%\end{theorem*}

\begin{proof}[Proof of predictive matching, \refThm{thmPredictiveMatching}]
Case (a) with \( n_{1} = \ldots = n_{K} = 1 \) implies that \( \nu_{1} s_{1}^{2} =  \ldots = \nu_{K} s_{K}^{2}  = 0 \) regardless of the data, which implies that the likelihood of the data \refEq{eqLikelihoodFull} is identical to the constant function 1, thus, independent of \( \bar{\vecB{\tau}} \) and \( \vec{\vartheta} \). Viewing the prior \( \bar{\vecB{\tau}} \propto \bar{\vecB{\tau}}^{-1} \) on the nuisance parameter that appears in both the numerator and the denominator of the Bayes factor as a limit of \( \bar{\vecB{\tau}} \sim \Gamma(u, u ) \) with \( u \downarrow 0 \) shows that without loss of generality we can set the Bayes factor to 1, whenever \( \pi_{1}(\vec{\vartheta}) \) is proper.

For case (b) and without loss of generality we consider the case with \( \nu_{K}=1 \) and \( \nu_{j}=0 \) for all \( j \in [K-1] \). The reduced likelihood \( h(\vecB{s^{2}} \, | \, \vec{\vartheta}) \) is then actually independent of \( s_{K}^{2} \), as we then get
\begin{align}
\BF_{10}(\vecB{s^{2}}) & = \frac{\int (s_{K}^{2})^{- \tfrac{1}{2}} ( 1 - \vec{\vartheta}_{+})^{ \tfrac{1}{2}} ( 1 - \vec{\vartheta}_{+})^{ - \tfrac{1}{2}} \pi_{1}(\vec{\vartheta}) \der \vartheta}{(s_{K}^{2})^{- \tfrac{1}{2}} ( \tfrac{1}{K})^{ \tfrac{1}{2}} ( 1 - \tfrac{ K -1}{K} )^{ - \tfrac{1}{2}}} = \int \pi_{1}(\vec{\vartheta}) \der \vec{\vartheta} .
\end{align}
%
%Thus, \( \BF_{10}(\vecB{s^{2}}) = 1 \) whenever \( \pi_{1}(\vec{\vartheta}) \) is proper.
%
Thus, for all data sets \( \vecB{s^{2}} \) of insufficient size \( \BF_{10}(\vecB{s^{2}}) = 1 \) whenever \( \pi_{1}(\vec{\vartheta}) \) is proper.
\end{proof}

\subsection{Information Consistency}
\label{appInformationConsistency}
\begin{proof}[Proof of information consistency, \refThm{thmInformationConsistent}]
Assuming labelling invariance we can let the \( s_{K}^{2} \) with fixed \( n_{K}  \) grow without loss of generality. For fixed \( \vecB{n} \) the order of integral and limit can be interchanged and reveals that %
\begin{align}
\lim_{s_{K}^{2} \rightarrow \infty} \BF_{10}(\vecB{s^{2}}) & = \Bc(\vecB{u})^{-1} \int \big ( \prod_{j=1}^{K-1} \vartheta_{j}^{\tfrac{\nu_{j}}{2} + u_{j} -1} \big ) (1 - \vec{\vartheta}_{+})^{\tfrac{\nu_{K}-\vecB{\nu}_{+}}{2} + u_{K} -1} \der \vec{\vartheta} .
\end{align}
%
%\begin{sloppypar}
The integrand becomes unbounded whenever \( u_{K} \leq \tfrac{\vecB{\nu}_{+}- \nu_{K}}{2} \). Recall that the minimal sample size has only two groups with two observations, say, \( \nu_{1} = 1 \) and \( \nu_{K} = 1 \). The requirement that \( \lim_{s_{K}^{2} \rightarrow \infty} \BF_{10}(\vecB{s^{2}}) \) should already diverge at the minimal sample sizes implies that \( u_{K} \leq 1/2 \). By symmetry we require this for all \( u_{j} \) for \( j \in [K] \).
%\end{sloppypar}
\end{proof}

\subsection{Model selection consistency}
\label{appModelSelectionConsistency}
For model selection consistency we note that the Bayes factor depends on the data via the statistic \( \vec{W} = ( W_{1}, \ldots, W_{K-1}) \) with
\begin{align}
W_{j} := \frac{ \nu_{j} s_{j}^{2}}{\nu_{K} s_{K}^{2}} = \frac{\sigma_{j}^{2} \nu_{j}  }{\sigma_{K}^{2} \nu_{K}}  \frac{  \Big (  \sum_{i=1}^{n_{j}} \tfrac{ ( Y_{ji} - \bar{Y}_{j})^{2}}{ \sigma_{j}^{2}} \Big ) / \nu_{j} }{ \Big ( \sum_{i=1}^{n_{K}} \tfrac{ ( Y_{Ki} - \bar{Y}_{K})^{2}}{ \sigma_{K}^{2}} \Big ) / \nu_{K}} =: \frac{\sigma_{j}^{2} \nu_{j}  }{\sigma_{K}^{2} \nu_{K}} X_{j} , \text{ for } j \in [K-1] ,
\end{align}
where \( X_{j} \sim F( \nu_{j} , \nu_{K}) \) is an \( F \)-distributed random variable with degrees of freedom \( \nu_{j} \) and \( \nu_{K} \) by virtue of the data being normally distributed.

Letting \( n_{j} := c_{j} n \) for \( c_{j} > 0 \), \( j \in [K] \), thus, \( c_{K}=1 \), and \( \sigma_{j}^{2} := \gamma_{j} \sigma_{K}^{2} \) where \( \gamma_{j} > 0 \) for \( j \in [K] \), thus, \( \gamma_{K}=1 \), note that \( W_{j} \approx c_{j} \gamma_{j} X_{j} \) for \( n \) large. Observe that since \( X_{j} \) is \( F \)-distributed we know that %for each \( j \in [K-1] \) the expectation and variance of \( X_{j} \) are %
\begin{align}
E (X_{j} ) = \frac{n}{n-2} = 1 + \Oc(1/n) \text{ and } \var ( X_{j}) = \frac{2 n^{2}( (1+c_{j}) n - 2)}{c_{j} n ( n - 2)^{2} (n -4)} = \Oc(1/n).
\end{align}
Hence, Chebyshev's inequality can be applied to show that \( X_{j} - 1 = \Oc_{P}(n^{-1/2}) \). The intuition to use the continuous mapping theorem and the replacement \( \vec{X} = \vec{1} \in \R^{K-1} \) in \( \BF_{10} \) forms the basis of the proof of \refThm{thmModelSelectionConsistency}. What needs taking care of is the dependence of the Bayes factor on \( n \).

\begin{proof}[Proof of model selection consistency, \refThm{thmModelSelectionConsistency}]
The proof relies on a Taylor approximation that holds with high probability and the subsequent asymptotic analysis of the Taylor terms. Key to this analysis is the large sample behavior of gamma functions. What is remarkable is that under the null the exponential growing terms cancelled out perfectly in all Taylor terms.

\paragraph{Notation for partial derivatives}
For the Taylor terms, we express the Bayes factor as follows \( \BF_{10}( \vecB{s}^{2}, n) = \tfrac{ \Bc( \tfrac{n}{2} \vecB{c} + \vecB{u} ) }{\Bc ( \vecB{u})} b(\vec{X}) G_{D}(\vec{X})  \), where with \( \vec{Z} \in \R^{K-1} \), \( Z_{j} = 1- c_{j} \gamma_{j} X_{j} \), the \( \vec{X} \)-dependent functions are
\begin{align}
b(\vec{X}) & := ( 1 + \sum_{j=1}^{K-1} c_{j} \gamma_{j} X_{j})^{ \tfrac{ \vecB{c}_{+} }{2} n}, \\
G_{D}(\vec{X}) & := F_{D} ( \tfrac{ \vecB{c}_{+} }{2} n \, ; \, \tfrac{ n}{2} \vec{c} + \vec{u} \, ; \, \tfrac{ \vecB{c}_{+} }{2} n + \vecB{u}_{+} \, ; \, \vec{Z}) .
\end{align}
For the Taylor series we employ multi-index notation to describe Leibniz's product rule for partial derivatives. The idea is to identify a partial derivative to a \( K-1 \)-dimensional vector of non-negative integers \( \vec{m} \in \N_{0}^{K-1} \). Each \( m_{j} \) represents the multiplicity of partial derivative with respect to the variable \( x_{j} \), thus, \( \partial^{\vec{m}} b(\vec{X}) := \tfrac{ \partial^{ \vec{m}_{+}}}{ \prod_{j=1}^{K-1} \partial x_{j}^{m_{j}}} b(\vec{X}) \) and more specifically
\begin{align}
\partial^{\vec{m}} b(\vec{X}) & =  (\tfrac{\vecB{c}_{+} }{2} n )_{-\vec{m}_{+}} \Big ( \prod_{j=1}^{ K- 1} (c_{j} \gamma_{j})^{m_{j}} \Big ) ( 1 + \sum_{j=1}^{K-1} c_{j} \gamma_{j} X_{j})^{ \tfrac{ \vecB{c}_{+} }{2} n - \vec{m}_{+}}  ,
\end{align}
where \( (a)_{-l} := \Gamma ( a + 1) / \Gamma ( a - l + 1) \) denotes the falling factorial, e.g., \( (a)_{-3} = a (a-1) (a-2) \) for \( a \in \N \). It can be shown that \( (a)_{-l} = (-1)^{l} (-a)_{l} \) and that \( (a)_{-l}/l! = \binom{a}{l}  \). Note that \( b(\vec{X}) \) also appears on the right-hand side. To simplify notation we write %As the Taylor series is going to be considered around the point \( \vec{X} = \vec{1} \) we write %
\begin{align}
\partial^{\vec{m}} b :=  \partial^{\vec{m}} b(\vec{X}) \Big |_{\vec{X} = \vec{1}} = \big ( \la \vecB{c}, \vecB{\gamma} \ra \big )^{\tfrac{ \vecB{c}_{+}}{2} n} \tfrac{ (\tfrac{\vecB{c}_{+} }{2} n )_{-\vec{m}_{+}} \Big ( \prod_{j=1}^{ K- 1} (c_{j} \gamma_{j})^{m_{j}} \Big ) }{\big ( \la \vecB{c}, \vecB{\gamma} \ra \big )^{\vec{m}_{+}}}.
\end{align}
Note that the first order partial derivatives are described by the vectors \( \vec{m} = \vec{e}_{k} \) for \( k \in [K-1] \).

Similarly, let \( \vec{l} \in \N_{0}^{K-1} \) with \( \vec{m} \preceq \vec{l} \), that is, \( 0 \leq m_{j} \leq l_{j} \) for \( j \in [K-1] \), then \( \vec{r} = \vec{l} - \vec{m} \in \Nc_{0}^{K-1} \) can be thought of as the remaining multiplicities of \( \vec{l} \) once the partial derivatives are taken with multiplicities \( \vec{m} \). This vector notation combined with differentiation under the integral sign shows that %\( G_{D, \partial \vec{m}} := \tfrac{ \partial^{ \vec{r}_{+}}}{ \prod_{j=1}^{K-1} \partial x_{j}^{m_{j}}} G_{D}(\vec{X}) \)
\begin{align}
\partial^{\vec{r}} G_{D} ( \vec{X}) & := \tfrac{ \partial^{ \vec{r}_{+}}}{ \prod_{j=1}^{K-1} \partial x_{j}^{r_{j}}} G_{D}(\vec{X}), \\
\label{eqGDPartialR}
& = ( - \tfrac{\vecB{c}_{+}}{2} n)_{- \vec{r}_{+}} \tfrac{ \prod_{j=1}^{K-1} ( \tfrac{c_{j}}{2} n + u_{j})_{r_{j}} }{ (\tfrac{\vecB{c}_{+}}{2} n + \vecB{u}_{+} )_{\vec{r}_{+}} } \Big ( \prod_{j=1}^{K-1} (c_{j} \gamma_{j})^{r_{j}} \Big ) G_{D, \vec{r}} ( \vec{X}) ,
\end{align}
where, formally by \refEq{eqGammaLarge} below,
\begin{align}
\label{eqApproxPochhammerRatio}
\tfrac{ \prod_{j=1}^{K-1} ( \tfrac{c_{j}}{2} n + u_{j})_{r_{j}} }{ (\tfrac{\vecB{c}_{+}}{2} n + \vecB{u}_{+} )_{\vec{r}_{+}} } = \tfrac{ \prod_{j=1}^{K-1} c_{j}^{r_{j}}}{ \vecB{c}_{+}^{\vec{r}_{+}}}  \big ( 1 + \Oc ( n^{-1}) \big ) ,
\end{align}
and where
\begin{align}
G_{D, \vec{r}} ( \vec{X}) = F_{D}( \tfrac{ \vecB{c}_{+}}{2} + \vec{r}_{+} \, ; \, \tfrac{ n}{2} \vec{c} + \vec{u} + \vec{r} \, ; \, \tfrac{ \vecB{c}}{2} n + \vecB{u}_{+} + \vec{r}_{+}  \, ; \, \vec{Z}) .
\end{align}
Observe that \( G_{D}( \vec{X}) = G_{D, \vec{0}}( \vec{X}) \). % and we let \( \partial^{\vec{r}} G_{D} := \partial^{\vec{r}} G_{D}( \vec{X}) |_{\vec{X} = \vec{1}} \).

With this notation the partial derivative of the Bayes factor accounting for multiplicities \( \vec{l} \) is
\begin{align}
\label{eqBfProductRule}
\partial^{\vec{l}} \BF_{10}( \vecB{s}^{2}, n) = \frac{ \Bc( \tfrac{ n}{2} \vecB{c} + \vecB{u} ) }{\Bc ( \vecB{u})} \left ( \sum_{ \vec{m} \preceq \vec{l}} \binom{\vec{l}}{\vec{m}} \partial^{\vec{m}} b ( \vec{X}) \partial^{\vec{l} - \vec{m}} G ( \vec{X}) \right ) ,
\end{align}
where \( \binom{\vec{l}}{\vec{m}} = \binom{l_{1}}{m_{1}} \cdots \binom{l_{K-1}}{m_{K-1}}= \prod_{j=1}^{K-1} \frac{l_{j}!}{ (l_{j} - m_{j})! m_{j}!} \) and where the sum is over all subvectors \( \vec{m} \) of \( \vec{l} \). For instance, with \( \vec{l}= \vec{e}_{k} \) this means \( \vec{m} = \vec{0} \) and \( \vec{m} = \vec{e}_{k} \). Note that \( \partial^{\vec{l}} \BF_{10}( \vecB{s}^{2}, n) \) only describes one entry of the \( \vec{l}_{+} \)-dimensional array of the total derivative of \( \BF_{10}( \vecB{s}^{2}, n) \) of order \( \vec{l}_{+} \).

\paragraph{Taylor approximation}
Because the samples variances of the \( X_{j} \)s are of order \( 1/n \), Chebyshev's inequality in conjunction with a union bound can be used to show that for any \( \epsilon \) there exists an \( N \) such that if \( n > N \) the following Taylor approximation holds with chance at least \( 1 - \epsilon \)
\begin{align}
\label{eqTaylorBf}
\BF_{10}( \vecB{s}^{2}, n ) \approx \frac{\Bc (\tfrac{ n }{2} \vecB{c}   + \vecB{u})}{\Bc ( \vecB{u}) }  \Big ( \sum_{ \vec{l} \in \N_{0}^{K-1}}  \partial^{\vec{l}} \big [ b G ( \vec{X}) \big ]_{\vec{X} = \vec{1}} \frac{Q^{\vec{l}} }{ \vec{l}!} \Big ) ,
\end{align}
where \( \partial^{\vec{l}} \big [ b G ( \vec{X}) \big ]_{\vec{X}} \) equals the sum on the right-hand side of \refEq{eqBfProductRule} evaluated at \( \vec{X} = \vec{1} \), \( \vec{Q} = ( \vec{X} - \vec{1}) \), \( \tfrac{\vec{Q}^{\vec{l}} }{ \vec{l} !} = \prod_{j=1}^{K-1} \tfrac{Q_{j}^{l_{j}} }{l_{j}!} \). Below we will show that for large \( n \) the Bayes factor behaves as
\begin{align}
\label{eqTaylorScaled}
\BF_{10}( \vecB{s}^{2}, n ) \approx \breve{T}^{(0)} \sum_{ \vec{l} \in \N_{0}^{K-1}} h_{\vec{l}}( \vecB{u}, \vecB{c}, \vecB{\gamma}) \frac{ \vec{Q}^{\vec{l}}}{ \vec{l} ! } ,
\end{align}
where under the null \( h_{\vec{l}}( \vecB{u}, \vecB{c}, \vecB{\gamma}, n) = \Oc(1) \) and under the alternative \( h_{\vec{l}}( \vecB{u}, \vecB{c}, \vecB{\gamma}, n) = \Oc( n^{ \vec{l}_{+} }) \), and where \( \breve{T}^{(0)} \) is the zeroth order term of the Taylor approximation studied in the next paragraph.

%as
%%
%\begin{align}
%\label{eqBfModelConsistencyT0}
%T^{(0)} := \BF_{10}( \vecB{s}^{2}, n ) \Big |_{\vec{X}=\vec{1}} = \frac{ \Bc (\tfrac{n }{2} \vecB{c} + \vecB{u}) }{\Bc ( \vecB{u}) } ( \la \vecB{c} , \vecB{\gamma} \ra )^{\tfrac{\vecB{c}_{+}}{2} n } G_{D} .
%\end{align}
%%
%This \( T^{(0)} \) term still depends on \( n \) and we denote its large sample version as \( \breve{T}^{(0)} \). Below we show that \( \breve{T}^{(0)} \) appears in each Taylor term. That is, we will show that for large \( n \) the Bayes factor behaves as
%%
%\begin{align}
%\BF_{10}( \vecB{s}^{2}, n ) \approx \breve{T}^{(0)} \sum_{ \vec{l} \in \N_{0}^{K-1}} h_{\vec{l}} \tfrac{ \vec{Q}^{\vec{l}}}{ \vec{l} ! } ,
%\end{align}
%%
%where under the null \( h_{\vec{l}} = \Oc(1) \) and under the alternative \( h_{\vec{l}} = \Oc( n^{ \vec{l}_{+} } \).
%
%Below we show that for a given \( \vec{l} \) that
%%
%\begin{align}
%\frac{\Bc (\tfrac{ n }{2} \vecB{c}   + \vecB{u})}{\Bc ( \vecB{u}) } \partial^{ \vec{l}} \big [ b G(\vec{X}) \big ]_{\vec{X} = \vec{1}} = \Oc(n^{\tfrac{1-K}{2}}), % \Oc_{P}(n^{ - \vec{l}_{+} / 2} ) ,
%\end{align}
%%
%under the null, and
%%
%\begin{align}
%\frac{\Bc (\tfrac{ n }{2} \vecB{c}   + \vecB{u})}{\Bc ( \vecB{u}) } \partial^{ \vec{l}} \big [ b G(\vec{X}) \big ]_{\vec{X} = \vec{1}} = \Oc(e^{\tilde{C}_{1} n}), %\Oc_{P} ( n^{-\vec{l}_{+}/2}) ,
%\end{align}
%%
%under the alternative for certain \( \tilde{C}_{1} = \tilde{C}_{1}(\vecB{u}, \vecB{\gamma}, \vecB{c}) \).

\paragraph{The \( T^{(0)} \) term}
The large sample behavior of the Bayes factor basically follows from gamma function asymptotics. The first object of interest is the deterministic term associated with \( \vec{l} = \vec{0} \), i.e., the Bayes factor evaluated at \( \vec{X} = \vec{1} \), but still dependent on the \( n \) term is
\begin{align}
\label{eqBfModelConsistencyT0}
T^{(0)} := \BF_{10}( \vecB{s}^{2}, n ) \Big |_{\vec{X}=\vec{1}} = \frac{ \Bc (\tfrac{n }{2} \vecB{c} + \vecB{u}) }{\Bc ( \vecB{u}) } ( \la \vecB{c} , \vecB{\gamma} \ra )^{\tfrac{\vecB{c}_{+}}{2} n } G_{D} .
\end{align}
The large sample behavior of the beta function follows that of gamma functions. Laplace's method implies that for \( v, b >0 \)
\begin{align}
\label{eqGammaLarge}
\Gamma ( v n + b) & = \sqrt{2 \pi} (v n)^{ v n + b - \tfrac{1}{2}} e^{-v n}  \big [ 1 + \tfrac{6 b^{2} - 6 b + 1}{12}  (v n)^{-1} + \Oc(n^{-2})]
\end{align}
as \( n \rightarrow \infty \). Hence, %Using this fact we consider the first order behavior of the Bayes factor in \refEq{eqTaylorBf},
%
%
%Gamma function asymptotics, i.e., \refEq{eqGammaLarge}, show that %the multinomial beta part of the Bayes factor behaves as
%C =
\begin{align}
\Bc (\tfrac{n}{2} \vecB{c} + \vecB{u}) & =  (4 \pi)^{\tfrac{K-1}{2}} n^{\tfrac{1-K}{2}}  \vecB{c}_{+}^{\tfrac{1}{2}} \Big ( \prod_{j=1}^{K-1} (c_{j})^{-\tfrac{1}{2}} \Big ) g(\vecB{c}, \vecB{u}, n) \big [ 1 + \Oc ( n^{-1}) \big ] ,
\end{align}
where the exponential behavior is captured by
\begin{align}
%g(\vec{c}, \vecB{u}, n) = (\vecB{c}_{+})^{- \tfrac{\vecB{c}_{+}}{2} n - \vecB{u}_{+}} \prod_{j=1}^{K-1} (c_{j})^{\tfrac{ c_{j}}{2} n + u_{j}} .
\label{eqBetaAsymptoticsExpontial}
g(\vecB{c}, \vecB{u}, n) = (\vecB{c}_{+})^{- \tfrac{\vecB{c}_{+} n }{2} - \vecB{u}_{+}  } \prod_{j=1}^{K-1} (c_{j})^{\tfrac{ c_{j} n }{2} + u_{j} } .
\end{align}
Note that the product only goes up to \( K-1 \), since \( c_{K} = 1\) by definition.

The hard part is to show consistency under the null. For this the exponential behavior of \( g(\vecB{c}, \vecB{u}, n) \) needs to be cancelled by that of \( G_{D} \), and we will show that it does so perfectly. To study the large \( n \) behavior of \( G_{D} \), and more generally \( G_{D, \vec{r}} \), we apply a Pfaff transform \parencite[p. 148]{lauricella1893sulle} yielding
\begin{align}
\label{eqGDr}
G_{D, \vec{r}} = \Big ( \prod_{j=1}^{K-1} c_{j} \gamma_{j}^{-\tfrac{c_{j}}{2} n - u_{j} - r_{j}} \Big )F_{D} \Big (\vecB{u}_{+} \, ; \, \tfrac{n}{2} \vec{c} + \vec{u} + \vec{r} \, ; \, \tfrac{\vecB{c}_{+}}{2} n + \vecB{u}_{+}  + \vec{r}_{+} \, ; \, \overrightarrow{ \tfrac{c \gamma - 1}{c \gamma}} \Big ) %\tfrac{ c_{1} \gamma_{1} - 1}{ c_{1} \gamma_{1} }, \ldots, \tfrac{ c_{K-1} \gamma_{K-1} - 1}{ c_{K-1} \gamma_{K-1} }  \Big ) .
\end{align}
where \( \overrightarrow{ \tfrac{c \gamma - 1}{c \gamma}} \in \R^{K-1} \) with \( (\overrightarrow{ \tfrac{c \gamma - 1}{c \gamma}} )_{j} = \tfrac{ c_{j} \gamma_{j} - 1}{ c_{j} \gamma_{j} } \). This rewrite of \( G_{D, \vec{r}} \) shows a cancellation of the \( (c_{j} \gamma_{j})^{r_{j}} \) terms in front of the \( G_{D, \vec{r}} \) in \refEq{eqGDPartialR}. Note that in the Lauricella function in \refEq{eqGDr} the lower term and the upper terms of the second kind depend on \( n \) in a linear fashion. The \( n \) dependence in these terms balance out as \( n \rightarrow \infty \) making the Lauricella function in \refEq{eqGDr} of order 1 as \( n \) grows. This is made rigorous by \refLem{lemFd}, which shows that the Lauricella function \refEq{eqGDr} converges to a (generalized) negative binomial series as \( n \rightarrow \infty \). Thus, % % and leads to %, that is, %
\begin{align}
\label{eqLauricellaLimit}
G_{D, \vec{r}} \approx \breve{G}_{D, \vec{r}} = \Big ( \prod_{j=1}^{K-1} (c_{j} \gamma_{j} )^{-\tfrac{c_{j}}{2} n - u_{j} - r_{j}} \Big ) \Big ( 1 - \tfrac{1}{\vecB{c}_{+}} \sum_{j=1}^{K-1} \tfrac{ c_{j} \gamma_{j} - 1}{\gamma_{j}} \Big )^{-\vecB{u}_{+}} ,
\end{align}
for \( n \) large. For \( T^{(0)} \) set \( \vec{r} = \vec{0} \), which shows that for large \( n \) %
\begin{align}
\label{eqBfLargeOrder1}
T^{(0)} \approx \breve{T}^{(0)} := C_{0}(K, \vecB{\gamma}, \vecB{c}, \vecB{u}) n^{\tfrac{1-K}{2}} \big ( \tfrac{ \la \vecB{c}, \vecB{\gamma} \ra}{ \vecB{c}_{+} } \big ) ^{ \tfrac{ \vecB{c}_{+}}{2} n } \Big ( \prod_{j=1}^{K-1} \gamma_{j}^{- \tfrac{c_{j}}{2} n} \Big ) ,
\end{align}
where the \( n \) independent term \( C_{0}(K, \vecB{\gamma}, \vecB{c}, \vecB{u}) \) is as asserted in \refEq{eqC0Constant}. A plugin of the null hypothesis \( \vecB{\gamma} = \vecB{1} \), thus, \( \la \vecB{c} , \vecB{\gamma} \ra = \vecB{c}_{+} \), in \refEq{eqBfLargeOrder1} shows that the exponentially growing terms are all equal to one, and therefore \( T^{(0)} = \Oc ( n^{ \tfrac{1-K}{2}}) \).

\paragraph{The \( T_{\vec{e}_{k}}^{(1)} \) terms}
The analysis of the gradient is similar to that of \( T^{(0)} \). It suffices to study the gradient coordinate wise. In particular, %the \( k \)th coordinate is
\begin{align}
T_{\vec{e}_{k}}^{(1)} := \frac{ \Bc (\tfrac{n}{2} \vecB{c}+ \vecB{u}) }{\Bc ( \vecB{u})} ( \la \vecB{c} , \vecB{\gamma} \ra )^{\tfrac{\vecB{c}_{+}}{2} n } c_{k} \gamma_{k} \tfrac{ \vecB{c}_{+}  }{2} n \Big [ \tfrac{G_{D}}{ \la \vecB{c} , \vecB{\gamma} \ra} - \tfrac{ c_{k} n + 2 u_{k}}{\vecB{c}_{+} n + 2 \vecB{u}_{+}} G_{D, \vec{e}_{k}} \Big ] .
\end{align}
The same operations as before, a Pfaff transform and \refEq{eqLauricellaLimit}, shows that %
\begin{align}
%\breve{T}_{\vec{e}_{k}}^{(1)} & = \breve{T}^{(0)} \tfrac{ \vecB{c}_{+}}{2} n  \big ( \tfrac{ c_{k} \gamma_{k} }{ \la \vecB{c} , \vecB{\gamma} \ra}  - \tfrac{c_{k} + 2 u_{k}/n }{\vecB{c}_{+} + 2 \vecB{u}_{+}/ n } \big ) , \\
\label{eqGradientKthCoordinate}
\breve{T}_{\vec{e}_{k}}^{(1)} & = \breve{T}^{(0)} \Big ( \tfrac{ \vecB{c}_{+} }{2}  \big ( \tfrac{c_{k} \gamma_{k}}{\la \vecB{c}, \vecB{\gamma} \ra} - \tfrac{ c_{k}}{ \vecB{c}_{+} } \big ) n + \tfrac{ c_{k} \vecB{u}_{+} - \vecB{c}_{+} u_{k} }{\vecB{c}_{+}} + \Oc ( n^{-1}) \Big ) ,
\end{align}
%
%\begin{sloppypar}
as \( n \rightarrow \infty \). Hence, under the alternative \( h_{\vec{e}_{k}}(\vecB{u}, \vecB{c}, \vecB{\gamma}, n) := \breve{T}_{\vec{e}_{k}}^{(1)} / \breve{T}^{(0)} = \Oc ( n) \) and accounting for the stochastic term \( Q_{k} = (X_{k} - 1) = \Oc_{P}(n^{-1/2}) \) leads to \( \sum_{k=1}^{K-1} h_{\vec{e}_{k}}(\vecB{u}, \vecB{c}, \vecB{\gamma}, n) Q_{k} = \Oc_{P}(n^{1/2}) \). On the other hand, under the null \( h_{\vec{e}_{k}}(\vecB{u}, \vecB{c}, \vecB{1}, n) = \breve{T}_{\vec{e}_{k}}^{(1)} / \breve{T}^{(0)} = \Oc (1) \), as then again \( \la \vecB{c}, \vecB{\gamma} \ra = \vecB{c}_{+} \) and \( \big ( \tfrac{c_{k} \gamma_{k}}{\la \vecB{c}, \vecB{\gamma} \ra} - \tfrac{ c_{k}}{ \vecB{c}_{+} } \big ) = 0 \), thus, a perfect cancellation of the \( \Oc(n) \) term. Consequently, \( \sum_{k=1}^{K-1} h_{\vec{e}_{k}}(\vecB{u}, \vecB{c}, \vecB{\gamma}, n) Q_{k}  = \Oc_{P}(n^{-1/2}) \). %
%\end{sloppypar}
%
%\( \sum_{j=1}^{K-1} (X_{k} - 1) \breve{T}_{\vec{e}_{k}}^{(1)} = U \big ( \tfrac{ \vecB{c}_{+} - \vecB{u}_{+} }{ \vecB{c}_{+}} \big ) n^{-1/2} + \Oc_{P}( n^{-3/2})  \), where \( U = \Oc_{P}(1) \). %and the asserted rate follows. %
%
%
%GUESS: Alternative \( \tfrac{ \vecB{c}_{+}}{2} \big ( \tfrac{(\la \vecB{c} , \vecB{\gamma} \ra -1)}{\la \vecB{c} , \vecB{\gamma} \ra} - \tfrac{ \vecB{c}_{+} - 1}{ \vecB{c}_{+} } \big )  \)
%
%\( U \tfrac{ \la \vecB{c}, \vecB{\gamma} \ra - \vecB{c}_{+}}{ 2 \la \vecB{c}, \vecB{\gamma} \ra } \)
%
%GUESS: Null \( \tfrac{\vecB{c}_{+} - \vecB{u}_{+}}{\vecB{c}_{+}} \)

\paragraph{Higher order terms}
The higher order terms exhibit the same behavior. %In particular, we show that for a given \( \vec{l} \), the large \( n \) behavior of the
%
%\( \breve{T}_{\vec{l}}^{\vec{l}_{+}} / \breve{T}^{(0)} = \Oc(1) \).
%
%
%
%\( \breve{T}_{\vec{l}}^{\vec{l}_{+}} / \breve{T}^{(0)} = \Oc(1) \). %
%
%under the null
%
%
%
%the partial derivative associated with \( \vec{l} \) divided by \( \breve{T}^{(0)} \) will also be of
%
%\( \vec{l}_{+} \)th order partial derivatives
%
%Similarly, under the null the \( \vec{l}_{+} \)th order partial derivative associated with \( \vec{l} \) will be of order \( \Oc(1) \)
%
Let \( \vec{l} \in \N^{K-1} \), then for \( n \) large the partial derivative associated to \( \vec{l} \) of the Bayes factor behaves as
\begin{align}
\nonumber
\breve{T}_{\vec{l}}^{(\vec{l}_{+})} & = \sum_{\vec{m} \preceq \vec{l}} \binom{ \vec{l}}{\vec{m}}
\breve{T}^{(0)} ( \tfrac{ \vecB{c}_{+}}{2} n)_{- \vec{m}_{+}} ( - \tfrac{ \vecB{c}_{+}}{2} n)_{- (\vec{l}_{+} - \vec{m}_{+})}  \tfrac{\prod_{j=1}^{K-1} (c_{j} \gamma_{j})^{m_{j}}}{ \la \vecB{c}, \vecB{\gamma} \ra^{\vec{m}_{+}}} \tfrac{ \prod_{j=1}^{K} c_{j}^{l_{j} - m_{j}} }{ \vecB{c}_{+}^{(\vec{l}_{+} - \vec{m}_{+})}} \big ( 1 + \Oc(n^{-1}) \big ) .
\end{align}
Note that \( ( \tfrac{ \vecB{c}_{+}}{2} n)_{- \vec{m}_{+}} ( - \tfrac{ \vecB{c}_{+}}{2} n)_{- (\vec{l}_{+} - \vec{m}_{+})} \) is a polynomial in \( n \) of order \( \vec{l}_{+} \). Hence, \( h_{\vec{l}}(\vecB{u}, \vecB{c}, \vecB{\gamma}, n) := \breve{T}_{\vec{l}}^{(\vec{l}_{+})} /\breve{T}_{0} = \Oc(n^{ \vec{l}_{+}}) \). %
We now show that under the null, the polynomial \( ( \tfrac{ \vecB{c}_{+}}{2} n)_{- \vec{m}_{+}} ( - \tfrac{ \vecB{c}_{+}}{2} n)_{- (\vec{l}_{+} - \vec{m}_{+})} \) is zero and \( h_{\vec{l}}(\vecB{u}, \vecB{c}, \vecB{1}, n) = \breve{T}_{\vec{l}}^{(\vec{l}_{+})} /\breve{T}_{0} = \Oc(1) \), where the constant term comes from the approximation of the ratio of Pochhammer symbols, i.e., \refEq{eqApproxPochhammerRatio}, e.g., \refEq{eqGradientKthCoordinate}. To see that there is no \( n \) contribution under the null, we plugin \( \vecB{\gamma} = \vecB{1} \) and rewrite the sum over \( \vec{m} \preceq \vec{l} \) as a sum over \( \vec{m}_{+} =p \) for \( p = 0, 1, \ldots, \vec{l}_{+} \) and a subsequent sum over all subvector \( \vec{m} \) that sum to \( p \), which yields %
\begin{align}
h_{\vec{l}}(\vecB{u}, \vecB{c}, \vecB{1}, n) & = \tfrac{ \prod_{j=1}^{K-1} c_{j}^{l_{j}}}{ \vecB{c}_{+}^{ \vec{l}_{+}}} \sum_{p=0}^{ \vec{l}_{+}} ( \tfrac{ \vecB{c}_{+} }{2} n)_{- p}  ( - \tfrac{ \vecB{c}_{+} }{2} n)_{- p} \sum_{ \substack{\vec{m} \preceq \vec{l} \\ \vec{m}_{+} = p}} \binom{ \vec{l}}{ \vec{m}} .
\end{align}
Next we apply the Chu-Vandermonde identity twice, once over the sum on the right-hand side of the previous display and once after using the identity \( (a)_{-l} / l! = \binom{a}{l} \), which leads to
\begin{align}
h_{\vec{l}}(\vecB{u}, \vecB{c}, \vecB{1}, n) & = \tfrac{ \prod_{j=1}^{K-1} c_{j}^{l_{j}}}{ \vecB{c}_{+}^{ \vec{l}_{+}}} \sum_{p=0}^{ \vec{l}_{+}} \binom{\vec{l}_{+}}{p} ( \tfrac{ \vecB{c}_{+} }{2} n)_{- p}  ( - \tfrac{ \vecB{c}_{+} }{2} n)_{- (\vec{l}_{+} - p)} \\
& = \tfrac{ \prod_{j=1}^{K-1} c_{j}^{l_{j}}}{ \vecB{c}_{+}^{ \vec{l}_{+}}} \vec{l}_{+} !\sum_{p=0}^{ \vec{l}_{+}} \binom{ \tfrac{ \vecB{c}_{+} }{2} n }{p} \binom{ - \tfrac{ \vecB{c}_{+} }{2} n }{\vec{l}_{+} -p} = 0  .
\end{align}
%
%This shows that the terms involving the falling Pochhammer symbols are zero under the null.
This shows that under the null, none of the Taylor terms lead to a growth in \( n \).

The stochastic terms in the assertion both under the null and the alternative follow from the definition of the exponential series by rewriting the sum of the Taylor approximation of interest, i.e., \refEq{eqTaylorScaled}, in terms of \( \tilde{p} \in \N_{0} \) and a subsequent sum over all subvectors \( \vec{l} \) such that \( \vec{l}_{+} = \tilde{p} \).

\paragraph{Model selection consistency under the alternative} To show that the Bayes factor increases under the alternative, %regardless of which \( \gamma_{j} \neq 1 \) and
irrespectively of \( \gamma_{j}  \) being larger or smaller than \( 1 \), we study the exponential term of \refEq{eqBfModelSelectionConsistency1}
\begin{align}
v(n) = \big ( \la \vecB{c}, \vecB{\gamma} \ra \big ) ^{ \tfrac{\vecB{c}_{+}}{2} n} \prod_{j=1}^{K-1} \gamma_{j}^{ - \tfrac{ c_{j}}{2} n}
\end{align}
The claim is that \( v \) monotonically increases in \( n \). Suppose that this is not true, then the ratio of subsequent terms
\begin{align}
v(n+1)/v(n) = \big ( \la \vecB{c}, \vecB{\gamma} \ra \big ) ^{ \tfrac{\vecB{c}_{+}}{2} } \prod_{j=1}^{K-1} \gamma_{j}^{ - \tfrac{ c_{j}}{2} }
\end{align}
would be less or equal to one. The gradient of \( v(n+1)/v(n) \) with respect to \( \gamma \) is of the form
\begin{align}
\tfrac{c_{k}}{2} \big ( \tfrac{ \vecB{c}_{+}}{ \la \vecB{c}, \vecB{\gamma} \ra } - \tfrac{1}{ \gamma_{k}} \big ) v(n+1)/v(n)
\end{align}
and this reveals a (global) minimum at \( \vecB{\gamma} = \vecB{1} \) at which \( v(n+1)/v(n) = 1 \). Hence, any \( \vecB{\gamma} \neq \vecB{1} \) leads to an exponentially increasing Bayes factor \( \BF_{10}(\vecB{s}^{2}, n) \). %
\end{proof}

The proof of the previous theorem relies on a particular Lauricella function \( G_{D} \) to be of order \( 1 \) as \( n \) increases as shown in the following lemma.

\begin{lemma}[Limit of a particular Lauricella function]
\label{lemFd}
For all \( v_{j}, b_{j} > 0  \), \( j \in [m] \) and \( | x_{j} | < 1  \), we have that %
\begin{align}
\lim_{n \rightarrow \infty} F_{D}(a \, ; \, n \vec{v} + \vec{b}  \, ; \, v_{+} n + b_{+} \, ; \, \vec{x}) = \Big ( 1 - \sum_{i=1}^{m} \tfrac{ v_{i}}{v_{+}} x_{i} \Big )^{-a},
\end{align}
as \( n \rightarrow \infty \). \( \hfill \diamond \)
\end{lemma}

\begin{proof}
The proof follows from the asymptotic behavior of the gamma function combined with repeated use of the (negative) binomial series.

Firstly, note that the \( n \) dependence occurs in the lower and the upper terms of the second type, which cancels out as \( n \) grows large. To show this consider the definition of the Pochhammer raising factorial that combined with the Laplace approximation \refEq{eqGammaLarge} for constants \( v, b > 0 \) leads to
\begin{align}
\label{eqLargePochhammer}
( v n + b)_{k}  = \frac{ \Gamma ( v n + b + k) } { \Gamma ( v n + b ) } = (v n )^{k} \big [ 1 +  k ( k + 2 b -1) (v n)^{-1} + \Oc ( (v n )^{-2}) \big ]
\end{align}
as \( n \rightarrow \infty \).

Secondly, to describe the large \( n \) behavior of the particular type D Lauricella hypergeometric series \( F_{D}:=F_{D}(a \, ; \, n \vec{v} + \vec{b}  \, ; \, v_{+} n + b_{+} \, ; \, \vec{x}) \) we use the notation \( i[k:m] = (i_{j}, \ldots, i_{m}) \in \N^{m-(k-1)} \) to denote the vector of indexes from \( k \) to \( m \). Based on this notation and by \refEq{eqLargePochhammer}, we have for \( n \) large that
\begin{align}
F_{D} & = \sum_{i[1:m]}\frac{ (a)_{i[1:m]_{+}} (v_{1} n + b_{1})_{i_{1}} \cdots (v_{m} n + b_{m})_{i_{m}}}{(v_{+} n + b_{+})_{i[1:m]_{+}}} \frac{ x_{1}^{i_{1}}}{i_{1}!} \cdots \frac{ x_{m}^{i_{m}}}{i_{m}!} \\
\nonumber
& \approx \sum_{i[1:m]} \frac{ (a)_{i[1:m]_{+}} v_{1}^{i_{1}} \cdots v_{m}^{i_{m}}}{v_{+}^{i[1:m]_{+}}}  \frac{ x_{1}^{i_{1}}}{i_{1}!} \cdots \frac{ x_{m}^{i_{m}}}{i_{m}!}  = \sum_{i = \vec{0}}^{\infty} (a)_{i[1:m]_{+}}  \frac{ (\tfrac{v_{1}}{v_{+}} x_{1})^{i_{1}}}{i_{1}!} \cdots \frac{ (\tfrac{v_{m}}{v_{+}} x_{m})^{i_{m}}}{i_{m}!}  .
\end{align}
The last equality defines the limit of \( F_{D} \) with respect to \( n \). It also captures the essence of the repeated use of the binomial series, namely, the redistribution of the scaling factor \( v_{+}^{-i[1:m]_{+}} \) over the variables \( x \).

Thirdly, with the notation \( i[2:m] \) it is simple to isolate the summation with respect to \( i_{1} \) only, which combined with the binomial series yields
\begin{align}
\lim F_{D} & =  \Big ( \sum_{i_{1} =0}^{\infty} (a)_{i[1:m]_{+}} \frac{ (\tfrac{v_{1}}{v_{+}} x_{1})^{i_{1}}}{i_{1}!}  \Big ) \sum_{i[2:m]}   \frac{ (\tfrac{v_{2}}{v_{+}} x_{2})^{i_{2}}}{i_{2}!} \cdots \frac{ (\tfrac{v_{m}}{v_{+}} x_{m})^{i_{m}}}{i_{m}!} \\
\nonumber
& = \Big ( \tfrac{ v_{+} - v_{1} x_{1}}{v_{+}} \Big )^{-a}  \sum_{i[2:m]}  (a)_{i[2:m]_{+}} \Big ( \tfrac{ v_{+} - v_{1} x_{1}}{v_{+}} \Big )^{- i[2:m]_{+}}  \frac{ (\tfrac{v_{2}}{v_{+}} x_{2})^{i_{2}}}{i_{2}!} \cdots \frac{ (\tfrac{v_{m}}{v_{+}} x_{m})^{i_{m}}}{i_{m}!} .
\end{align}
Note that, as before, the scaling factor \( \Big ( \tfrac{ v_{+} - v_{1} x_{1}}{v_{+}} \Big )^{- i[2:m]_{+}} \) can be redistributed over the variables resulting in \( (\tfrac{v_{k}}{v_{+} - v_{1} x_{1}} x_{k})^{i_{k}} / i_{k}! \) for \( k=2, \ldots, m \). The summation with respect to \( i_{2} \) is again a binomial series and yields %
\begin{align}
\lim F_{D} & = \Big ( \tfrac{ v_{+} - v_{1} x_{1}}{v_{+}} \Big )^{-a} \Big ( \tfrac{v_{+} - v_{1} x_{1} - v_{2} x_{2}}{ v_{+} - v_{1} x_{1}} \Big )^{-a}  \\
\nonumber
\times &  \sum_{i[3:m]}  (a)_{i[3:m]_{+}} \Big ( \tfrac{v_{+} - v_{1} x_{1} - v_{2} x_{2}}{ v_{+} - v_{1} x_{1}} \Big )^{-i[3:m]_{+}}  \frac{ (\tfrac{v_{3}}{{v_{+} - v_{1} x_{1} }} x_{3})^{i_{3}}}{i_{3}!} \cdots \frac{ (\tfrac{v_{m}}{{v_{+} - v_{1} x_{1} }} x_{m})^{i_{m}}}{i_{m}!} .
\end{align}
Observe that the numerator and denominator of the first and second \( -a \) exponentiated terms in the previous display are equal and thus cancel. Repeating this procedure to \( m \) and telescoping through the \( -a \) exponentiated terms yields the results.
\end{proof}

\subsection{Limit and across-sample consistency}
\label{appAcrossSampleConsistency}

\begin{proof}[Proof of across-sample consistency, \refThm{thmAcrossSampleConsistency}]
To simplify notation we write \( n:= n_{K} \) and \( \vec{\texttt{ss}}= \overrightarrow{\nu s^{2}} \), where \( \texttt{ss}_{j} = \nu_{j} s_{j}^{2} \) is the sum of squares of the \( j \)th sample. Since \( S_{K}^{2}  \) is \( \sqrt{n} \)-consistent we can find an \( N \) such that for all \( n > N \) the following statement holds with chance at least \( 1 - \epsilon \) %the Bayes factor can be written as % \( \BF_{10}(\vecB{s^{2}}) \) then with high probability
% \
\begin{align}
\label{eqBfLimitSampleLaplaceGeneral}
\BF_{10}^{[K]}(\vec{s^{2}}, S_{K}^{2}, n) = \BF_{10}^{[K]}(\vec{s^{2}}, \sigma_{0}^{2}, n) + \tfrac{h_{n}}{\sqrt{n}} \tilde{T}_{1}(n) + o_{P}(n^{-\tfrac{1}{2}}) \tilde{T}_{2}(n),
\end{align}
where \( h_{n}  \) is a bounded sequence of random variables due to \( S_{K}^{2} - \sigma_{0}^{2} = \Oc_{P}(n^{-\tfrac{1}{2}}) \) and where
\begin{align}
\tilde{T}_{1}(n) & = \Big ( \tfrac{ \partial}{ \partial x }\BF_{10}^{[K]}(\vec{s^{2}}, x, n) \Big ) \Big |_{x = \sigma_{0}^{2} },  \\
\tilde{T}_{2}(n) & = \Big ( \tfrac{ \partial^{2}}{ \partial x^{2} }\BF_{10}^{[K]}(\vec{s^{2}}, x, n) \Big ) \Big |_{x = \sigma_{0}^{2} } .
\end{align}
To prove the theorem we have to show that \( \lim_{n \rightarrow \infty} \BF_{10}^{[K]}(\vec{s^{2}}, \sigma_{0}^{2}, n) \) exists, is equal to \refEq{eqKMinusOneTricomi}, and that both \( \tilde{T}_{1}(n) \) and \( \tilde{T}_{2}(n) \) are bounded in \( n \). To this end, we want to first take the limit and then integrate. To see that this is permissible we first show that the integrand of \( \BF_{10}^{[K]}(\vec{s^{2}}, \sigma_{0}^{2}, n) \) as a sequence in \( n \) is uniformly bounded in \( \vec{\vartheta} \).

%Recall that the vectors \( \vec{\texttt{ss}}, \vec{a}, \vec{\nu} \) are of length \( K-1 \), which implies that the upper bound in the sums and products in this proof is \( K-1 \). In particular, \( \vec{\nu}_{+} := \sum_{j=1}^{K-1} \nu_{j} \).

\paragraph{Uniformly boundedness of the integrand}
To further simplify notation we introduce the vectors \( \vec{a}, \vec{c} \in \R^{K-1}  \) with \( a_{j} = \tfrac{\nu_{j}}{2}  \) for \( j \in [K-1] \) and \( b = \tfrac{n}{2} \). By definition of \( \BF_{10}^{[K]}(\vec{s^{2}}, \sigma_{0}^{2}, n)  \), the innocuous replacement \( n= \nu_{K} \) we have that %
% \(  \)
\begin{align}
\label{eqKToKMinOne0}
\BF_{10}^{[K]}(\vec{s^{2}}, \sigma_{0}^{2}, n)  & = (1 + \tfrac{\vec{\texttt{ss}}_{+}}{n \sigma_{0}^{2}} )^{a_{+}+b} \int \tilde{h}(\vec{s^{2}}, \sigma_{0}^{2}, n \, | \, \vec{\vartheta}) \pi_{1}(\vec{\vartheta}) \der \vec{\vartheta} ,
\end{align}
where \( \pi_{1}(\vec{\vartheta}) \) is the Dirichlet prior with parameters \( \vecB{u} \) and where
\begin{align}
\tilde{h}(\vec{s^{2}}, \sigma_{0}^{2}, n \, | \, \vec{\vartheta}) = \Big ( \prod_{j=1}^{K-1} \vartheta_{j}^{ a_{j}} \Big ) ( 1-\vec{\vartheta}_{+})^{ b } (1- \sum_{j=1}^{K-1} [1 - \tfrac{ \texttt{ss}_{j}}{n \sigma_{0}^{2}}] \vartheta_{j})^{-(a_{+}+b)} ,
\end{align}
is the marginalized likelihood with \( \sigma_{0}^{2} \) in place of \( s_{K}^{2} \), thus, \( \tilde{h}(\vec{s^{2}}, \sigma_{0}^{2}, n \, | \, \vec{\vartheta}_{0}) = (1 + \tfrac{\vec{\texttt{ss}}_{+}}{n \sigma_{0}^{2}} )^{- (a_{+}+b)} \). %The fact that the sum of all the powers in the last display is zero is exploited in the
By definition of the exponential function as a series, the first term in \refEq{eqKToKMinOne0} remains bounded, that is, % we have that
\begin{align}
\lim_{n \rightarrow \infty} (1 + \tfrac{\vec{\texttt{ss}}_{+}}{n \sigma_{0}^{2} } )^{ \tfrac{\vec{\nu}_{+} + n}{2}} =  e^{ \tfrac{\vec{\texttt{ss}}_{+}}{2\sigma_{0}^{2}} } \Big ( 1-  \tfrac{\vec{\texttt{ss}}_{+}}{4 n \sigma_{0}^{2}}  ( \tfrac{\vec{\texttt{ss}}_{+}}{\sigma_{0}^{2}} - 2 \vec{\nu}_{+}) + \Oc ( n^{-2}) \Big ) .
\end{align}
The prior does not play a role in the asymptotics for \( n \rightarrow \infty \), as we will show that %it suffices to show that %\( h(\vec{s^{2}}, \sigma_{0}^{2}, n \, | \, \vec{\vartheta})  \) is bounded, since for any \( \vecB{u} \) we have %
\begin{align}
\int \tilde{h}(\vec{s^{2}}, \sigma_{0}^{2}, n \, | \, \vec{\vartheta})  \pi_{1}(\vec{\vartheta}) \der \vec{\vartheta} \leq C( \vecB{u}) \int \tilde{h}(\vec{s^{2}}, \sigma_{0}^{2}, n \, | \, \vec{\vartheta}) \der \vec{\vartheta}.
\end{align}
for a certain constant \( C(\vecB{u}) \) independent of \( n \).

\paragraph{Case (i)} The case with \( \vecB{u} \) all at least 1, we can take \( C( \vecB{u}) \) to be the maximum of the prior \( \Dir ( \vec{\vartheta} \, ; \, \vecB{u}) \) on \( \vec{\vartheta} \) in the \( K-1 \) simplex. The maximum of the marginalized likelihood \( \tilde{h}(\vec{s^{2}},  \sigma^{2}_{0}, n \, | \, \vec{\vartheta}) \) at each \( n \) can be found by setting the partial derivatives to zero. At each fixed \( n \) \refLem{lemMaxReducedLikelihood} can be used to find the maximum \( \hat{\vartheta} \) as a function of \( \vec{a}, b, \vec{c} \). By definition of \( \vec{a}, b, \vec{c} \) and by denoting the observed precisions \( \vec{t} \in \R^{K-1} \) by \( t_{j} := (s_{j}^{2})^{-1} \), it then follows that \( \hat{\vartheta}_{k}=\frac{ t_{k}}{ \sigma_{0}^{-2} + \vec{t}_{+}} \), which is free of \( n \). A plugin and a direct calculation show that the maximum value of the marginalized likelihood at each \( n \) is %attained at %
\begin{align}
f_{\max,n} & := \Big ( \prod_{k=1}^{K-1}  t_{k}^{\tfrac{\nu_{k}}{2}} \Big )  (\sigma_{0}^{2})^{\tfrac{\vec{\nu}_{+}}{2}}  e^{- \tfrac{\vec{\nu}_{+}}{2}} [ 1 - \tfrac{ \vec{\nu}_{+}^{2}}{4n} + \Oc( n^{-2})] . %\Big ( 1 + \frac{\vec{\nu}_{+}}{n} \Big )^{- \tfrac{ \vec{\nu}_{+}}{2} - \tfrac{n}{2}}
\end{align}
Hence, as a sequence in \( n \) the integrand is uniformly bounded by a constant. % and integrable on the \( K-1 \) simplex.

\paragraph{Case (ii)} For any \( u_{j} < 1 \), \( j \in [K-1] \) the prior diverges at \( \vartheta_{j} = 0 \) and \( C(\vecB{u})\) cannot be taken to be the maximum value of the prior on the \( K-1 \) simplex. Instead, \( C(\vecB{u}) \) can be the maximum of \( \pi_{1}(\vec{\vartheta}) \) for \( \vec{\vartheta} \) in a subset \( R \) containing \( \hat{\vartheta} \). Since the true variances are assumed to be non-zero, finite and the data continuous, we can take \( R \) with high probability to be a compact subset that intersects with \( \bigoplus_{j=1}^{K-1}[ \epsilon_{j}, 1-\epsilon_{j}] \subset [0,1]^{K-1} \) for \( \epsilon_{j}  \) depending on \( u_{j} \). On \( R \) the proof of Case (i) can be repeated to show that that the integrand is bounded. For any \( u_{j} < 1 \), \( j \in [K-1] \) the integrand over \( \vartheta_{j} \in [0, \epsilon_{j}) \) behaves as \( \vartheta_{j}^{\tfrac{\nu_{j}}{2} + u_{j} - 1}  + \Oc(|\vartheta_{j}|)  \). %A potential problem occurs when \( \tfrac{\nu_{j}}{2} + u_{j} < 1 \), as this would imply a si
On this domain the integrand remains integrable whenever \( u_{j} > - \tfrac{\nu_{j}}{2} \), which is true by assumption. The same arguments extend to the case with \( u_{K} < 1 \).

\paragraph{Identifying the \( K-1 \)-sample Bayes factor}
Uniform boundedness allows us to interchange the limit and integral and conclude that the limiting integral exist, and implies that \( \BF^{[K]}_{10}(\vec{s^{2}}, \sigma_{0}^{2}, n)  \) converges to
\begin{align}
%\nonumber
%\lim_{n \rightarrow \infty} \BF_{10}^{[K]}(\vec{s^{2}}, \sigma_{0}^{2}, n)   %&  = \lim e^{\tfrac{m \tau}{2}}  \theta^{\tfrac{\nu_{1}}{2} + u_{1} - 1} ( 1 - \vartheta)^{u_{2} -1} \big [ \tfrac{1- \vartheta}{1 - [1- \tfrac{m}{n} \tau] \vartheta} \big ]^{\tfrac{n}{2}} \big [ 1 - [1- \tfrac{m}{n} \tau] \vartheta \big ]^{-\tfrac{\nu_{1}}{2}} \\
%& =  \frac{\int_{0}^{1}  \big ( \tfrac{\vartheta}{1-\vartheta} \big )^{\tfrac{\nu_{1}}{2}}  e^{  -\tfrac{m \tau}{2} \big ( \tfrac{ \vartheta }{1-\vartheta } \big ) } \vartheta^{u_{1} - 1} ( 1 - \vartheta)^{ u_{2} - 1} \der \vec{\vartheta}}{ e^{ - \tfrac{m \tau}{2} } \Bc( u_{1}, u_{2}) } .
 \frac{ \int \Big ( \prod  \vartheta_{j}^{ \tfrac{\nu_{j}}{2} + u_{j} - 1} \Big )  ( 1 - \vec{\vartheta}_{+}  )^{ u_{K} - \tfrac{ \vec{\nu}_{+}}{2} -1} \exp \big (  -   \sum  \tfrac{ \texttt{ss}_{j} }{2 \sigma_{0}^{2}} ( \tfrac{ \vartheta_{j}}{1 - \vec{\vartheta}_{+}} )   \big )  \der \vec{\vartheta}}{ \Bc ( \vecB{u}) \exp ( - \tfrac{ \vec{\texttt{ss}}_{+} }{2 \sigma_{0}^{2} }  ) }.  %,  \\
%& = \frac{ \Big ( \prod_{j=1}^{K-1} \Gamma( \tfrac{\nu_{j}}{2} + u_{j}) \Big ) \Uc \Big ( \tfrac{ \vec{\nu}}{2} + \vec{u} \, ; \, \tfrac{ \vec{\nu}_{+}}{2} - w + 1 \, ; \, \tfrac{ \overrightarrow{\nu s^{2}} }{2 \sigma_{K}^{2}} \Big )}{ \Bc ( u_{1}, \ldots, u_{K-1}, w) \exp ( - \tfrac{\tau}{2} (\nu s^{2})_{+} ) },
\end{align}
From the change of variables \( \vartheta_{j} = \tfrac{\xi_{j}}{1+ \xi_{+}} \), thus, \( \der \vec{\vartheta} = (1 + \xi_{+})^{-K} \der \vec{\xi} \), and by definition of the integral representation of the multivariable Tricomi function \( \Uc \), see for instance \parencite{ng2011dirichlet,phillips1988characteristic}, we have that the resulting \( K-1 \) sample Bayes factor is given by %has the representation %
\begin{align}
\nonumber
\BF_{10 \, ; \, \sigma_{0}^{2}}^{[K-1]}(\vec{s^{2}}) & = \frac{ \int \Big ( \prod_{j=1}^{K-1} \tau_{j}^{\tfrac{ \nu_{j}}{2}} \Big ) \exp ( - \tfrac{1}{2} \sum_{j=1}^{K-1} \nu_{j} s^{2}_{j} \tau_{j} ) \pi_{\sigma_{0}^{2}}(\vec{\tau} \, | \, \Mc_{1}^{[K-1]}) \der \vec{\tau}   }{ (\sigma_{0}^{2})^{ - \tfrac{ \vec{\nu}_{+}}{2}} \exp ( - \tfrac{ (\overrightarrow{\nu s^{2}})_{+} }{2 \sigma_{0}^{2} }  ) } , \\
\label{eqKMinusOneTricomi}
& = \frac{ \Big ( \prod_{j=1}^{K-1} \Gamma( \tfrac{\nu_{j}}{2} + u_{j}) \Big ) \Uc \Big ( \tfrac{ \vec{\nu}}{2} + \vec{u} \, ; \, \tfrac{ \vec{\nu}_{+}}{2} - u_{K} + 1 \, ; \, \tfrac{ \overrightarrow{\nu s^{2}} }{2 \sigma_{0}^{2}} \Big )}{ \Bc ( \vec{u}, w) \exp ( - \tfrac{ (\overrightarrow{\nu s^{2}})_{+} }{2 \sigma_{0}^{2} }  ) }  ,
\end{align}
where \( \overrightarrow{\nu s^{2}} = ( \nu_{1} s_{1}^{2}, \ldots, \nu_{K-1} s_{K-1}^{2}) \) denotes the vector of sums of squares, \( (\overrightarrow{\nu s^{2}})_{+} = \sum_{j=1}^{K-1} \nu_{j} s_{j}^{2} \), and \( \vec{\nu}_{+} := \sum_{j=1}^{K-1} \nu_{j} \), as before. This Bayes factor is based on uniform priors on the nuisance parameters \( \vec{\mu} \in \R^{K-1} \), and an inverse Dirichlet distribution on the precisions \( \vec{\tau} = (\tau_{1}, \ldots, \tau_{K-1}) \in \R^{K-1} \) scaled by \( 1/\sigma_{0}^{-2} \), that is,
\begin{align}
\label{eqScaledInverseDirichletPrior}
\pi_{\sigma_{0}^{2}}(\vec{\tau} \, | \, \Mc_{1}^{[K-1]}) = \frac{(\sigma_{0}^{2} )^{K-1} \prod_{j=1}^{K-1} (\sigma_{0}^{2} \tau_{j})^{u_{j} - 1}}{ \Bc ( \vec{u}, w)  (1+ \sigma_{0}^{2} \vec{\tau}_{+})^{\vec{u}_{+} + w}} ,
\end{align}
where we wrote \( w = u_{K} \) so the statement only involves vectors of length \( K-1 \).

%Note that the lower dimensional Bayes factor \( \BF_{10 \, ; \, \sigma_{0}^{2}}^{[K-1]}( \vec{s^{2}}) \) is in general hard to compute, because the Tricomi function \( \Uc(\vec{b} \, ; \, c \, ; \, \vec{x})  \) defines a \( K-1 \)-dimensional integral. \textcite{phillips1988characteristic} showed that if \( c < 1 \), the following simplification holds %
%
% \begin{align}
% \Uc(\vec{b} \, ; \, c \, ; \, \vec{x}) = \int_{0}^{\infty} e^{-t} t^{\vec{b}_{+} - c} \prod_{j=1}^{K-1} (t + x_{j})^{-b_{j}} \der t .
% \end{align}
%
%For \( \BF_{10 \, ; \, \sigma_{0}^{2}}^{[K-1]}( \vec{s^{2}}) \) this simplification holds whenever \( \vec{\nu}_{+} < 2 u_{K} \), which will be of little practical use when, for instance, \( u_{K} = 1/2 \). \refThm{thmAcrossSampleConsistency} shows that for the case with \(  \vec{\nu}_{+} \geq 2 u_{K} \) the lower dimensional Bayes factor \( \BF_{10 \, ; \, \sigma_{0}^{2}}^{[K-1]}( \vec{s^{2}}) \) can be well approximated by a one-dimensional integral, because the type D Lauricella function in \( \BF_{10}( \vecB{s^{2}}) \) has a simplified one-dimensional integral representation due to \( \vecB{u}_{+} > 0 \).

Recall that \( \texttt{ss}_{j}= \nu_{j} s_{j}^{2} \) summarizes the observations of the \( j \)th sample. Observe also that the numerator of this limiting Bayes factor resembles the marginalized likelihood, i.e., \refEq{eqMarginalTau}, of the \( K-1 \) samples with their respective precisions \( \vec{\tau} =(\tau_{1}, \ldots, \tau_{K-1}) \) all fixed at \( 1/\sigma_{0}^{2} \). Hence, up to the factor \( (\sigma_{0}^{2})^{-\tfrac{\vec{\nu}_{+}}{2}} \) the denominator defines the marginal likelihood of the lower-dimensional null hypothesis \( \Hc_{0}^{K-1}: \tau_{j} = \sigma_{0}^{-2} \) for \( j \in [K-1] \) with \( \mu_{j} \propto 1 \). The missing factor is retrieved from the numerator by the change of variable \( \tau_{j} =  \tfrac{\vartheta_{j}}{ \sigma_{0}^{2} (1 - \vec{\vartheta}_{+}) } \) and yields the assertion above \refEq{eqScaledInverseDirichletPrior}.

The lower dimensional Bayes factor \( \BF_{10 \, ; \, \sigma_{0}^{2}}^{[K-1]}( \vec{s^{2}}) \) is in general hard to compute, because the Tricomi function \( \Uc(\vec{b} \, ; \, c \, ; \, \vec{x})  \) defines a \( K-1 \)-dimensional integral. \textcite{phillips1988characteristic} showed that if \( c < 1 \), the following simplification holds %
\begin{align}
\Uc(\vec{b} \, ; \, c \, ; \, \vec{x}) = \int_{0}^{\infty} e^{-t} t^{\vec{b}_{+} - c} \prod_{j=1}^{K-1} (t + x_{j})^{-b_{j}} \der t .
\end{align}
For \( \BF_{10 \, ; \, \sigma_{0}^{2}}^{[K-1]}( \vec{s^{2}}) \) this simplification holds whenever \( \vec{\nu}_{+} < 2 u_{K} \), which will be of little practical use when, for instance, \( u_{K} = 1/2 \). \refThm{thmAcrossSampleConsistency} now shows that for the case with \(  \vec{\nu}_{+} \geq 2 u_{K} \) the lower dimensional Bayes factor \( \BF_{10 \, ; \, \sigma_{0}^{2}}^{[K-1]}( \vec{s^{2}}) \) can be well approximated by a one-dimensional integral, because the type D Lauricella function in \( \BF_{10}( \vecB{s^{2}}) \) has a simplified one-dimensional integral representation due to \( \vecB{u}_{+} > 0 \).

\paragraph{Residual terms} To show that the convergence is at rate \( 1/\sqrt{n} \), we show that both \( \tilde{T}_{1}(n) \) and \( \tilde{T}_{2}(n) \) in \refEq{eqBfLimitSampleLaplaceGeneral} are of order 1. The analysis is analogous to showing the existence of \( \BF_{10}^{[K-1]} \). %Indeed, we show that the derivative with respect to \( x:=s^{2}_{K} \) as a sequence of \( n \) is uniformly bounded, which allows us to change the order of integral and limit. As before, the asymptotics do not depend on the prior.

For \( \tilde{T}_{1}(n) \) we study the derivative of the Bayes factor \( \BF_{10}^{[K]}(\vec{s^{2}}, x, n) \) with respect to \( x \). For this we swap the order of integration and differentiation and consider %
%
%As before the prior does not play a role in the asymptotics. %Let \( a_{j} = \tfrac{\nu_{j}}{2} \), \( b = \tfrac{n}{2} \) as before, \( \tilde{c}_{j} = \tfrac{\texttt{ss}_{j}}{n} \), \( \tilde{c}_{+} = \tfrac{\vec{\texttt{ss}}_{+}}{n} \), and consider
%
\begin{align}
%g & := \tfrac{ \partial }{ \partial x} \Big [ (1 + \tfrac{\vec{\texttt{ss}}_{+}}{n x} )^{a_{+}+b} \Big ( \prod_{j=1}^{K-1} \vartheta_{j}^{ a_{j}} \Big ) ( 1-\vec{\vartheta}_{+})^{ b } (1- \vec{\vartheta}_{+} + \sum_{j=1}^{K-1}  \tfrac{\tilde{c}_{j}  \vartheta_{j}}{x} )^{-(a_{+}+b)} \Big ] \Bigg |_{x= \sigma_{0}^{2}} .
g & := \tfrac{ \partial }{ \partial x} \tfrac{h(\vec{s^{2}}, x, n \, | \, \vec{\vartheta})}{h(\vec{s^{2}}, x, n \, | \, \vec{\vartheta}_{0})} \Big |_{x = \sigma_{0}^{2}} = g_{1} + g_{2} %
%\Big [ (1 + \tfrac{\tilde{c}_{+}}{x} )^{a_{+}+b} \Big ( \prod_{j=1}^{K-1} \vartheta_{j}^{ a_{j}} \Big ) ( 1-\vec{\vartheta}_{+})^{ b } (1- \vec{\vartheta}_{+} +   \tfrac{\sum_{j=1}^{K-1} \tilde{c}_{j}  \vartheta_{j}}{x} )^{-(a_{+}+b)} \Big ] \Bigg |_{x= \sigma_{0}^{2}} .
\end{align}
where
\begin{align}
\label{eqG1a}
g_{1} & := - \tfrac{\vec{\texttt{ss}}_{+}}{ n \sigma_{0}^{4} }  ( a_{+} + b)  (1 + \tfrac{\vec{\texttt{ss}}_{+}}{n \sigma_{0}^{2}} )^{a_{+}+b - 1} \\
\label{eqG1b}
& \times \Big ( \prod_{j=1}^{K-1} \vartheta_{j}^{ a_{j}} \Big ) ( 1-\vec{\vartheta}_{+})^{ b } (1- \sum_{j=1}^{K-1} [1 - \tfrac{\texttt{ss}_{j}}{n \sigma_{0}^{2}}] \vartheta_{j})^{-(a_{+}+b)} ,  \\
\label{eqG2a}
g_{2} & :=  \tfrac{1 }{n \sigma_{0}^{4}}  ( a_{+} + b) (1 + \tfrac{\vec{\texttt{ss}}_{+}}{n \sigma_{0}^{2} } )^{a_{+}+b } \\
\label{eqG2b}
& \sum_{k=1}^{K-1} \texttt{ss}_{k} \vartheta_{k} \Bigg [  \Big ( \prod_{j=1}^{K-1} \vartheta_{j}^{ a_{j}} \Big ) ( 1-\vec{\vartheta}_{+})^{ b } (1- \sum_{j=1}^{K-1} [1 - \tfrac{\texttt{ss}_{j}}{n \sigma_{0}^{2}}] \vartheta_{j})^{-a_{+}-1-b} \Bigg ] . %\\
\end{align}
Note that by definition of \( \vec{a}, b, \vec{\texttt{ss}} \) the terms \refEq{eqG1a} and \refEq{eqG2a} converge to \( - \tfrac{ \vec{\texttt{ss}}_{+} }{2 \sigma_{0}^{4} } e^{ \tfrac{\vec{\texttt{ss}}_{+}}{2 \sigma_{0}^{2}} } \) and \( \tfrac{ 1 }{2 \sigma_{0}^{4} } e^{ \tfrac{\vec{\texttt{ss}}_{+}}{2 \sigma_{0}^{2} } }  \), respectively. The proof that \refEq{eqG1b} is uniformly bounded in \( n \) is exactly as before. The same proof holds for each member in the sum of \refEq{eqG2b} by relabelling the power corresponding to \( \vartheta_{k} \) to \( a_{k} + 1 \). Hence, limit and integral can be interchanged and we conclude that the limiting integral exists. A computation as before shows that
\begin{align}
\label{eqT1}
\breve{T}_{1} := \lim_{n \rightarrow \infty} \Big ( \tfrac{ \partial}{ \partial x } \BF_{10}(\vec{s^{2}}, x, n) \Big ) \Big |_{x = \sigma_{0}^{2}} & = \frac{  \prod_{j=1}^{K-1} \Gamma ( \tfrac{ \nu_{j}}{2} + u_{j}) }{ 2 \Bc ( \vecB{u})  \sigma_{0}^{4} \exp(  - \tfrac{\vec{\texttt{ss}}_{+}}{2 \sigma_{0}^{2}} ) } G_{2} ,
\end{align}
where
\begin{align}
G_{2} & := \sum_{k=1}^{K-1} (\tfrac{ \nu_{k}}{2} + u_{k}) \Uc ( \tfrac{ \vec{\nu}}{2} + \vec{u} + \vec{e}_{k} \, ; \, \tfrac{ \vec{\nu}_{+} + 1}{2} - w + 1 \, ; \, \tfrac{ \vec{\texttt{ss}} } {2 \sigma_{0}^{2} }  )  \\
&   - \vec{\texttt{ss}}_{+} \Uc ( \tfrac{ \vec{\nu}}{2} + \vec{u}  \, ; \, \tfrac{ \vec{\nu}_{+} }{2} - w + 1 \, ; \, \tfrac{ \vec{\texttt{ss}} }{2 \sigma_{0}^{2}}  ) ,
\end{align}
%
%\begin{sloppypar}
where \( \vec{e}_{k} \in \R^{K-1} \) denotes the \( k \)th basis vector that is one at the \( k \)th entry and zero elsewhere. The analysis of the third order term %\( T_{3} := \lim_{n \rightarrow \infty} \Big ( \tfrac{ \partial^{2}}{ \partial x^{2} } \BF_{10}(\vec{\texttt{ss}},x, n) \Big ) \Big |_{x = \sigma_{0}^{2}} \) %
is a repeat of that of \( \breve{T}_{1} \) and implies that the last term in \refEq{eqBfLimitSampleLaplaceGeneral} is indeed \( o_{P}(n^{-\tfrac{1}{2}}) \) and the result follows.
\end{proof}

If \( Y_{Ki} \) has four moments, then \( S_{K}^{2} \) is asymptotically normal. In particular, for normal data this explicitly means \( \sqrt{n} ( S_{K}^{2} - \tau^{-1}) \inDist \Nc(0, 2 \tau^{-2}) \) and implies the following result.

%\begin{corr}[Asymptotic normality across samples]
%For \( Y_{ki} \sim \Nc(\mu_{k}, \sigma_{K}^{-1}) \) then as \( n_{K} \rightarrow \infty \) %converges of the Bayes factor
%%
%\begin{align}
%\sqrt{n_{K}} \Big ( \BF_{10}^{[K]}(y^{[K]}) - \BF_{10 \, ; \, \sigma_{K}^{2}}^{[K-1]}(y^{[K-1]}) \Big ) \inDist \Nc \Big (0, \frac{2 T_{2}^{2}}{\sigma_{K}^{4}} \Big )
%\end{align}
%%
%where \( T_{2} \) is given in \refEq{eqT2}. \( \hfill \diamond \)
%\end{corr}

\begin{proof}[Proof of asymptotic normality across-samples]
A rewrite of \refEq{eqKToKMinOne0} shows that \( n_{K} \)
\begin{align}
\sqrt{n_{K}} \Big ( \BF_{10}^{[K]}(y^{[K]}) - \BF_{10 \, ; \, \sigma_{K}^{2}}^{[K-1]}(y^{[K-1]}) \Big ) & = \sqrt{n_{K}} \Big ( S_{K}^{2} - \tau^{-1} \Big ) \tilde{T}_{2}(n_{K}) \\
& + o_{P}(1) \tilde{T}_{3}(n_{K}) .
\end{align}
A series expansion of \( \tilde{T}_{2}(n_{K}) \) in \( n_{K} \) shows that \( \tilde{T}_{2}(n_{K}) = T_{2} + \tfrac{1}{n} \breve{T}_{2} + \Oc(\tfrac{1}{n_{K}^{2}}) \) and the result follows. The term \( \breve{T}_{2} \) can be derived explicitly as was done in the proof of the previous theorem, but does not matter for the assertion, but its presence reveals a finite sample \( \Oc(n_{K}^{-1/2}) \) bias that vanishes as \( n_{K} \rightarrow \infty \).
\end{proof}

The proof of across sample consistency relies on the following lemma.

\begin{lemma}[Maximum of the marginalized likelihood]
\label{lemMaxReducedLikelihood}
If \( \vec{a}, \vec{c}, \vec{\vartheta} \in \R^{K-1} \) and \( b \in \R \) all positive and \( \vec{\vartheta}_{+} < 1 \), then
\begin{align}
f(\vec{a}, b, \vec{c} \, | \, \vec{\vartheta}) = \Big ( \prod_{j=1}^{K-1} \vartheta_{j}^{ a_{j}} \Big ) ( 1-\vec{\vartheta}_{+})^{ b } (1- \sum_{j=1}^{K-1} [1 - c_{j}] \vartheta_{j})^{-(a_{+}+b)} ,
\end{align}
attains its maximum at
\begin{align}
\hat{\vartheta}_{k} & = \frac{ a_{k} \prod_{j \neq k}^{K-1} c_{j} }{ b \prod_{j = 1}^{K-1} c_{j}  + \sum_{i=1}^{K-1} a_{i} \prod_{j \neq i}^{K-1} c_{j}} ,
%
%c_{\times}[-i] }, %= \frac{ \nu_{k}
\end{align}
%
%%
%\begin{align}
%\hat{\vartheta}_{k} & = \frac{ a_{k} c_{\times}[-k]}{ b c_{\times}  + \sum_{i=1}^{K-1} a_{i} c_{\times}[-i] }, %= \frac{ \nu_{k} m_{\times}[-k]}{ \tau m_{\times}  + \sum_{i=1}^{K-1} \nu_{i} m_{\times}[-i] }
%\end{align}
%
where \( \prod_{j \neq k}^{K-1} c_{j}  \) denotes the product of the elements of \( \vec{c} \) with the \( k \)th element taken out. \( \hfill \diamond \)
%
%is except for the \( k \)th element of \( c \) is the product of all elements of \( c \). \( \hfill \diamond \)
%which reveals that the numerator is product of \( K \) terms and the numerator is a sum of \( K \).
%
%where we use a subscript \( \times \) to abbreviate the product over the elements of the vector, and where a negative index in the rectangular brackets denotes all elements of the vector except for that index. For instance, for \( c \in \R^{K-1} \) we have \( c_{\times} = \prod_{j=1}^{K-1} c_{j} \) and \( c_{\times}[-k] = \prod_{j \neq k}^{K-1} c_{j} \). \( \hfill \diamond \)
\end{lemma}

\begin{proof}
Recall that the maximum is invariant under smooth transformations, which allows us to study the problem in the parametrisation \( \vec{\xi} = (\xi_{1}, \ldots, \xi_{K-1}) \), where \( \vartheta_{j} = \tfrac{\xi_{j}}{1 + \vec{\xi}_{+}} \). The target function becomes
\begin{align}
f(\vec{a}, b, \vec{c} \, | \, \vec{\xi}) = \Big ( \prod_{j=1}^{K-1} \xi_{j}^{ a_{j}} \Big )  (1 +  \sum_{j=1}^{K-1}  c_{j} \xi_{j})^{-(a_{+}+b)}  ,
\end{align}
and a direct computation shows that its gradient consists of elements
\begin{align}
\tfrac{ \partial}{\partial \xi_{k}} f(\vec{a}, b, \vec{c} \, | \, \vec{\xi}) = f(\vec{a}, b, \vec{c} \, | \, \vec{\xi})  \Big [ \tfrac{ a_{k}}{\xi_{k}} - \tfrac{ (a_{+} + b) c_{k} }{1 + \sum_{j=1}^{K} c_{j} \xi_{j} } \Big ] .
\end{align}
It is now easy to verify that for \( \hat{\xi}= ( \hat{\xi}_{1}, \ldots, \hat{\xi}_{K-1}) \) with \( \hat{\xi}_{k} = \tfrac{ a_{k}}{b c_{k}} \) the vector of partial derivatives is zero. Straightforward calculations show that for \( k \neq l \in [K-1] \) that
\begin{align}
\tfrac{ \partial^{2}}{\partial \xi_{k} \partial \xi_{l}} f(\vec{a}, b, \vec{c} \, | \, \vec{\xi}) = f(\vec{a}, b, \vec{c} \, | \, \vec{\xi})  \Big [ \tfrac{ a_{k}}{\xi_{k}} - \tfrac{ (a_{+} + b) c_{k} }{1 + \sum_{j=1}^{K} c_{j} \xi_{j} } \Big ]_{\vec{\xi} = \hat{\xi}} = \tfrac{ b^{2} c_{k} c_{l}}{a_{+} + b}
\end{align}
and for \( k \in [K-1] \)
\begin{align}
\tfrac{ \partial^{2}}{\partial \xi_{k}^{2}} f(\vec{a}, b, \vec{c} \, | \, \vec{\xi}) = f(\vec{a}, b, \vec{c} \, | \, \vec{\xi})  \Big [ \tfrac{ a_{k}}{\xi_{k}} - \tfrac{ (a_{+} + b) c_{k} }{1 + \sum_{j=1}^{K} c_{j} \xi_{j} } \Big ]_{\vec{\xi} = \hat{\xi}} = - \tfrac{ (b c_{k})^{2} (a[-k]_{+} +b)}{a_{k} (a_{+} + b)} ,
\end{align}
from which we conclude that \( \hat{\xi} \) is a maximum. The transformation \( \hat{\vartheta} = \tfrac{ \hat{\xi}_{k}}{1 + \hat{\xi}_{+}} \) yields the results.
\end{proof}

\color{black}
\section{Analysis Code} \label{sec:analysis-code}
Here, we provide the code for all examples given in the main text.

%\begin{minted}{R}
\begin{lstlisting}[style=mystyle]
devtools::install_github('fdabl/bfvartest', build_vignettes = TRUE)
library('bfvartest')

# 5.1 Sex Differences in Personality
twosd_test(n1 = 969, n2 = 716, sd1 = sqrt(15.6), sd2 = sqrt(19.9), u = 0.50)

# 5.2 Testing Against a Single Value
x <- c(6.2, 5.8, 5.7, 6.3, 5.9, 5.8, 6.0)
n <- length(x)
sd_x <- sd(x) # use rounded 0.22 in the paper

## (i) BF_{+0}
onesd_test(
    n = n, s = sd_x, popsd = sqrt(0.10),
    u = 0.50, alternative_interval = c(1, Inf), log = FALSE
)

## (ii) BF_{10}
onesd_test(
    n = n, s = sd_x, popsd = sqrt(0.10),
    u = 0.50, alternative_interval = c(0, Inf), log = FALSE
)

## (iii) BF_{+0} informed
onesd_test(
    n = n, s = sd_x, popsd = sqrt(0.10),
    u = 2.16, alternative_interval = c(1, Inf), log = FALSE
)

# 5.3 Comparing Measurement Precision
n <- 990
sdigit <- 0.98
slaser <- 0.89

## (i) BF_{+0}
twosd_test(
    n1 = n, n2 = n, sd1 = slaser, sd2 = sdigit,
    u = 0.50, alternative_interval = c(1, Inf), log = FALSE
)

## (ii) BF'_{0+} non-overlapping interval
1 / twosd_test(
    n1 = n, n2 = n, sd1 = slaser, sd2 = sdigit, u = 0.50, log = FALSE,
    null_interval = c(0.90, 1.10), alternative_interval = c(1.10, Inf)
)

# 5.4 The "Standardization" Hypothesis in Archeology
ns <- c(117, 171, 55)
sds <- c(12.74, 8.13, 5.83)
hyp <- c('1=2=3', '1,2,3', '1>2>3')
res <- ksd_test(hyp = hyp, ns = ns, sds = sds, u = 0.50, iter = 6000)
res$BF

# 5.5 Increased Variability in Mathematical Ability
ns <- c(3280, 6007, 7549, 9160, 9395, 6410)
sds <- c(5.99, 5.39, 4.97, 4.62, 3.69, 3.08)
hyp <- c('1=2=3=4=5=6', '1,2,3,4,5,6', '1>2>3>4>5>6')
res <- ksd_test(hyp = hyp, ns = ns, sds = sds, u = 0.50, iter = 6000)
res$BF
\end{lstlisting}

\end{document}